\documentclass[%
 reprint,
nofootinbib,
 amsmath,amssymb,
 aps,
]{revtex4-2}
\usepackage{graphicx}
\usepackage{dcolumn}
\usepackage{bm}
\usepackage{hyperref}

\usepackage{orcidlink}

\usepackage{physics}

\usepackage[normalem]{ulem} 

\usepackage{aas_macros}

\usepackage{subfigure}

\usepackage{booktabs} 
\usepackage{siunitx}  
\usepackage{caption} 
\usepackage{enumerate}

\usepackage{tikz}
\usepackage{xcolor}
\definecolor{RED}{rgb}{1,0,0} 

\usepackage{ragged2e} 

\usepackage{ulem} 

\usepackage[utf8]{inputenc}
\newcommand{\rei}[1]{}

\newcommand{\proof}[1]{\textcolor{black}{#1}}

\begin{document}

\preprint{APS/123-QED}

\title{Induced Compton scattering in magnetized electron and positron pair plasma}

\author{Rei Nishiura\orcidlink{0009-0003-8209-5030}}
 \email{nishiura@tap.scphys.kyoto-u.ac.jp}
 \affiliation{%
 Department of Physics, Kyoto University, Kyoto 606-8502, Japan}%
\author{Shoma F. Kamijima\orcidlink{0000-0002-4821-170X}}%
 \email{shoma.kamijima@yukawa.kyoto-u.ac.jp}
\affiliation{%
 Center for Gravitational Physics and Quantum Information, 
 Yukawa Institute for Theoretical Physics, Kyoto University, Kyoto 606-8502, Japan}%
\author{Masanori Iwamoto\orcidlink{0000-0003-2255-5229}}%
 \email{masanori.iwamoto@yukawa.kyoto-u.ac.jp}
\affiliation{%
 Center for Gravitational Physics and Quantum Information, 
 Yukawa Institute for Theoretical Physics, Kyoto University, Kyoto 606-8502, Japan}%
 \author{Kunihito Ioka\orcidlink{0000-0002-3517-1956}}%
 \email{kunihito.ioka@yukawa.kyoto-u.ac.jp}
\affiliation{%
 Center for Gravitational Physics and Quantum Information, 
 Yukawa Institute for Theoretical Physics, Kyoto University, Kyoto 606-8502, Japan}%

%
%
\date{\today}

\begin{abstract}
A formulation for the parametric instability of electromagnetic (EM)
waves in magnetized pair plasma is developed. The
linear growth rate of induced Compton scattering is derived
analytically for 
frequencies below the cyclotron frequency for the first time.
We identify three modes of
density fluctuation: \proof{\textit{ordinary, charged, and neutral modes}}. In the
charged mode, the ponderomotive force separates charges (electrons and
positrons) longitudinally, in contrast to the nonmagnetized case. 
We also
recognize
two effects
that significantly reduce the scattering rate for 
waves polarized perpendicular to the magnetic field: (1) the
\proof{\textit{gyroradius effect}} due to 
the magnetic suppression of particle orbits
and (2) \rei{Debye screening} \proof{\it{Debye screening effect}} for
wavelengths larger than the Debye length. Applying this to fast radio
bursts (FRBs), we find that these effects facilitate the escape of
X-mode waves from 
the magnetosphere and outflow of a magnetar and neutron star,
enabling 100\% polarization as observed. Our
formulation provides a foundation for consistently addressing the
nonlinear interaction of EM waves with magnetized plasma in
astrophysics and laser physics.
\end{abstract}

\maketitle


\section{INTRODUCTION} 
Nonlinear interactions between electromagnetic (EM) waves and plasma are actively studied in various fields. The stability of these phenomena is often determined by parametric instability \citep{1974PhFl...17..778D,1979PhFl...22.1115C}. Parametric instability refers to a general class of instabilities where an incident wave (or pump wave) decays into multiple new waves (daughter waves). This process transfers energy from the incident wave to the daughter waves \proof{and plasma}. As a result, the incident wave loses energy.
Parametric instabilities include induced (or stimulated) Raman scattering, induced Brillouin scattering, induced Compton scattering, modulation instability, filamentation instability, etc. These interactions are relevant to astrophysical contexts, including the Sun \citep{1963SPhD....7..988G,1966PhFl....9.1483B,1972JPlPh...8..197B,1978ApJ...224.1013D,1990JGR....9510525I,1993JGR....9813247J,1994JGR....9923431H,2001A&A...367..705D,2006PhPl...13l4501N,2015JPlPh..81a3202D,2017ApJ...842...63S,2022RvMPP...6...22N}, the Earth's ionosphere \citep{1972JGR....77.1242C,1974JGR....79.1478P,1982PhRvL..49.1561T,1989PhR...179...79R,1990JGR....9521221D}, active galactic nuclei \citep{1971SvA....15..190S,1993MNRAS.262..603C}, pulsars \citep{1973PhFl...16.1480M,1976MNRAS.174...59B,1978MNRAS.185..297W,1982MNRAS.200..881W,1996AstL...22..399L}, and fast radio bursts (FRBs) \citep{2008ApJ...682.1443L,2023MNRAS.522.2133I,2024PhRvE.110a5205I}. They are also crucial in experimental plasma research such as laser nuclear fusion \citep{1973PhFl...16.1522K,1974PhRvL..33..209M,1975PhFl...18.1002F,1994PhPl....1.1626T,1996PhRvL..77.2483D} and laser acceleration \citep{1979PhRvL..43..267T,2004Natur.431..541F,2009RvMP...81.1229E}.

FRBs, first discovered in 2007~\citep{2007Sci...318..777L,2013Sci...341...53T,2019A&ARv..27....4P}, are the brightest radio transients in the Universe. Since their discovery, extensive research has been conducted both observationally and theoretically. FRBs are also used for probing cosmology \citep{2003ApJ...598L..79I,2004MNRAS.348..999I,2020Natur.581..391M}. Most FRBs originate from distant galaxies, and their progenitors remain largely unidentified. However, a significant breakthrough occurred in 2020 when an FRB 20200428 was detected from the magnetar (strongly magnetized neutron star) SGR 1935+2154 within our Galaxy, coinciding with an X-ray bursts \citep{2020Natur.587...54C,2020Natur.587...59B,2020ATel13687....1Z,2020ApJ...898L..29M,2020ATel13686....1T,2020ATel13688....1R}. This discovery provided compelling evidence linking magnetars to FRBs.

The emission region is proposed in two different models: within the magnetar's magnetosphere \citep{2014PhRvD..89j3009K,2016MNRAS.457..232C,2016MNRAS.462..941L,2017MNRAS.468.2726K,2017ApJ...836L..32Z,2018MNRAS.477.2470L,2018A&A...613A..61G,2018ApJ...868...31Y,2020ApJ...893L..26I,2020MNRAS.494.2385K,2021MNRAS.508L..32C,2021ApJ...922..166L,2024ApJ...972..124Q}
and shocks outside the magnetosphere \citep{2014MNRAS.442L...9L,2016MNRAS.461.1498M,2017ApJ...842...34W,2020MNRAS.494.4627M,2019MNRAS.485.4091M,2020ApJ...896..142B,2021MNRAS.500.2704Y}, leading to great controversy \citep{2020MNRAS.498.1397L,2020arXiv200505093L,2020ApJ...900L..21Y,2021MNRAS.500.2704Y,2024PhRvE.110a5205I}. Observations imply some issues of the gamma-ray burst-like shock model \citep{2022Natur.607..256C,2022NatAs...6..393N}. Although the pulsar-like magnetosphere model is gaining observational support, it is argued that coherent radiation, such as FRBs and pulsar emissions generated within the magnetar magnetosphere, undergoes scattering with magnetospheric plasma. Identifying the emission region of FRBs has become one of the most critical issues in astrophysics.

One of the most essential scattering processes for FRBs from magnetars is induced Compton scattering with magnetospheric $e^\pm$ plasma. The scattering attenuates radiation, preventing the escape from the magnetosphere, unless the Lorentz factor of the scattering particles is highly relativistic \citep{1975Ap&SS..36..303B,1978MNRAS.185..297W,1982MNRAS.200..881W,2008ApJ...682.1443L}. The scattering also affects the spectrum and polarization.

Induced Compton scattering can be understood in two ways: a classical interpretation as a kind of kinetic parametric instability \citep{1974PhFl...17..778D}, and a quantum mechanical interpretation with stimulated emission in Compton scattering \citep{1975Ap&SS..36..303B,1978MNRAS.185..297W,1982MNRAS.200..881W,2008ApJ...682.1443L}. Both interpretations are recognized as the same fundamental process \citep{2022ApJ...930..106G}. Classically, the interaction of incident and scattered EM waves generates plasma density fluctuations, transferring the incident wave's energy into scattered waves and density fluctuations via Landau resonance\footnote{In this study, induced Compton scattering is treated as a distinct process from induced Brillouin scattering. Induced Brillouin scattering is interpreted as the interaction in the beat of the incident and scattered waves producing an acoustic wave, and the density fluctuation of the acoustic wave scatters the incident wave. In $e^\pm$ plasma, the acoustic wave is a quasi-particle, intrinsically unstable via Landau resonance, so induced Brillouin scattering occurs simultaneously with induced Compton scattering. Therefore, induced Compton scattering is sometimes referred to as the kinetic effect of induced Brillouin scattering \citep{2017PhRvE..96e3204S}.}. Quantum mechanically, when the occupation number of incident photons is large (coherent light), stimulated emission amplifies the growth rate of Compton scattering. In this study, we discuss induced Compton scattering using the classical interpretation.

 It is essential to accurately account for the effects of strongly magnetized $e^\pm$ plasma, when considering parametric instability, including induced Compton scattering, in a neutron star's magnetosphere. In a strong \proof{background }magnetic field, the trajectories of charged particles change significantly, fundamentally altering plasma diagnostics. 
 Previous
 studies have examined the propagation of FRBs using parametric instability theories that do not consider \proof{background }magnetic fields. It has been shown that, in Thomson scattering, the motion of scattering particles is constrained by the \proof{background }magnetic field, resulting in the suppression of the scattering cross section for EM waves polarized perpendicular to the \proof{background }magnetic field by the inverse square of the cyclotron frequency \citep{1971PhRvD...3.2303C,1979PhRvD..19.2868H,1979PhRvD..19.1684V,1992herm.book.....M,2000ApJ...540..907G} \footnote{In this study, we refer to this as the gyroradius effect.}. This gyroradius effect is suggested to apply similarly to induced Compton scattering \citep{1976MNRAS.174...59B,2020MNRAS.494.1217K,2024PhRvD.109d3048N}. However, Thomson scattering is based on the scattering of \rei{individual} \proof{test} particles, and it is not sure that the same effects apply to induced Compton scattering, where the collective motion of the plasma becomes significant. This aspect still needs to be thoroughly discussed.

\citet{2024PhRvD.109d3048N}
has shown that Thomson scattering occurs even in highly magnetized $e^\pm$ plasma as collective Thomson scattering
. In a strong \proof{background }magnetic field, the drift motion of charged particles becomes dominant compared to their motion in the direction of the electric field of the incident EM wave. At first glance, it appears that the currents of electrons and positrons cancel each other out, resulting in no scattering. However, since each particle in the plasma statistically has a different initial phase, this cancellation does not occur. Therefore, the Thomson cross section per particle is of the same order as single-particle scattering.

The parametric instabilities, including induced Compton scattering, in magnetized $e^\pm$ plasma have yet to be formulated considering the collective motion
(e.g., \citep{1979PhFl...22.1089K,1980PhFl...23.2376F,1986PhR...130..143L,2021ApJ...922..166L,2024ApJ...972..124Q}). Related kinetic parametric instability in magnetic fields has only been discussed under limited conditions. For example, in solar coronal heating and solar wind acceleration, parametric instability has been discussed using the drift-kinetic equation (or gyro-kinetic equation), which averages out particle motions perpendicular to the magnetic field \citep{1990JGR....9510525I}. However, this approximate kinetic parametric instability theory, used in solar physics, cannot describe X-mode wave decay or scattering of Alfvén waves at arbitrary scattering angles for FRB contexts \footnote{In $e^{\pm}$ plasma, an X-mode wave is a linearly polarized EM wave whose electric field component is perpendicular to the plane formed by the background magnetic field and the wave's propagation direction. The other O-mode wave and Alfvén wave have electric field components within this plane.}.

This study formulates induced Compton scattering in magnetized $e^\pm$ plasma. Specifically, we analytically and quantitatively derive how much the linear growth rate is suppressed compared to the nonmagnetized case. Our formulation also has wide applications: the density fluctuation caused by the beat of the incident and scattered waves, in equation \eqref{eq:density_fluctuation_pondero}, can be applied to other parametric instabilities such as induced Brillouin scattering, induced Raman scattering, and so on. 

In Section \ref{sec:Thomson_free}, we derive equation \eqref{eq:density_fluctuation_pondero}, which describes the density fluctuations caused by the ponderomotive force (time-averaged nonlinear force of EM fields) generated by the beat of incident and scattered waves in magnetized $e^\pm$ plasma. Next, in Section \ref{sec:induced_Compton_parallel}, we analytically determine the maximum linear growth rate of induced Compton scattering for waves polarized parallel to the \proof{background }magnetic field (e.g., O-mode waves). Furthermore, in Section \ref{sec:induced_Compton_perpendicular}, we analytically derive the maximum linear growth rate of induced Compton scattering for waves polarized perpendicular to the \proof{background }magnetic field (e.g., X-mode waves) under strong \proof{background }magnetic field conditions.

Throughout this paper, the notation \( A = 10^n A_n \) and the Centimeter-Gram-Second (CGS) system of units are consistently employed. Additionally, it is essential to note that the italic symbol $e$ represents the magnitude of the electron charge. At the same time, the roman type $\mathrm{e}$ denotes the base of the natural logarithm, $\mathrm{exp}(1)$.

\section{INDUCED COMPTON SCATTERING IN MAGNETIZED ELECTRON-POSITRON PLASMA}
\label{sec:INDUCED_COMPTON_SCATTERING_IN_MAGNETIZED_ELECTRON-POSITRON_PLASMA}
\subsection{Basic equations}
\label{sec:Thomson_free}
We calculate induced Compton scattering rate in strongly magnetized $e^\pm$ plasma. We impose the following assumptions.
\begin{enumerate}[(i)]
\item A uniform magnetic field $\bm{B}_0=(B_0,0,0)$ exists in the $x$-axis direction.
\item The vector potential of EM wave $\bm{A}_0$ is incident with a wave vector \(\bm{k}_0\) and an angular frequency \(\omega_0\), and the scattered wave $\bm{A}_1$ has a wave vector \(\bm{k}_1\) and an angular frequency \(\omega_1\).
\item The incident wave and scattered waves generated through parametric instability are assumed to be monochromatic. The case where the incident and scattered waves have a broadband spectrum will be discussed in Section \ref{sec:discussion_induced_Compton_magnetized}.
Additionally, the Fourier coefficient \(\widetilde{X}(\bm{k}, \omega)\) of a physical quantity \(X(\bm{r}, t)\) is defined as follows:
\begin{equation}
    X(\bm{r}, t) = \mathrm{e}^{\mathrm{i}\left\{(\bm{k} \cdot \bm{r}-(\omega+\mathrm{i}\epsilon) t)\right\}} \widetilde{X}(\bm{k}, \omega)+\text{c.c.},
    \label{eq:Fourier_transformation_definition2}
\end{equation}
by taking the Landau contour, \(\epsilon\). For simplicity, the notation of the Landau contour will be omitted in the following discussion.
\item The magnetic field of the incident EM wave is assumed to be sufficiently small compared to the background magnetic field, that is $|\bm{B}_{\text{wave}}|\ll|\bm{B}_0|.$
\item Particles are non-relativistic. As will be derived later, the condition for incident EM waves polarized parallel to the background magnetic field is described by equation \eqref{eq:strength_parameter_constraint_nonrela_para}. For EM waves polarized perpendicular to the background magnetic field, the condition is given by equation \eqref{eq:strength_parameter_constraint_nonrela_perp}.
\item The initial one-particle distribution functions for electrons and positrons are assumed to be uniform, denoted as \( f_{0\pm} = f_{0\pm}(\bm{v}) \), before scattering. Additionally, we assume that the uniform electron density $n_{0-}$ and the uniform positron density $n_{0+}$ are equal. The $e^\pm$ plasma uniform density is defined as
\begin{equation}
n_{\text{e}0}\equiv n_{0+}=n_{0-}.     
\end{equation}
\end{enumerate}
\begin{figure*}
\centering
\includegraphics[width=\textwidth]{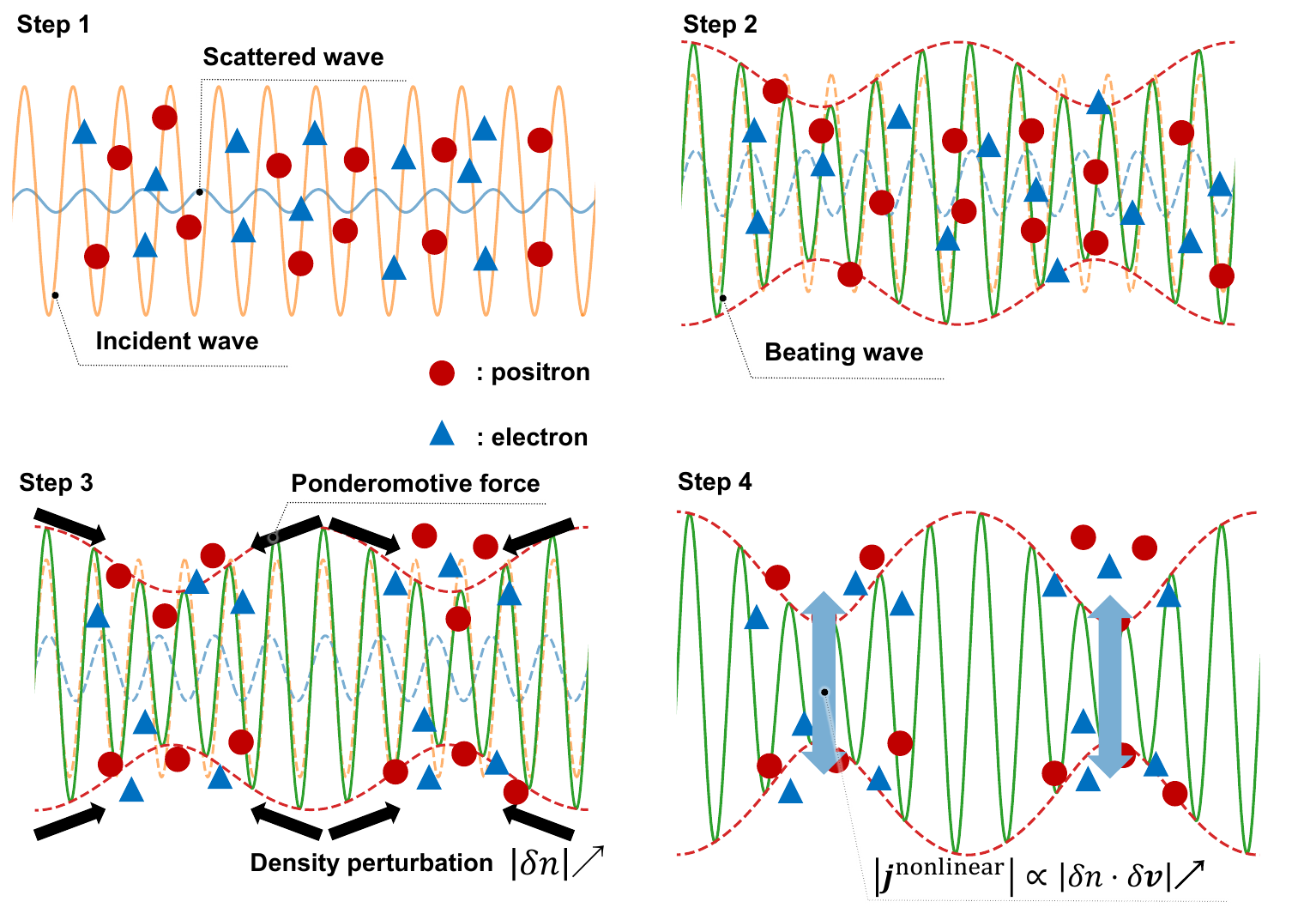}
\caption[Concept of Parametric Decay Instability]{\justifying Schematic picture of parametric instability in $e^\pm$ plasma without a background magnetic field. Step 1: An incident wave (orange solid line) enters the plasma. Within the plasma, there are positrons (red circles) and electrons (blue triangles), as well as scattered (or initially background) waves (blue solid line). Step 2: The incident wave (orange dashed line) interferes with the scattered wave (blue dashed line), producing a beating wave (green solid line). This interference causes periodic changes in the beating wave amplitude. Step 3: The ponderomotive force acts on electrons and positrons in the beating wave. This force, independent of the charge sign, pushes particles toward a weaker field. Consequently, positrons and electrons accumulate in regions of lower wave amplitude, amplifying the density fluctuations. Step 4: The amplified density fluctuations enhance the nonlinear current, amplifying the scattered waves. This further intensifies the wave beating. Throughout these processes, the energy of the incident wave is efficiently transferred to the scattered waves and density fluctuations. For detailed physical interpretations with a background magnetic field, refer to Figures \ref{fig:density_perturbation_nomg}, \ref{fig:charged_mode}, and \ref{fig:neutral_mode}.}
\label{fig:parametric_instability}
\end{figure*}

EM waves are described by the wave equation using the vector potential $\bm{A}$,
\begin{equation}
\frac{\partial^{2} \bm{A}}{\partial t^{2}} - c^{2} \Delta \bm{A} = 4 \pi c \bm{j}.
\label{eq:wave_eq_vector}
\end{equation}
Here, we use the Lorenz gauge, represented by \(\nabla\cdot\bm{A}+(1/c)(\partial\phi_{\text{e}}/\partial t)=0\). The following discussion considers transverse EM waves, denoted by \(\bm{k}\cdot\bm{A}=0\), i.e., \(\phi_{\text{e}}=0\).\footnote{\proof{
Our theoretical framework cannot describe parametric instability for longitudinal incident or scattered waves. It is fully applicable when both the incident and scattered waves are in the X-mode. For the O-mode, it applies to nearly parallel and nearly perpendicular propagation, while for the Alfvén wave, it is limited to nearly parallel propagation. Therefore, we here focus on scattering of transverse components for these waves.}
}
The vector potential is expressed as the superposition of the incident wave and the scattered wave,
\begin{equation}
\bm{A}(\bm{r}, t) = \bm{A}_{0} \mathrm{e}^{\mathrm{i}\left(\bm{k_{0}} \cdot \bm{r} - \omega_{0} t\right)} + \bm{A}_{1} \mathrm{e}^{\mathrm{i}\left(\bm{k_{1}} \cdot \bm{r} - \omega_{1} t\right)} + \text{c.c.},
\label{eq:incident_and_scattered_wave}
\end{equation}
where the term $\text{c.c.}$ denotes the complex conjugate.
$\bm{A}_{0}$ and $\bm{A}_{1}$ are defined as the complex amplitudes. The source current of the wave equation \eqref{eq:wave_eq_vector} is produced by particles in the $e^\pm$ plasma,
\begin{equation}
\bm{j} = \sum_{q= \pm e} q n_{\pm}(\bm{r}, t) \bm{v}_{\pm}(\bm{r}, t),
\label{eq:current_oomoto}
\end{equation}
where $q = -e$ represents the charge of an electron.  The equation of motion for charged particles under the background magnetic field is represented by
\begin{equation}
\frac{\dd \bm{v}_{\pm}}{\dd t} = \pm \frac{e}{m_{\text{e}}}\left(\bm{E} + \frac{\bm{v}_{\pm} \times \bm{B}_{0}}{c}\right).
\label{eq:EOM_oomoto_induced_Compton_magnetized}
\end{equation}
The oscillatory solution of the charged particles due to the incident wave is expanded up to the first order in $v/c$. \rei{Here, we consider the case where the amplitude of the scattered wave $\bm{A}_1$ is sufficiently smaller than the amplitude of the incident wave $\bm{A}_0$.} $\bm{v}_{0\pm}^{(1)}$ is defined as the Fourier coefficient of $\bm{v}_{\pm}$ with respect to $(\bm{k}_0, \omega_0)$. \rei{$\bm{v}_{\pm}$ can be written by using $\bm{v}_{0\pm}^{(1)}$,}
Then, $\bm{v}_{0\pm}^{(1)}$ is given by
\begin{equation}
\begin{aligned}
\bm{v}_{0\pm}^{(1)} = &\mp \frac{e}{m_{\text{e}} c} \bm{A}_{0\|} \mp \frac{e}{m_{\text{e}} c} \frac{\omega_{0}^{2}}{\omega_{0}^{2} - \omega_{\text{c}}^{2}} \bm{A}_{0\perp} \\
& -\mathrm{i} \frac{e}{m_{\text{e}} c} \frac{\omega_{0} \omega_{\text{c}}}{\omega_{0}^{2} - \omega_{\text{c}}^{2}} \bm{A}_0 \times \hat{\bm{B}_{0}},
\end{aligned}
\label{eq:general_velocity}
\end{equation}
where $\bm{A}_{0\|}$ and $\bm{A}_{0\perp}$ denote the components of the vector potential parallel and perpendicular to the \proof{background }magnetic field, respectively, such that $\bm{A}_{0} = \bm{A}_{0\|} + \bm{A}_{0\perp}$. Moreover, the cyclotron frequency is defined as
\begin{equation}
\omega_{\text{c}\pm} \equiv \pm \frac{e B_0}{m_{\text{e}} c}, \quad \omega_{\text{c}} \equiv \omega_{\text{c}+} = -\omega_{\text{c}-}.
\end{equation}
The non-relativistic conditions under which equation \eqref{eq:general_velocity} is applicable will be discussed in a later section (The conditions are represented by equations \eqref{eq:strength_parameter_constraint_nonrela_para}, \eqref{eq:strength_parameter_constraint_nonrela_perp}, and \eqref{eq:strength_parameter_constraint_nonrela_neutral}.). In equation \eqref{eq:general_velocity}, the first term corresponds to the oscillation of particles by the electric field component parallel to the background magnetic field. The second term represents the particles' polarization drift (a kind of inertial drift). The third term means the particles' $\bm{E}\times\bm{B}$ drift.

As shown in Figure \ref{fig:parametric_instability}, through parametric instability, the beat between the incident and the scattered EM wave\proof{s} produce\rei{s} plasma density fluctuations $\delta n$. Induced Compton scattering is described as the process in which the energy of the incident wave is converted into the scattered wave and the density fluctuation through Landau resonance, involving particles moving at velocities comparable to the phase velocity of the beat. The angular frequency and the wave vector of the density fluctuation must satisfy the following,
\begin{equation}
\omega = \omega_{1} - \omega_{0}, \quad \bm{k} = \bm{k}_{1} - \bm{k}_{0}.
\label{eq:energy_momentum_conservation_Compton}
\end{equation}
In the following discussion, we consider the case where the scattered wave is a Stokes wave (\(\omega < 0\))\footnote{A scattered wave with \(\omega > 0\) is referred to as an anti-Stokes wave.}.
The plasma density fluctuation is described by the Vlasov equation,
\begin{equation}
\frac{\partial f_{\pm}}{\partial t} + \bm{v} \cdot \bm{\nabla} f_{\pm} + \bm{F} \cdot \frac{\partial f_{\pm}}{\partial \bm{p}} = 0,
\end{equation}
where the plasma distribution function is normalized such that $\int_{-\infty}^{+\infty} f_{\pm}(\bm{r}, \bm{v}, t) \dd^{3} \bm{v} = 1$. The fluctuation $\delta f_{\pm}$ from the uniform distribution function is defined as
\begin{equation}
f_{\pm}(\bm{r}, \bm{v}, t) = f_{0\pm}(\bm{v}) + \delta f_{\pm}(\bm{r}, \bm{v}, t).
\end{equation}
The external force acting on the particles is the slowly varying ponderomotive force due to the beating wave and the Lorentz force produced by the charge density fluctuations originating from the ponderomotive force,
\begin{equation}
\bm{F} = -\bm{\nabla} \phi_{\pm} \pm e \left( \bm{E} + \frac{\bm{v} \times \bm{B}_{0}}{c} \right).
\end{equation}
The electric field fluctuations in the plasma are described by Maxwell’s equation,
\begin{equation}
\bm{\nabla} \cdot \bm{E} = \sum_{q = \pm e} 4 \pi q n_{\text{e} 0} \int \delta f_{\pm} \dd^{3} \bm{v}.
\end{equation}
Assuming $\omega_{1} \sim \omega_{0} \gg |\omega|$, the ponderomotive potential in a uniform magnetic field in the $x$ direction can be expressed as \citep{osti_4516607,1968CzJPh..18.1280K,1977PhRvL..39..402C,1981PhFl...24.1238C,1981PhRvL..46..240H,10.1063/1.864196,1996GeoRL..23..327L}
\begin{equation}
\phi_{\pm} = \frac{e^{2}}{2 m_{\text{e}}} \left\langle \frac{E_{\|}^{2}}{\omega_{0}^{2}} - \frac{E_{\perp}^{2}}{\omega_{\text{c}}^{2} - \omega_{0}^{2}} + \mathrm{i} \frac{\omega_{\text{c}\pm} \left(E_{z}^{*} E_{y} - E_{y}^{*} E_{z}\right)}{\omega_{0} \left(\omega_{\text{c}}^{2} - \omega_{0}^{2}\right)} \right\rangle,
\label{eq:ponderomotive_potential}
\end{equation}
where the time average is taken over a period longer than $\omega_0^{-1}$ but shorter than $|\omega|^{-1}$, \footnote{The high frequency forces resulting from the incident and scattered waves ($ |\omega|\ll \omega_0,\omega_1$) are time-averaged to zero at the first order. The second-order terms yield the ponderomotive force due to the beating waves.}.

The density fluctuations are derived from the Vlasov equation for plasma. By expanding the distribution function using the fluctuating electric field $\bm{E}$ in the plasma as a perturbation, the equation for the fluctuation of the distribution function is given by
\begin{equation}
\begin{aligned}
    &\frac{\partial \delta f_{ \pm}}{\partial t}+\bm{v} \cdot \frac{\partial \delta f_{ \pm}}{\partial \bm{r}}+\left(-\frac{1}{m_{e}} \bm{\nabla} \phi_{ \pm} \pm \frac{e}{m_{e}} \bm{E}\right) \cdot \frac{\partial f_{0\pm}}{\partial \bm{v}}\\
    &\pm \frac{e}{m_{\text{e}} c}\left(\bm{v} \times \bm{B}_{0}\right) \cdot \frac{\partial \delta f_{ \pm}}{\partial \bm{v}}=0.
\end{aligned}
\label{eq:Vlasov_pondero}
\end{equation}
The fluctuation of the distribution function is Fourier transformed as equation \eqref{eq:Fourier_transformation_definition2}.
As shown in Appendix \ref{Ap:pondero_fluctuation}, we can solve the equation for the fluctuation of the distribution function \eqref{eq:Vlasov_pondero}. We obtain the Fourier components of the density fluctuations \eqref{eq:density_fluctuation_pondero_Appendix},
\begin{equation}
\begin{aligned}
    &\widetilde{\delta n_{ \pm}}(\bm{k},\omega) \\
    &= n_{\text{e}0} \int \dd^{3} \bm{v}~ \widetilde{\delta f_{ \pm}}(\bm{k},\bm{v}, \omega) \\
    &=-\frac{n_{\text{e} 0}}{m_{\text{e}}}\left\{\widetilde{\phi_{ \pm}}(\bm{k},\omega) \sum_{\ell=-\infty}^{+\infty} \int \dd^{3} \bm{v} \frac{J_{\ell}^{2}\left(k_{\perp} r_{\text{L}\pm}\right)\bm{k} \cdot \frac{\partial f_{0\pm}}{\partial \bm{v}^{*}}}{\omega-k_{\|} v_{\|} + \ell \omega_{\text{c}\mp}} \right\} \\
    &\pm \frac{n_{\text{e} 0}H_{ \pm}}{m_{\text{e}}\varepsilon_{\text{L}}}\left\{\widetilde{\phi_{+} }(\bm{k},\omega)\sum_{\ell=-\infty}^{+\infty} \int \dd^{3} \bm{v} \frac{J_{\ell}^{2}\left(k_{\perp} r_{\text{L}+}\right)\bm{k} \cdot \frac{\partial f_{0\pm}}{\partial \bm{v}^{*}}}{\omega-k_{\|} v_{\|}+\ell \omega_{\text{c}-}} \right. \\
    &\left.\quad -\widetilde{\phi_{-}}(\bm{k},\omega)\sum_{\ell=-\infty}^{+\infty} \int \dd^{3} \bm{v} \frac{J_{\ell}^{2}\left(k_{\perp} r_{\text{L}-}\right)\bm{k} \cdot \frac{\partial f_{0\pm}}{\partial \bm{v}^{*}}}{\omega-k_{\|} v_{\|}+\ell \omega_{\text{c}+}} \right\},
\end{aligned}
\label{eq:density_fluctuation_pondero}
\end{equation}
where $J_\ell(z)$ is a Besel function, and
\begin{equation}
    k_{\|}\equiv k_x,\quad k_{\perp}\equiv\sqrt{k_y^2+k_z^2}.
\end{equation} 
The gyroradius is given by
\begin{equation}
 r_{\text{L}\pm}\equiv\frac{v_{\perp}}{\omega_{\text{c}\pm}},\quad  r_{\text{L}}\equiv r_{\text{L}+} =-r_{\text{L}-},
\end{equation}
and the differential operator is defined as
\begin{equation}
    \bm{k} \cdot \frac{\partial f_{0\pm}}{\partial \bm{v}^{*}} \equiv k_{\|} \frac{\partial f_{0\pm}}{\partial v_{\| }} + \frac{\ell \omega_{\text{c}\pm}}{v_{\perp}} \frac{\partial f_{0\pm}}{\partial v_{\perp}}.
\end{equation}
$H_{\pm}$ is the longitudinal electric susceptibility for positrons/electrons, and $\varepsilon_{\text{L}}$ is the longitudinal dielectric constant, defined respectively as
\begin{equation}
    H_{ \pm} \equiv \int \dd^{3} \bm{v} \frac{4 \pi e^{2} n_{\text{e} 0}}{m_{\text{e}} k^{2}} \sum_{\ell=-\infty}^{+\infty} \frac{J_{\ell}^{2}\left(k_{\perp} r_{\text{L}\pm}\right)\bm{k} \cdot \frac{\partial f_{0\pm}}{\partial \bm{v}^{*}}}{\omega-k_{\|} v_{\|} + \ell \omega_{\text{c}\mp}},
    \label{eq:longitudinal_electric_susceptibility_induced_Compton}
\end{equation}
\begin{equation}
    \varepsilon_{\text{L}}(\bm{k}, \omega)=1+H_{+}(\bm{k}, \omega)+H_{-}(\bm{k}, \omega).
    \label{eq:longitudinal_dielectric_constant_induced_Compton}
\end{equation}

Assuming a Maxwellian distribution for the zeroth-order distribution function, the velocity integral \(\int \dd^3\bm{v}\) in the density fluctuation \eqref{eq:density_fluctuation_pondero}, and the longitudinal electric susceptibility \eqref{eq:longitudinal_electric_susceptibility}, can be performed analytically.
We assume the following zeroth-order distribution function,
\proof{\begin{equation}
f_{0+}=f_{0-}= \frac{1}{(\pi v_{\text{th}}^2)^{\frac{3}{2}}} \exp\left(-\frac{v_{\|}^2 + v_{\perp}^2}{v_{\text{th}}^2}\right)\equiv f_0,
\label{eq:Maxwellian_distribution_induced_Compton}
\end{equation}}
\begin{equation}
v_{\text{th}} \equiv \sqrt{\frac{2 k_{\text{B}} T_{\text{e}}}{m_{\text{e}}}},
\label{eq:thermal_velocity}
\end{equation}
where the particle motion is assumed to be non-relativistic and isotropic.
Then, the longitudinal electric susceptibilities for electrons and positrons are the same and expressed as an infinite sum of special functions,
\begin{equation}
    H_{+}(\bm{k}, \omega)=H_{-}(\bm{k}, \omega)=H(\bm{k}, \omega),
\end{equation}
\begin{equation}
\begin{aligned}
    H(\bm{k}, \omega) &= \frac{\omega_{\text{p}}^{2}}{k^{2} v_{\text{th}}^{2}}\left\{1+\frac{\omega}{k_{\|} v_{\text{th}}}\sum_{\ell=-\infty}^{+\infty} I_{\ell}\left[\frac{1}{2}\left(\frac{k_{\perp} v_{\text{th}}}{\omega_{\text{c}}}\right)^{2}\right] \right. \\
    &\quad \times  Z\left(\frac{\omega-\ell \omega_{\text{c}}}{k_{\|} v_{\text{th}}}\right) \mathrm{e}^{-\frac{1}{2}\left(\frac{k_{\perp} v_{\text{th}}}{\omega_{\text{c}}}\right)^{2}}\bigg\},
\end{aligned}
\label{eq:longitudinal_electric_susceptibility}
\end{equation}
where $I_\ell(z)$ is a modified Bessel function, $Z(\xi)$ is a plasma dispersion function,
\begin{equation}
    Z(\xi) \equiv \frac{1}{\sqrt{\pi}} \int_{-\infty}^{\infty} \frac{1}{z-\xi} e^{-z^2} \mathrm{~d} z,
\end{equation}
and
\begin{equation}
  \omega_{\text{p}}\equiv\sqrt{\frac{8\pi e^2n_{\text{e}0}}{m_{\text{e}}}},
  \label{eq:plasma_frequency_induced_Compton_magnetized}
\end{equation}
is the plasma frequency of $e^\pm$ plasma. For detailed derivations of equation \eqref{eq:longitudinal_electric_susceptibility}, see Chapter 10 of \citet{2012FuST...61..104F}.

In the next section, we will consider induced Compton scattering for the cases where the electric field component of the incident wave is polarized parallel to the \proof{background }magnetic field ($\bm{A}_{0\perp} = \bm{0}$) and perpendicular to the \proof{background }magnetic field ($\bm{A}_{0\|} = \bm{0}$).

\subsection{Scattering of EM waves polarized parallel to the background magnetic field}
\label{sec:induced_Compton_parallel}
\begin{figure*}
\centering
\includegraphics[width=\textwidth]{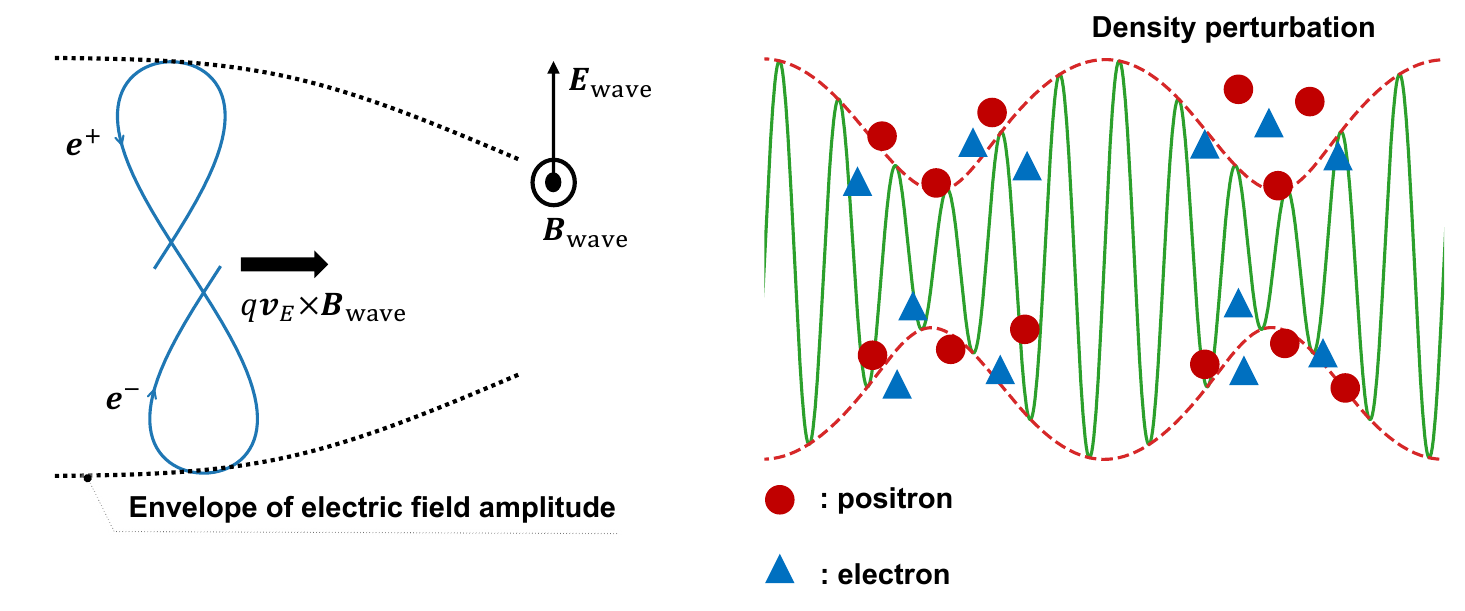}
\caption[Concept of Parametric Decay Instability]{\justifying The physical interpretation of ponderomotive force (left panel) and schematic picture of parametric instability (right panel) for the ``ordinary mode" for an incident wave polarized parallel to the background magnetic field, $\bm{A}_{0\perp} = \bm{0}$. In the left panel, the solid blue line represents the trajectories of electrons and positrons as they are perturbed by the beating EM wave. The black arrow depicts the motion of the oscillation center of these particle trajectories. First, consider the trajectory of a charged particle under the 
EM field of the incident wave. At the lowest order of \(v/c\), the particles experience a force in the direction of the electric field, proportional to \( q \bm{E}_{\text{wave}}\propto\bm{v}_E \) (\(\bm{v}_E\) corresponds to $\bm{v}_{0\pm}^{(1)}$ in equation \eqref{eq:motion_in_no_magnetic}), oscillating at the angular frequency \(\omega_0\). Additionally, at the first order of \(v/c\), the particles feel a Lorentz force from the oscillating magnetic field, given by \((q/c) (\bm{v}_E \times \bm{B}_{\mathrm{wave}}) \propto q^2 (\bm{E}_{\text{wave}} \times \bm{B}_{\mathrm{wave}})\), which induces an oscillation at \(2\omega_0\) and results in a figure-eight motion. Next, when considering the superposition of the incident wave with a perturbed scattered wave of small amplitude, the amplitude of the EM wave slowly varies in space and time with an angular frequency of $|\omega_1-\omega_0|$. In this case, the Lorentz force in the direction of \(\bm{E}_{\text{wave}} \times \bm{B}_{\mathrm{wave}}\) toward the oscillation center becomes smaller when the particle is located in a weak field and larger when it is in a strong field. As a result, the time-averaged force over one period drives the particle toward the region of smaller amplitude. As shown in the right panel, this causes density fluctuations, and particles accumulate in the direction of smaller amplitude, regardless of charge sign.}
\label{fig:density_perturbation_nomg}
\end{figure*}

The ponderomotive force and Lorentz force on particles are the same as in the absence of a \proof{background }magnetic field, when the electric field component of the incident wave is polarized parallel to the \proof{background }magnetic field.
From parallel electric field components in equation \eqref{eq:general_velocity}, the first-order velocity of the particles is expressed as
\begin{equation}
\bm{v}_{0\pm}^{(1)}=\mp \frac{e}{m_{\mathrm{e}} c} \bm{A}_0.
\label{eq:motion_in_no_magnetic}
\end{equation}
Since only the first term of the ponderomotive potential \eqref{eq:ponderomotive_potential} contributes, it can be expressed as follows, using equation \eqref{eq:incident_and_scattered_wave}:
\begin{equation}
\begin{aligned}
\phi_{\pm} & = \frac{e^{2}}{2 m_{\mathrm{e}} c^{2}}\left\langle\left|\bm{A}_{\|}\right|^{2}\right\rangle \\
           & = \text{const.} + \frac{e^{2}}{m_{\mathrm{e}} c^{2}} \bm{A}_{1} \cdot \bm{A}_{0}^{*} \mathrm{e}^{\mathrm{i}(\bm{k} \cdot \bm{r}-\omega t)}+\text{c.c.}\\
           & = \text{const.} +\widetilde{\phi_{\pm}}\mathrm{e}^{\mathrm{i}(\bm{k} \cdot \bm{r}-\omega t)}+\text{c.c.}.
\end{aligned}
\label{eq:pondero_parallel}
\end{equation}
This ponderomotive \rei{force} \proof{potential}, as illustrated by Figure \ref{fig:density_perturbation_nomg}, generates density fluctuations of ``ordinary mode" in $e^\pm$ plasma without charge separation. Substituting equation \eqref{eq:pondero_parallel} into equation \eqref{eq:density_fluctuation_pondero}, only the first term of equation \eqref{eq:density_fluctuation_pondero} remains because the ponderomotive potential \eqref{eq:pondero_parallel} is independent of the charge sign. Therefore, the density fluctuation is given by
\begin{equation}
\widetilde{\delta n}_{\pm}(\bm{k}, \omega)=-\frac{n_{\mathrm{e} 0}e^{2}}{m_{\mathrm{e}}^2c^{2}}  \mu A_{1} A_{0}^{*} \sum_{\ell=-\infty}^{+\infty} \int \dd^{3} \bm{v} \frac{J_{\ell}^{2}\left(k_{\perp} r_{\mathrm{L}}\right)\bm{k} \cdot \frac{\partial f_{0}}{\partial \bm{v}^{*}}}{\omega-k_{\|} v_{\|}-\ell \omega_{\text{c}}}, 
\label{eq:density_perturbation_momg_kirei}
\end{equation}
where
\begin{equation}
\mu \equiv \frac{\bm{A}_{1} \cdot \bm{A}_{0}^{*}}{A_{1} A_{0}}, 
\label{eq:definition_mu_induced_magnetized}
\end{equation}
represents the cosine between the electric fields of the incident and scattered waves.
The \((\bm{k}_1, \omega_1)\) component of the source current \eqref{eq:current_oomoto} can be separated into the following linear term \(\widetilde{\bm{j}_1}^{\text{linear}}\) and nonlinear term \(\widetilde{\bm{j}_1}^{\text{nonlinear}}\), using equation \eqref{eq:motion_in_no_magnetic},
\begin{equation}
\begin{aligned}
\widetilde{\bm{j}_1}^{\text{linear}}(\bm{k}_1, \omega_1) &\equiv en_{\mathrm{e0}}\bm{v}_{1+}^{(1)} - en_{\mathrm{e0}}\bm{v}_{1-}^{(1)} \\
&=-\frac{2 e^2 n_{\mathrm{e0}}}{m_{\mathrm{e}} c} \bm{A}^{}_1,
\end{aligned}
\label{eq:definition_of_linear_current_induced_Compton}
\end{equation}
\begin{equation}
\begin{aligned}
\widetilde{\bm{j}_1}^{\text{nonlinear}}(\bm{k}_1, \omega_1) &\equiv e\widetilde{\delta n_{+}}\bm{v}_{0+}^{(1)} - e\widetilde{\delta n_{-}}\bm{v}_{0-}^{(1)}\\
&=-\frac{e^2}{m_{\mathrm{e}} c}\left(\widetilde{\delta n_{+}} + \widetilde{\delta n}_{-}\right) \bm{A}^{}_0.
\end{aligned}
\label{eq:definition_of_nonlinear_current_induced_Compton}
\end{equation}
Here, \(\bm{v}_{1\pm}^{(1)}\) is represented by replacing the subscript 0 with 1 in equation \eqref{eq:motion_in_no_magnetic}.
The scattered wave component of the EM wave equation \eqref{eq:wave_eq_vector} can be written as
\begin{equation}
    (c^2 k_1^2 - \omega_1^2) A_1 = 4\pi c\frac{\bm{A}^{*}_1}{A_1}\cdot\left(\widetilde{\bm{j}_1}^{\text{linear}} + \widetilde{\bm{j}_1}^{\text{nonlinear}}\right).
    \label{eq:wave_eq_for_scattered_wave_nonlinear}
\end{equation}
We can calculate equation \eqref{eq:wave_eq_for_scattered_wave_nonlinear} by substituting equation\proof{s} \eqref{eq:definition_of_linear_current_induced_Compton} and \eqref{eq:definition_of_nonlinear_current_induced_Compton},
\begin{equation}
\left(c^{2} k_{1}^{2} - \omega_{1}^{2} + \omega_{\text{p}}^{2}\right)\left|\bm{A}_{1}\right|^{2} = -\frac{\omega_{\text{p}}^{2}}{2 n_{\mathrm{e0}}} \bm{A}_{1}^{*} \cdot \bm{A}_{0}\left(\widetilde{\delta n}_{+} + \widetilde{\delta n}_{-}\right).
\label{eq:wave_eq_scattered_pondero}
\end{equation}
Substituting the density fluctuation equation \eqref{eq:density_perturbation_momg_kirei} into equation \eqref{eq:wave_eq_scattered_pondero}, we obtain the dispersion relation for the scattered wave,
\begin{equation}
\begin{aligned}
c^{2} k_{1}^{2} - \omega_{1}^{2} + \omega_{\text{p}}^{2} = & \frac{1}{4} c^{2}\left(\omega_{\text{p}} a_{\mathrm{e}} \mu\right)^{2} \\
& \times \sum_{\ell=-\infty}^{+\infty} \int \dd^{3} \bm{v} \frac{J_{\ell}^{2}\left(k_{\perp} r_{\mathrm{L}}\right)\bm{k} \cdot \frac{\partial f_{0}}{\partial \bm{v}^{*}}}{\omega - k_{\| } v_{\| } - \ell \omega_{\text{c}}}.
\end{aligned}
\label{eq:dispersion_relation_no_magnetic}
\end{equation}
Here,
\begin{equation}
a_{\mathrm{e}} \equiv \frac{2 e\left|\bm{A}_{0}\right|}{m_{\mathrm{e}} c^2},
\label{eq:strength_parameter_no_magnetic}
\end{equation}
is the dimensionless strength parameter of the incident EM wave. The factor of 2 in equation \eqref{eq:strength_parameter_no_magnetic} arises because the 
EM
field amplitude is defined in equation \eqref{eq:incident_and_scattered_wave}, which gives that the incident wave amplitude is represented as $2A_0$. Using the dimensionless strength parameter in equation \eqref{eq:strength_parameter_no_magnetic}, the applicability condition derived from the non-relativistic motion of charged particles in equation \eqref{eq:motion_in_no_magnetic} is
\begin{equation}
\left|\frac{\bm{v}_{0\pm}^{(1)}}{c}\right|=\frac{e\left|\bm{A}_0\right|}{m_{\text{e}} c^2} \ll 1 \Rightarrow a_{\text{e}}\ll 1.
\label{eq:strength_parameter_constraint_nonrela_para}
\end{equation}

The linear growth rate of the scattered wave energy, $\Gamma_{\mathrm{C}}$, is maximized when the beating EM wave experiences Landau resonance with particles moving at velocities comparable to the phase velocity of the beat. The following discussion is valid under the weak coupling condition\footnote{When the weak coupling condition \(\left|\text{Im}~ \omega_{1}\right| \ll k_{\|} v_{\mathrm{th}} \) is violated, the induced Brillouin scattering begins to exhibit a growth rate comparable to that of the induced Compton scattering. The condition \(\left|\text{Im}~ \omega_{1}\right| \geq k_{\|} v_{\mathrm{th}} \) is referred to as the strong coupling condition. Under the strong coupling condition, the induced Brillouin scattering becomes dominant over the induced Compton scattering  \citep{2016PhRvL.116a5004E,2017PhRvE..96e3204S,2023MNRAS.522.2133I}.},
\begin{equation}
    \left|\text{Im}~ \omega_{1}\right| \ll k_{\|} v_{\mathrm{th}}.
    \label{eq:weak_coupling_condition}
\end{equation} 
For sufficiently \rei{large} \proof{strong background} magnetic fields, $k_{\perp} v_{\mathrm{th}} / \omega_{\mathrm{c}} \ll k_{\perp} v_{\mathrm{th}} / |\omega|< k v_{\mathrm{th}} / |\omega|\sim 1$ holds. Under this condition, and assuming a Maxwellian distribution for the uniform $e^\pm$ plasma \eqref{eq:Maxwellian_distribution_induced_Compton}, we can retain only the $\ell=0$ terms of equation \eqref{eq:longitudinal_electric_susceptibility} in the infinite sum of the modified Bessel functions,
\begin{equation}
\begin{aligned}
&\sum_{\ell=-\infty}^{+\infty} \int \dd^{3} \bm{v} \frac{J_{\ell}^{2}\left(k_{\perp} r_{\mathrm{L}}\right)\bm{k} \cdot \frac{\partial f_{0}}{\partial \bm{v}^{*}}}{\omega-k_{\|} v_{\|}-\ell \omega_{\mathrm{c}}}= \frac{m_{\text{e}}k^2}{4\pi e^2n_{\text{e}0}}H_{+}\\
&\overset{\ell=0}{\sim} \frac{2}{v_{\mathrm{th}}^{2}}\left\{1+\frac{\omega}{k_{\|} v_{\text{th}}}I_{0}\left[\frac{1}{2}\left(\frac{k_{\perp} v_{\text{th}}}{\omega_{\text{c}}}\right)^{2}\right] \right. \\
& \left. \quad \times \mathrm{e}^{-\frac{1}{2}\left(k_{\perp} v_{\text{th}} / \omega_{\text{c}}\right)^{2}} Z\left(\frac{\omega}{k_{\|} v_{\text{th}}}\right) \right\} \\
&\sim \frac{2}{v_{\mathrm{th}}^{2}}\left\{1+\frac{\omega}{k_{\|} v_{\text{th}}} Z\left(\frac{\omega}{k_{\|} v_{\text{th}}}\right) \right\}
\end{aligned}
\label{eq:sekibun_no_keisan_induced_Compton}
\end{equation}
Here, we used $I_{0}\left[\frac{1}{2}\left(k_{\perp} v_{\text{th}} / \omega_{\text{c}}\right)^{2}\right]\mathrm{e}^{-\frac{1}{2}\left(k_{\perp} v_{\text{th}} / \omega_{\text{c}}\right)^{2}}\simeq 1$. For sufficiently small arguments of the plasma dispersion function, the following asymptotic expansion holds \citep{1961pdf..book.....F},
\begin{equation}
Z(\xi)=\mathrm{i} \sqrt{\pi} \mathrm{e}^{-\xi^{2}}-2 \xi+\frac{4}{3} \xi^{3}+\cdots\quad \text{for}~|\xi| \ll 1.
\label{eq:plasma_dispersion_expansion_small}
\end{equation}
The asymptotic expansion \eqref{eq:plasma_dispersion_expansion_small} can be performed to equation \eqref{eq:sekibun_no_keisan_induced_Compton} under the weak coupling condition \eqref{eq:weak_coupling_condition}.
Therefore, taking the imaginary part of the dispersion relation \eqref{eq:dispersion_relation_no_magnetic}, the growth rate $\Gamma_{\text{C}}$ is given by
\begin{equation}
\Gamma_{\text{C}}\equiv2 \operatorname{Im} \omega_1=-\frac{\sqrt{\pi}}{2}\left(\frac{c}{v_{\text{th}}}\right)^2 \frac{\left(\omega_{\text{p}} a_{\text{e}} \mu\right)^2}{\omega_1} \frac{\omega}{k_{\|} v_{\text{th}}} \mathrm{e}^{-\left(\frac{\omega}{k_{\|} v_{\text{th}}}\right)^2}.
\label{eq:maximum_growth_rate_no_magnetic_maedankai}
\end{equation}
The growth rate is maximized for
\begin{equation}
\frac{\omega}{k_{\|} v_\text{th}}=-\frac{1}{\sqrt{2}},\quad \text{and}\quad \mu=\pm1.
\label{eq:max_growth_dispersion}
\end{equation}
Under this condition, the maximum growth rate, assuming $\omega_1 \sim \omega_0$, is given by
\begin{equation}
\Gamma_{\mathrm{C}}^{\max }=\sqrt{\frac{\pi}{32 \mathrm{e}}} \frac{\omega_{\mathrm{p}}^{2} a_{\mathrm{e}}^{2}}{\omega_{0}} \frac{m_{\mathrm{e}} c^{2}}{k_{\mathrm{B}} T_{\mathrm{e}}},
\label{eq:maximum_growth_rate_no_magnetic}
\end{equation}
where we used equation \eqref{eq:thermal_velocity}. The growth rate of the ordinary mode is the same as \proof{that} without a \proof{background }magnetic field \citep{2022ApJ...930..106G}.

The scattered wave has the following property at the maximum growth rate. Denote the cosine of the wave vector angles between the incident and scattered waves by
\begin{equation}
\nu \equiv \frac{\bm{k_{0}} \cdot \bm{k_{1}}}{k_{0} k_{1}}.
\label{eq:definition_of_nu_induced_Compton}
\end{equation}
When $\omega_{1} \sim \omega_{0} \gg |\omega|$, from energy-momentum conservation \eqref{eq:energy_momentum_conservation_Compton} and equation \eqref{eq:max_growth_dispersion}, we have
\begin{equation}
\begin{aligned}
    \omega_1-\omega_0 & = -\frac{1}{\sqrt{2}}\left|(\bm{k}_1-\bm{k}_0)\cdot\hat{\bm{B}_0}\right|v_{\text{th}} \\
    & \simeq -\sqrt{\frac{2(1-\nu)\omega_0\omega_1}{c^2} \cos^2\theta_{kB} \frac{k_{\text{B}}T_{\text{e}}}{m_{\text{e}}}}.
\end{aligned}
\label{eq:relation_omega1_and_omega_0}
\end{equation}
Here, $\theta_{kB}$ is the angle between the wave vector of the density fluctuations and the \rei{uniform} \proof{background} magnetic field. The angular frequency of the most amplified scattered wave is expressed, from equation \eqref{eq:relation_omega1_and_omega_0}, as
\begin{equation}
\omega_1(\nu,\theta_{kB}) \simeq \omega_0\left(1-\sqrt{2(1-\nu) \cos ^2 \theta_{kB} \frac{k_\text{B} T_{\text{e}}}{m_{\mathrm{e}} c^{2}}}\right).
\label{eq:relation_omega1_omega0}
\end{equation}

The maximum linear growth rate of induced Compton scattering, given by equation \eqref{eq:maximum_growth_rate_no_magnetic}, is subject to several conditions. First, EM waves polarized parallel to the \proof{background }magnetic field cannot propagate with an angular frequency \(\omega_0 < \omega_{\text{p}}\) (the cutoff frequency). Therefore, the condition 
\begin{equation}
    \omega_{\text{p}} < \omega_0,
    \label{eq:cutoff_frequency_induced_Compton}
\end{equation}
must be satisfied. Second, the energy imparted by the ponderomotive potential to particles in the plasma, \(2n_{\text{e}0}\phi_{\pm}\), must not exceed the energy of the original 
EM
field, \rei{\(\langle U_{\text{EM}}\rangle = 2E_{0\|}^2\)} \proof{\(\langle U_{\text{EM}}\rangle = E_{0\|}^2/(2\pi)\)}. This yields, using equation \eqref{eq:pondero_parallel}, the condition \((1/\sqrt{2})\omega_{\text{p}} < \omega_0\), which is automatically satisfied by the cutoff frequency condition \eqref{eq:cutoff_frequency_induced_Compton}. Furthermore, using equation \eqref{eq:relation_omega1_omega0}, the condition \(\omega_{1} \sim \omega_{0} \gg |\omega|\) imposes the following constraint on the plasma temperature,
\begin{equation}
\frac{k_{\mathrm{B}} T_{\mathrm{e}}}{m_{\mathrm{e}} c^2} \ll \frac{1}{2(1 - \nu) \cos^2 \theta_{kB}}.
\label{eq:temperature_constraint_induced_Compton}
\end{equation}
Moreover, another condition for the maximum growth rate is derived from the weak coupling condition \eqref{eq:weak_coupling_condition}, using equation \eqref{eq:maximum_growth_rate_no_magnetic},
\begin{equation}
a_{\mathrm{e}} \ll 4\left(\frac{2\text{e}}{\pi}\right)^{\frac{1}{4}} \frac{\omega_0}{\omega_{\mathrm{p}}}\left(\frac{k_{\mathrm{B}} T_{\mathrm{e}}}{m_{\mathrm{e}} c^2}\right)^{\frac{3}{4}}\left\{(1-\nu) \cos ^2 \theta_{kB}\right\}^{\frac{1}{4}}.
\label{eq:strength_parameter_limit_para2}
\end{equation}


\subsection{Scattering of EM waves polarized perpendicular to the background magnetic field}
\label{sec:induced_Compton_perpendicular}
Let us consider the case where the electric field component of the incident wave is perpendicular to the \proof{background} magnetic field. This corresponds to the induced Compton scattering of 
an X-mode wave or
a circularly polarized EM wave propagating in the direction of the magnetic field (Alfvén wave). As shown in Figure \ref{fig:nonlinear_current}, there are two types of density fluctuations that can invoke nonlinear currents \eqref{eq:definition_of_nonlinear_current_induced_Compton}, $e\widetilde{\delta n_{+}}\bm{v}_{0+}^{(1)}-e\widetilde{\delta n_{-}}\bm{v}_{0-}^{(1)}$:
\begin{enumerate}[(i)]
\item Charged mode: The \rei{second} \proof{third} term in equation \eqref{eq:general_velocity} multiplied by the density fluctuation created by the \rei{second} \proof{third} term in equation \eqref{eq:ponderomotive_potential}. This mode is of the lowest order in $\omega_0/\omega_{\text{c}}$ in a strong \proof{background} magnetic field.
\item Neutral mode: The \rei{third} \proof{second} term in equation \eqref{eq:general_velocity} multiplied by the density fluctuation created by the \rei{third} \proof{second} term in equation \eqref{eq:ponderomotive_potential}. This mode is of the next order in $\omega_0/\omega_{\text{c}}$ in a strong \proof{background }magnetic field.
\end{enumerate}
Other combinations, such as the second term in equation \eqref{eq:general_velocity} with the density fluctuation created by the third term in equation \eqref{eq:ponderomotive_potential}, or the third term in equation \eqref{eq:general_velocity} with the density fluctuation created by the second term in equation \eqref{eq:ponderomotive_potential}, do not generate a nonlinear current. This is because the nonlinear currents produced by positrons, $e\widetilde{\delta n_{+}}\bm{v}_{0+}^{(1)}$, and electrons, $-e\widetilde{\delta n_{-}}\bm{v}_{0-}^{(1)}$, completely cancel each other out in equation \eqref{eq:definition_of_nonlinear_current_induced_Compton} for the scattered wave. 

We calculate the linear growth rate of induced Compton scattering for each mode under arbitrary background magnetic fields and plasma densities. Specifically, we focus on a detailed analysis of the behavior in a strong \proof{background }magnetic field,
\begin{equation}
    \omega_0 \ll \omega_{\text{c}}.
\end{equation}
We examine both the high plasma density case ($\omega_{\text{p}} > \omega_0$) and the low plasma density case ($\omega_{\text{p}} < \omega_0$).
\subsubsection{Charged mode}
\begin{figure*}
\centering
\includegraphics[width=\textwidth]{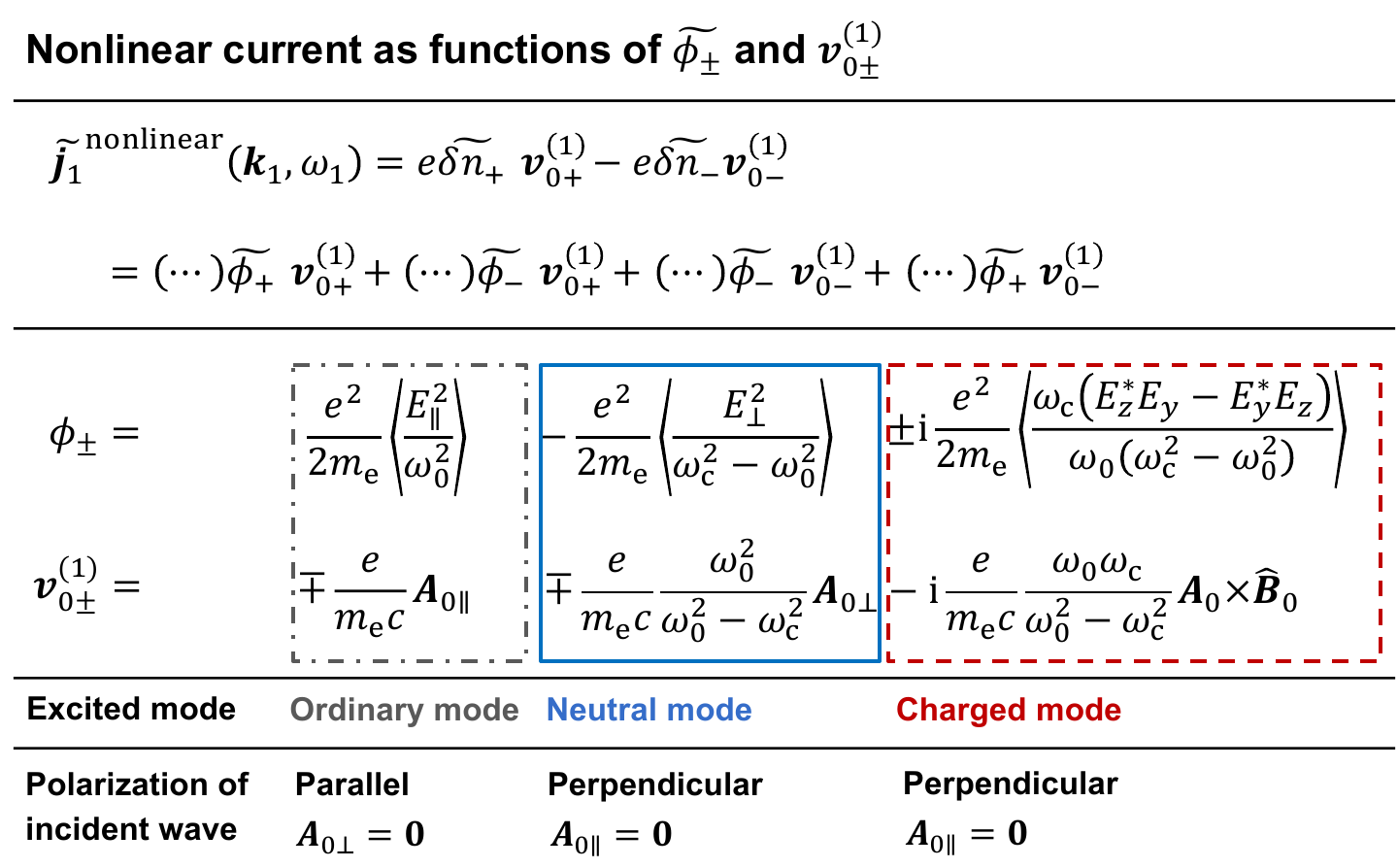}
\caption[Modes of Nonlinear Current as Functions of $\phi$ and $v_0$.]{\justifying 
Illustration of the nonlinear current generated by the density fluctuations \eqref{eq:density_fluctuation_pondero} induced by the beating EM wave. The first row represents the nonlinear current for the \((\bm{k}_1, \omega_1)\) component \eqref{eq:definition_of_nonlinear_current_induced_Compton}, which are represented by \(\phi_{\pm}\) (equation \eqref{eq:ponderomotive_potential}) and \(\bm{v}_{0\pm}^{(1)}\) (equation \eqref{eq:general_velocity}). The second row shows that both \(\phi_{\pm}\) and \(\bm{v}_{0\pm}^{(1)}\) are composed of three terms each. These terms are related to the polarization direction of the incident wave relative to the background magnetic field, with the first term aligned parallel to the \proof{background }magnetic field and the remaining two terms perpendicular. The nonlinear current can excite three unstable modes: ordinary mode (the black dot-dashed square), neutral mode (the blue solid square), and charged mode (the red dashed square). The ordinary mode is defined in Section \ref{sec:induced_Compton_parallel}. The neutral mode and charged mode are defined in Section \ref{sec:induced_Compton_perpendicular}. When the incident wave is polarized parallel to the \proof{background }magnetic field, the nonlinear current generates ordinary mode. When the incident wave is polarized perpendicular to the \proof{background }magnetic field, it generates both the neutral and charged mode\proof{s}. The final row illustrates how each term in \(\phi_{\pm}\) and \(\bm{v}_{0\pm}^{(1)}\) corresponds to the polarization direction of the incident wave.}
\label{fig:nonlinear_current}
\end{figure*}
\begin{figure*}
\centering
\includegraphics[width=\textwidth]{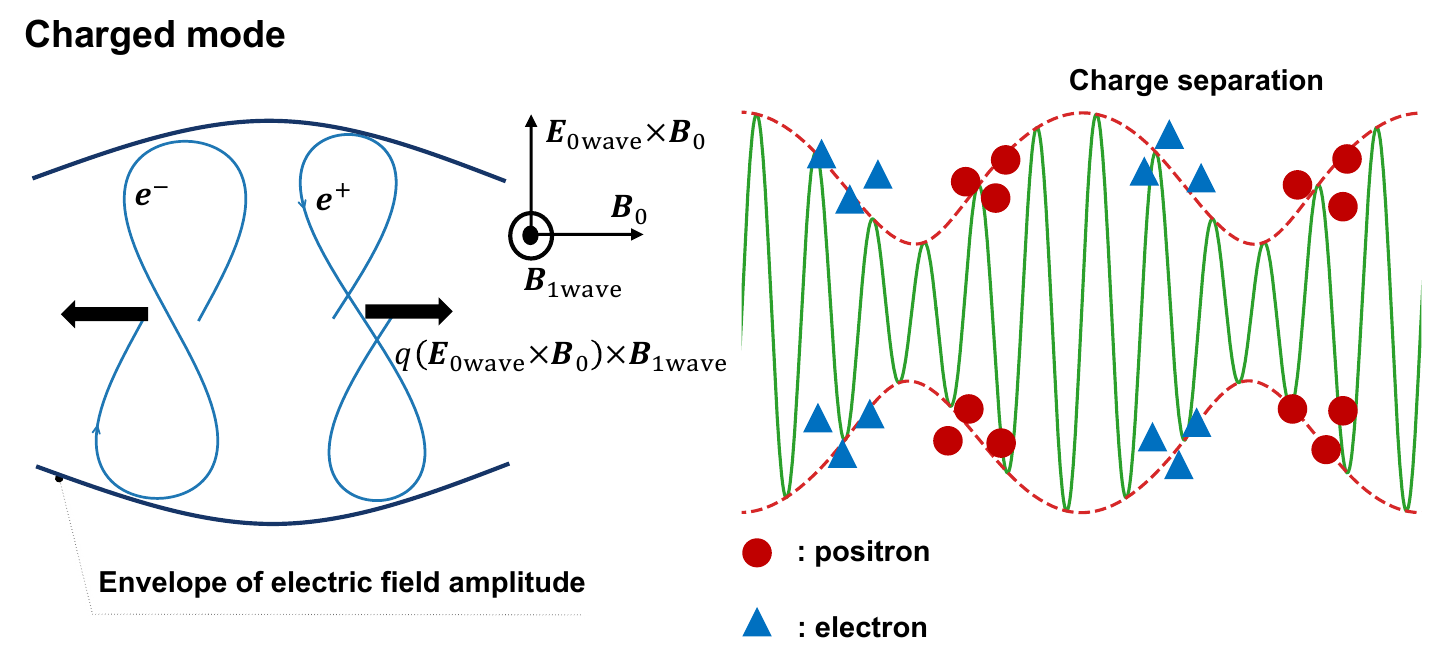}
\caption[Concept of Parametric Decay Instability]{\justifying The physical interpretation of ponderomotive force (left panel) and schematic picture of parametric instability (right panel) for the ``charged mode" for an incident wave polarized perpendicular to the background magnetic field, $\bm{A}_{0\|} = \bm{0}$. In the left figure, the solid blue lines represent the trajectories of electrons and positrons perturbed by the beating EM wave. The black arrows depict the motion of the oscillation centers of these particle trajectories. At the lowest order in \(v/c\), the particles experience a force in the direction of the \(\bm{E}_{0\text{wave}}\times\bm{B}_{0}\) drift, oscillating at the frequency $\omega_0$. Additionally, at the next order of \(v/c\), the particles feel a Lorentz force from the magnetic field component of the scattered wave, given by \(q(\bm{E}_{0\text{wave}}\times\bm{B}_{0})\times\bm{B}_{1\text{wave}}\), which induces an oscillation at $\omega_0+\omega_1\simeq2\omega_0$ and results in a figure-eight motion. Note that if $\bm{B}_{0\text{wave}}$ and $\bm{B}_{1\text{wave}}$ are exactly parallel or antiparallel, the charged mode is not excited. Substituting $\mu = \pm 1$ into equation \eqref{eq:growth_rate_perp_keisanmae} confirms this. This Lorentz force for electrons and positrons differ from each other in phase by $\pi$ due to the difference in charge sign (see equation \eqref{eq:pondero_drift}). Therefore, electrons and positrons experience forces in opposite directions. In the right figure, charge separation arises because positive and negative charges gather in different regions with respect to each other. The density fluctuations and the beating wave differ in phase by $\pi/2$.}
\label{fig:charged_mode}
\end{figure*}
Let us derive the density fluctuations of the charged mode. The ponderomotive \rei{force} \proof{potential} acting on charged particles reverses direction with the charge sign,
\begin{equation}
    \widetilde{\phi_{ \pm}}=\mathrm{i} \frac{e^{2}}{m_{\text{e}} c^{2}} \frac{\omega_{0} \omega_{\text{c}\pm}}{\omega_{\text{c}}^{2}-\omega_{0}^{2}}\left(\bm{A_{0}^{*}} \times \hat{\bm{B_{0}}}\right) \cdot \bm{A}_{1},
    \label{eq:pondero_drift}
\end{equation}
which comes from the third term of equation \eqref{eq:ponderomotive_potential}.
As shown in Figure \ref{fig:nonlinear_current}, the nonlinear current \(\widetilde{\bm{j}_1}^{\text{nonlinear}}\) for the charged mode\rei{, as given in equation \eqref{eq:definition_of_nonlinear_current_induced_Compton},} is obtained by combining the ponderomotive potential \eqref{eq:pondero_drift} with the third term of the particle velocity in equation \eqref{eq:general_velocity}. This velocity is expressed as
\begin{equation}
\bm{v}_{0 \pm}^{(1)}=-\mathrm{i} \frac{e}{m_{\text{e}} c} \frac{\omega_{0} \omega_{\mathrm{c}}}{\omega_{0}^{2}-\omega_{\mathrm{c}}^{2}} \bm{A}_0 \times \hat{\bm{B}_{0}}.
\label{eq:drift_velocity_lowest_order}
\end{equation}
Using the dimensionless strength parameter from equation \eqref{eq:strength_parameter_no_magnetic}, the non-relativistic condition for the charged particle motion \eqref{eq:drift_velocity_lowest_order} imposes the following constraint,
\begin{equation}
\left|\frac{\bm{v}_{0\pm}^{(1)}}{c}\right|\simeq\frac{e\left|\bm{A}_0\right|}{m_{\text{e}} c^2}\left|\frac{\omega_{0} \omega_{\mathrm{c}}}{\omega_{0}^{2}-\omega_{\mathrm{c}}^{2}}\right| \ll 1 \Rightarrow a_{\text{e}}\left|\frac{\omega_{0} \omega_{\mathrm{c}}}{\omega_{0}^{2}-\omega_{\mathrm{c}}^{2}}\right|\ll 1.
\end{equation}
In particular, the condition in a strong \proof{background }magnetic field becomes
\begin{equation}
    a_{\text{e}}\frac{\omega_0}{\omega_{\mathrm{c}}}\ll 1.
    \label{eq:strength_parameter_constraint_nonrela_perp}
\end{equation}
Substituting equation \eqref{eq:pondero_drift} into equation \eqref{eq:density_fluctuation_pondero}, the density fluctuations are expressed as
\proof{\begin{equation}
\begin{aligned}
\widetilde{\delta n_{ \pm}}(\bm{k}, \omega)= &-\mathrm{i} \frac{e^{2} n_{\text{e} 0}}{\varepsilon_{\mathrm{L}} m_{\text{e}}^{2} c^{2}} \frac{\omega_{0} \omega_{\mathrm{c}\pm}}{\omega_{\mathrm{c}}^{2}-\omega_{0}^{2}}\left(\bm{A_{0}^{*}} \times \hat{\bm{B}_{0}}\right) \cdot \bm{A}_{1} \\
&\times \sum_{\ell=-\infty}^{+\infty} \int \dd^{3} \bm{v} \frac{J_{\ell}^{2}\left(k_{\perp} r_{\mathrm{L}}\right)\bm{k} \cdot \frac{\partial f_{0}}{\partial \bm{v}^{*}}}{\omega-k_{\|} v_{\|}-\ell \omega_{\mathrm{c}}}, 
\end{aligned}
\label{eq:density_fluctuation_perp}
\end{equation}}where we used equation \eqref{eq:longitudinal_dielectric_constant_induced_Compton}.
As shown in equation \eqref{eq:density_fluctuation_perp}, the density fluctuations of electrons and positrons have different phase by \(\pi\) (between $\mathrm{i}$ and $-\mathrm{i}$). As illustrated in Figure \ref{fig:charged_mode}, these density fluctuations induce charge separation in $e^\pm$ plasma. Notably, in contrast to induced Raman scattering, which occurs in the regime where \(\omega_0 > \omega_{\text{p}}\) and leads to charge separation, charged mode can occur even in the regime where \(\omega_0 < \omega_{\text{p}}\).

Using equation \eqref{eq:density_fluctuation_perp}, we derive the dispersion relation for the scattered wave. The \((\bm{k}_1, \omega_1)\) component of the source current \eqref{eq:current_oomoto} is given, using equations \proof{\eqref{eq:general_velocity} and} \eqref{eq:drift_velocity_lowest_order}, by
\proof{\begin{equation}
\begin{aligned}
\widetilde{\bm{j}_1}^{}(\bm{k}_1,\omega_1) &= e\left(n_{\mathrm{e0}}\bm{v}_{1+}^{(1)}+\widetilde{\delta n_{+}}\bm{v}_{0+}^{(1)}\right)\\
&\quad-e\left(n_{\mathrm{e0}}\bm{v}_{1-}^{(1)}+\widetilde{\delta n_{-}}\bm{v}_{0-}^{(1)}\right) \\
&=-\frac{2e^2n_{\mathrm{e0}}}{m_{\text{e}} c}\frac{\omega_1^2}{\omega_1^2-\omega_{\text{c}}^2}\bm{A}_{1\perp} \\
&\quad-\mathrm{i} \frac{e^2}{m_{\text{e}} c} \frac{\omega_{0} \omega_{\mathrm{c}}}{\omega_{0}^{2}-\omega_{\mathrm{c}}^{2}}\left(\widetilde{\delta n_{+}}-\widetilde{\delta n}_{-}\right) \bm{A}_0^{} \times \hat{\bm{B}}_{0} \\
&=\widetilde{\bm{j}_1}^{\text{linear}}+\widetilde{\bm{j}_1}^{\text{nonlinear}},
\end{aligned}
\label{eq:current1_perp}
\end{equation}}\proof{where we first examine the case where \(\bm{A}_{1\|} = \bm{0}\). As will be shown later, if \(\bm{A}_{1\perp} = \bm{0}\), the scattered wave does not grow.} Substituting equation \eqref{eq:density_fluctuation_perp} into equation \eqref{eq:current1_perp} gives the source current. Then, substituting this current into the wave equation for the scattered wave \eqref{eq:wave_eq_for_scattered_wave_nonlinear}, the dispersion relation for the scattered wave is represented by
\proof{\begin{equation}
\begin{aligned}
c^{2} k_{1}^{2}-&\omega_{1}^{2}+\omega_{\text{p}}^2\frac{\omega_1^2}{\omega_1^2-\omega_{\text{c}}^2}=\frac{c^{2}}{4 \varepsilon_{\mathrm{L}}}\left(\frac{\omega_{0} \omega_{\mathrm{c}}}{\omega_{\mathrm{c}}^{2}-\omega_{0}^{2}}\right)^{2} \\
\times &\left(\bar{a}_{\mathrm{e}} \omega_{\mathrm{p}}\right)^{2}\sum_{\ell=-\infty}^{+\infty} \int \dd^{3} \bm{v} \frac{J_{\ell}^{2}\left(k_{\perp} r_{\mathrm{L}}\right)\bm{k} \cdot \frac{\partial f_{0}}{\partial \bm{v}^{*}}}{\omega-k_{\|} v_{\|}-\ell \omega_{\mathrm{c}}}. 
\end{aligned}
\label{eq:dispersion_relation_pondero_drift}
\end{equation}}Here, $\bar{a}_{\mathrm{e}}$ is defined, using equation \eqref{eq:definition_mu_induced_magnetized}, as
\begin{equation}
\begin{aligned}
 \bar{a}_{\mathrm{e}}&\equiv\frac{2 e}{m_{\text{e}} c^{2}}\left|\left(\bm{A_{0}^{*}} \times \hat{\bm{B_{0}}}\right) \cdot \hat{\bm{A}}_{1}\right| \\
& =a_{\text{e}} \sqrt{1-\mu^2}\left|\bm{n} \cdot \hat{\bm{B}}_0\right|, 
\end{aligned}
\end{equation}
where
\begin{equation}
\bm{n} \equiv \frac{\bm{A}_1 \times \bm{A}_0^{*}}{\left|\bm{A}_1 \times \bm{A}_0^{*}\right|},
\label{eq:definition_of_n_induced_Compton}
\end{equation}
is the unit normal vector to the plane formed by $\bm{A}_0^{*}$ and $\bm{A}_1$. 

From the dispersion relation \eqref{eq:dispersion_relation_pondero_drift}, the linear growth rate $\Gamma_{\mathrm{C}}$ for the scattered wave energy can be obtained. The growth rate is maximized when the beating wave experiences Landau resonance with particles moving at velocities comparable to the phase velocity of the beat. For sufficiently \rei{large} \proof{strong background} magnetic fields, where $k_{\perp} v_{\mathrm{th}} / \omega_{\mathrm{c}} \ll k v_{\mathrm{th}} / |\omega| \sim 1$, only the $\ell=0$ term of equation \eqref{eq:dispersion_relation_pondero_drift} is relevant. 
\proof{So, we can apply equation \eqref{eq:sekibun_no_keisan_induced_Compton} to \eqref{eq:dispersion_relation_pondero_drift}. 
The imaginary part of the left-hand side of the dispersion relation \eqref{eq:dispersion_relation_pondero_drift} can be approximated by assuming $|\text{Im}~\omega_1|\ll\omega_1,~\omega_{\text{c}}$,
\begin{equation}
\begin{aligned}
    \operatorname{Im}&\left[c^2 k_1^2 - \omega_1^2 
    + \omega_{\mathrm{p}}^2 \frac{\omega_1^2}{\omega_1^2 - \omega_{\mathrm{c}}^2}\right] \\
    &\simeq  -2 \omega_1\operatorname{Im} \omega_1 \left\{1 
    + \frac{\omega_{\mathrm{c}}^2 \omega_{\mathrm{p}}^2}
    {\left(\omega_{\mathrm{c}}^2 - \omega_1^2\right)^2}\right\}.
\end{aligned}
\end{equation}}The linear growth rate is then expressed\proof{, using equations \eqref{eq:thermal_velocity} and \eqref{eq:plasma_dispersion_expansion_small},} as
\proof{\begin{equation}
\begin{aligned}
&\Gamma_{\mathrm{C}} = 2 \operatorname{Im} \omega_{1} \\
=& -\frac{1}{4} \frac{m_{\text{e}} c^{2}}{ k_{\mathrm{B}} T_{\mathrm{e}}} 
\left(\frac{\omega_{0} \omega_{\mathrm{c}}}{\omega_{\mathrm{c}}^{2} - \omega_{0}^{2}}\right)^{2}
\left\{1 + \frac{\omega_{\mathrm{c}}^2 \omega_{\mathrm{p}}^2}{\left(\omega_{\mathrm{c}}^2 - \omega_1^2\right)^2}\right\}^{-1} 
 \\
 &\times \frac{\left(a_{\mathrm{e}} \omega_{\mathrm{p}}\right)^{2}}{\omega_{1}}(1-\mu^2) 
\left|\bm{n} \cdot \hat{\bm{B}}_0\right|^2\\
&\times\operatorname{Im}\left[
\frac{1 + \zeta \left(\text{i} \sqrt{\pi} \text{e}^{-\zeta^2} - 2 \zeta\right)}{
1 + \frac{2 \omega_{\mathrm{p}}^{2}}{k^{2} v_{\mathrm{th}}^{2}} 
\left\{1 + \zeta \left(\text{i} \sqrt{\pi} \text{e}^{-\zeta^2} - 2 \zeta\right)\right\}}
\right],
\end{aligned}
\label{eq:growth_rate_perp_keisanmae}
\end{equation}}
where $\zeta$ is defined as
\begin{equation}
    \zeta \equiv \frac{\omega}{k_{\parallel} v_{\text{th}}}.
\end{equation}
The linear growth rate is maximized when
\begin{equation}
|\bm{n} \cdot \hat{\bm{B}}_0|=\pm1,
\quad {\rm and} \quad
\mu = 0,
\label{eq:cond_max}
\end{equation}
as defined in equation \eqref{eq:definition_mu_induced_magnetized}, i.e., $\bm{A}_0^{*}\perp\bm{A}_1$. This corresponds to the case where the directions of the background magnetic field, the electric field of the incident wave, and that of the scattered wave form an orthonormal basis in three-dimensional space. \proof{When \(\bm{A}_{1\perp} = \bm{0}\), the dispersion relation for the scattered wave can be derived in the same manner as in the case of \(\bm{A}_{1\|} = \bm{0}\). In this case, \(|\bm{n} \cdot \hat{\bm{B}}_0|\) always vanishes, leading to a zero source term in the dispersion relation. Consequently, the scattered wave does not grow.}

We consider the limits (a) when the wavelength of the density fluctuations is much larger or (b) much smaller than the Debye length. Here, Debye length in $e^\pm$ plasma is defined as
\begin{equation}
\lambda_{\text{De}}\equiv\left(\frac{k_{\text{B}}T_{\text{e}}}{8 \pi e^2 n_{\text{e}0}}\right)^{\frac{1}{2}}=\frac{v_{\text{th}}}{\sqrt{2} \omega_{\text{p}}}.
\label{eq:definition_of_Debye_length_induced_Compton}
\end{equation}

(a) First, considering the case when the wavelength of the density fluctuations is much larger than the Debye length,
\begin{equation}
    \frac{\omega_{\mathrm{p}}^{2}}{k^{2} v_{\mathrm{th}}^{2}}\sim \left(\frac{\lambda}{\lambda_{\text{De}}}\right)^2\gg 1.
    \label{eq:Debye_screening_limit_induced_Compton}
\end{equation}
In this case, the maximum growth rate $\Gamma_{\text{C}}^{\max}$ is given by,
\proof{\begin{equation}
\begin{aligned}
& \Gamma_{\text{C}}^{\max }(\mu=0)=-\frac{1}{2}\left(\frac{\omega_0 \omega_\text{c}}{\omega_{\text{c}}^2-\omega_0^2}\right)^2\left\{1 + \frac{\omega_{\mathrm{c}}^2 \omega_{\mathrm{p}}^2}{\left(\omega_{\mathrm{c}}^2 - \omega_1^2\right)^2}\right\}^{-1} \\
& \times \left(\frac{c}{v_{\text{th}}}\right)^2\frac{a_{\mathrm{e}}^2 \omega_{\text{p}}^2}{\omega_1}\left(\frac{k v_{\text{th}}}{\omega_{\text{p}}}\right)^4\frac{\zeta \mathrm{e}^{\zeta^2} \sqrt{\pi}}{4\left(1-2 \zeta^2\right)^2 \mathrm{e}^{2 \zeta^2}+4 \zeta^2 \pi}.
\end{aligned}
\label{eq:charged_mode_totyuunosiki}
\end{equation}}This maximum occurs when $\zeta = -1/\sqrt{2}$,\footnote{Without relying on the analytical expansion of the plasma dispersion function \eqref{eq:plasma_dispersion_expansion_small}, a numerical calculation shows that the maximum growth occurs at \(\zeta = \zeta_{\text{p}} \equiv -1.47\cdots\). The corresponding maximum linear growth rate is approximately 34\% higher than the value given by equation \eqref{eq:growth_rate_induced_Compton_debye_sheelding}.
}
 i.e.,
\begin{equation}
\frac{\omega}{k_{\|} v_\text{th}}=-\frac{1}{\sqrt{2}}.
\label{eq:condition_for_maximum_growth_rate_charged}
\end{equation}
At this condition, the maximum linear growth rate is calculated as 
\proof{\begin{equation}
\begin{aligned}
&\left.\Gamma_{\mathrm{C}}^{\max }(k) \equiv 2 \operatorname{Im} \omega_{1}\right|_{k_{\|} v_{\mathrm{th}}=-\sqrt{2}\omega}  \\
=& \sqrt{\frac{\mathrm{e}}{32\pi}} \left(\frac{\omega_0 \omega_{\text{c}}}{\omega_{\text{c}}^2 - \omega_0^2}\right)^2\left\{1 + \frac{\omega_{\mathrm{c}}^2 \omega_{\mathrm{p}}^2}{\left(\omega_{\mathrm{c}}^2 - \omega_1^2\right)^2}\right\}^{-1}\\ &\times\left(\frac{c}{v_{\text{th}}}\right)^2\frac{a_{\mathrm{e}}^2 \omega_{\text{p}}^2}{\omega_1} \left(\frac{k v_{\text{th}}}{\omega_{\text{p}}}\right)^4.
\end{aligned}
\end{equation}}When $\omega_{1} \sim \omega_{0} \gg |\omega|$, \rei{we can approximate by using equation \eqref{eq:definition_of_nu_induced_Compton},} \proof{assuming that the source term (right-hand side) in the dispersion relation \eqref{eq:dispersion_relation_pondero_drift} is negligible, we obtain by using equations \eqref{eq:energy_momentum_conservation_Compton} and \eqref{eq:definition_of_nu_induced_Compton},}
\proof{\begin{equation}
    \begin{aligned}
        k^{2}&=\left|\bm{k}_1-\bm{k}_0\right|^2 \\
        &=\frac{1}{c^2} \left( \omega_1 \sqrt{1 - \frac{\omega_{\mathrm{p}}^2}{\omega_1^2 - \omega_{\mathrm{c}}^2}} 
- \omega_0 \sqrt{1 - \frac{\omega_{\mathrm{p}}^2}{\omega_0^2 - \omega_{\mathrm{c}}^2}} \right)^2\\
 &+ 2 (1 - \nu) \frac{\omega_0 \omega_1}{c^2} 
\sqrt{\left(1 - \frac{\omega_{\mathrm{p}}^2}{\omega_1^2 - \omega_{\mathrm{c}}^2}\right) 
\left(1 - \frac{\omega_{\mathrm{p}}^2}{\omega_0^2 - \omega_{\mathrm{c}}^2}\right)}\\
        &\simeq2 (1 - \nu) \frac{\omega_0 \omega_1}{c^2}\left(1+\frac{\omega_{\text{p}}^2}{\omega_{\text{c}}^2}\right),
    \end{aligned}
    \label{eq:approximate_wave_number_induced_Compton}
\end{equation}
for $\omega_{\text{c}}\gg\omega_0$.}
Then, the growth rate is expressed, for $\omega_{\text{c}}\gg\omega_0$, as
\proof{\begin{equation}
\Gamma_{\mathrm{C}}^{\max}(\nu) = \sqrt{\frac{2\mathrm{e}}{\pi}} \frac{k_{\text{B}}T_{\text{e}}}{m_{\text{e}}c^2} \left(\frac{\omega_0}{\omega_{\text{c}}}\right)^2\left(1+\frac{\omega_{\text{p}}^2}{\omega_{\text{c}}^2}\right) \left(\frac{\omega_0}{\omega_{\text{p}}}\right)^4 \frac{a_{\mathrm{e}}^2 \omega_{\text{p}}^2}{\omega_0} (1 - \nu)^2.
\label{eq:growth_rate_induced_Compton_debye_sheelding_maedankaidesu}
\end{equation}}This maximum occurs for $\nu=-1$, i.e., for backward scattering, yielding,
\proof{\begin{equation}
\Gamma_{\mathrm{C}}^{\max} = \sqrt{\frac{32\mathrm{e}}{\pi} }\frac{k_{\text{B}}T_{\text{e}}}{m_{\text{e}}c^2} \left(\frac{\omega_0}{\omega_{\text{c}}}\right)^2\left(1+\frac{\omega_{\text{p}}^2}{\omega_{\text{c}}^2}\right) \left(\frac{\omega_0}{\omega_{\text{p}}}\right)^4 \frac{a_{\mathrm{e}}^2 \omega_{\text{p}}^2}{\omega_0}.
\label{eq:growth_rate_induced_Compton_debye_sheelding}
\end{equation}}The conditions 
\begin{equation}
\nu = -1,\quad 
\mu = 0,\quad {\rm and} 
\quad
\left|\bm{n} \cdot \hat{\bm{B}}_0\right| = \pm1,
\label{eq:maximum_charged_Debye_induced_Compton}
\end{equation}
mean backward scattering of an EM wave incident along the direction of the background magnetic field, where the angle between the electric field components of the incident and scattered waves is perpendicular. The angular frequency of the scattered wave at this maximum growth rate is given by equations \rei{\eqref{eq:relation_omega1_and_omega_0}}\proof{\eqref{eq:energy_momentum_conservation_Compton}, \eqref{eq:condition_for_maximum_growth_rate_charged}, \eqref{eq:approximate_wave_number_induced_Compton}, and \eqref{eq:maximum_charged_Debye_induced_Compton}} as
\proof{\begin{equation}
\begin{aligned}
\omega_1 &= \omega_0-\frac{1}{\sqrt{2}}k\cos\theta_{kB}v_{\text{th}}\\
&\simeq \omega_0\left(1-\sqrt{ 2(1-\nu)\cos^2\theta_{kB}\frac{k_\text{B} T_{\text{e}}}{m_{\mathrm{e}} c^{2}}\left(1+\frac{\omega_{\text{p}}^2}{\omega_{\text{c}}^2}\right)}\right)\\
&\xrightarrow{\nu=-1,~\cos\theta_{kB}=\pm1} \omega_0\left(1-2\sqrt{ \frac{k_\text{B} T_{\text{e}}}{m_{\mathrm{e}} c^{2}}\left(1+\frac{\omega_{\text{p}}^2}{\omega_{\text{c}}^2}\right)}\right).
\end{aligned}
\label{eq:saidaiseichou_no_sannrannsyuuhasuu_Debye}
\end{equation}}

The applicability conditions can be derived from weak coupling condition \eqref{eq:weak_coupling_condition} represented, using equation \eqref{eq:growth_rate_induced_Compton_debye_sheelding}, by
\proof{\begin{equation}
a_{\mathrm{e}}\frac{\omega_0}{\omega_{\mathrm{c}}} \ll \left(\frac{\pi}{\text{e}}\right)^{\frac{1}{4}} \frac{\omega_{\mathrm{p}}}{\omega_0} \left(\frac{m_{\mathrm{e}} c^2}{k_{\mathrm{B}} T_{\mathrm{e}}}\right)^{\frac{1}{4}}\left(1+\frac{\omega_{\text{p}}^2}{\omega_{\text{c}}^2}\right)^{-\frac{1}{4}}.
\label{eq:strength_parameter_limit_perp2-1}
\end{equation}}Additionally, the condition $\omega_{1} \sim \omega_{0} \gg |\omega|$ imposes a restriction on the plasma temperature. By using equation \eqref{eq:saidaiseichou_no_sannrannsyuuhasuu_Debye}, we get
\proof{\begin{equation}
\frac{k_{\mathrm{B}} T_{\mathrm{e}}}{m_{\mathrm{e}} c^2} \ll \frac{1}{4}\left(1+\frac{\omega_{\text{p}}^2}{\omega_{\text{c}}^2}\right)^{-1}.
\label{eq:tempreature_constraint_perp1}
\end{equation}}

For an incident X-mode wave, the growth rate is reduced to $1/4$ of that given by equation \eqref{eq:growth_rate_induced_Compton_debye_sheelding}. Assuming that the incident wave’s electric field direction and wave vector are perpendicular to the background magnetic field, maximum growth occurs for perpendicular scattering ($\nu = 0$). This is because, under the conditions $\mu = 0$ and $|\bm{n} \cdot \hat{\bm{B}}_0| = \pm1$, fixing the incident wave vector uniquely determines the scattered wave vector that achieves maximum growth. Consequently, the maximum linear growth rate of the charged mode for the X-mode wave is obtained by substituting $\nu = 0$ into equation \eqref{eq:growth_rate_induced_Compton_debye_sheelding_maedankaidesu}.

(b) Next, we consider the case where the wavelength of the density fluctuations is much smaller than the Debye length \eqref{eq:definition_of_Debye_length_induced_Compton},
\begin{equation}
    \frac{\omega_{\mathrm{p}}^{2}}{k^{2} v_{\mathrm{th}}^{2}}\sim \left(\frac{\lambda}{\lambda_{\text{De}}}\right)^2\ll 1.
    \label{eq:Debye_length_is_very_large}
\end{equation}
In this case, the maximum linear growth rate of the scattered wave, \rei{represented by $\Gamma_{\mathrm{C}}$} \proof{$\Gamma_{\mathrm{C}}$ in equation \eqref{eq:growth_rate_perp_keisanmae}}, is maximized when $\omega / (k_{\parallel} v_{\text{th}}) = -1 / \sqrt{2}$, given by
\proof{\begin{equation}
\begin{aligned}
\Gamma_{\mathrm{C}}^{\max }(\mu) 
& = \sqrt{\frac{\pi}{32 \mathrm{e}}} \left( \frac{\omega_0}{\omega_{\mathrm{c}}} \right)^2 
\left( 1 + \frac{\omega_{\mathrm{p}}^2}{\omega_{\mathrm{c}}^2} \right)^{-1} 
\frac{\omega_{\mathrm{p}}^{2} a_{\mathrm{e}}^{2}}{\omega_{1}} \\
& \quad \times \frac{m_{\mathrm{e}} c^{2}}{k_{\mathrm{B}} T_{\mathrm{e}}} 
(1 - \mu^2) \left| \bm{n} \cdot \hat{\bm{B}}_0 \right|^2,
\end{aligned}
\label{eq:charged_mode_growth_rate_without_debye_screening}
\end{equation}for $\omega_{\text{c}}\gg\omega_0$.} The growth rate is maximized when $\mu=0$ and $|\bm{n} \cdot \hat{\bm{B}}_0|=\pm1$ in equation (\ref{eq:cond_max}). By approximating $\omega_1\sim\omega_0$ \rei{and $\omega_{\text{c}}\gg\omega_0$}, we obtain
\proof{\begin{equation}
\Gamma_{\mathrm{C}}^{\max } =\sqrt{\frac{\pi}{32 \mathrm{e}}} \left(\frac{\omega_0}{\omega_{\text{c}}}\right)^2\left(1+\frac{\omega_{\text{p}}^2}{\omega_{\text{c}}^2}\right)^{-1}\frac{\omega_{\mathrm{p}}^{2} a_{\mathrm{e}}^{2}}{\omega_{0}} \frac{m_{\mathrm{e}} c^{2}}{k_{\mathrm{B}} T_{\mathrm{e}}}.
\label{eq:maximum_growth_rate_magnetic2}
\end{equation}}

Let us determine the angular frequency of the scattered wave. We define
\begin{equation}
\bm{k}_0 \cdot \bm{\hat{B}}_0 \equiv k_0 \cos \theta_0, \quad \bm{k}_1 \cdot \bm{\hat{B}}_0 \equiv k_1\cos \theta_1.
\end{equation}
Then,
\proof{\begin{equation}
\cos \theta_{k B}\simeq\frac{\cos \theta_1-\cos \theta_0}{\sqrt{2(1-\nu)}}.
\end{equation}}According to equation \rei{\eqref{eq:relation_omega1_and_omega_0}}\proof{ \eqref{eq:saidaiseichou_no_sannrannsyuuhasuu_Debye}}, we have
\proof{\begin{equation}
\omega_1\simeq \omega_0\left(1-\left|\cos \theta_1-\cos \theta_0\right| \sqrt{\frac{k_{\text{B}} T_{\text{e}}}{m_{\text{e}} c^2}\left(1+\frac{\omega_{\text{p}}^2}{\omega_{\text{c}}^2}\right)}\right).
\end{equation}}By using equation \eqref{eq:maximum_growth_rate_magnetic2}, the applicability condition derived from weak coupling condition \eqref{eq:weak_coupling_condition} is
\proof{\begin{equation}
a_{\mathrm{e}}\frac{\omega_0}{\omega_{\text{c}}} \ll 4\left(\frac{\text{e}}{\pi}\right)^{\frac{1}{4}} \frac{\omega_{0}}{\omega_{\mathrm{p}}}\left(1+\frac{\omega_{\text{p}}^2}{\omega_{\text{c}}^2}\right)^{\frac{3}{4}}\left(\frac{k_{\mathrm{B}} T_{\mathrm{e}}}{m_{\mathrm{e}} c^2}\right)^{\frac{3}{4}}\left|\cos \theta_{1}-\cos \theta_0\right|^{\frac{1}{2}}.
\label{eq:strength_parameter_limit_perp2-2}
\end{equation}}Additionally, the condition $\omega_{1} \sim \omega_{0} \gg |\omega|$ imposes the following restriction on plasma temperature,
\proof{\begin{equation}
\frac{k_{\mathrm{B}} T_{\mathrm{e}}}{m_{\mathrm{e}} c^2} \ll \frac{1}{\left(\cos \theta_{1} - \cos \theta_0\right)^2}\left(1+\frac{\omega_{\text{p}}^2}{\omega_{\text{c}}^2}\right)^{-1}.
\label{eq:tempreature_constraint_perp2}
\end{equation}}
\begin{figure}
\RaggedRight
\includegraphics[width=\columnwidth]{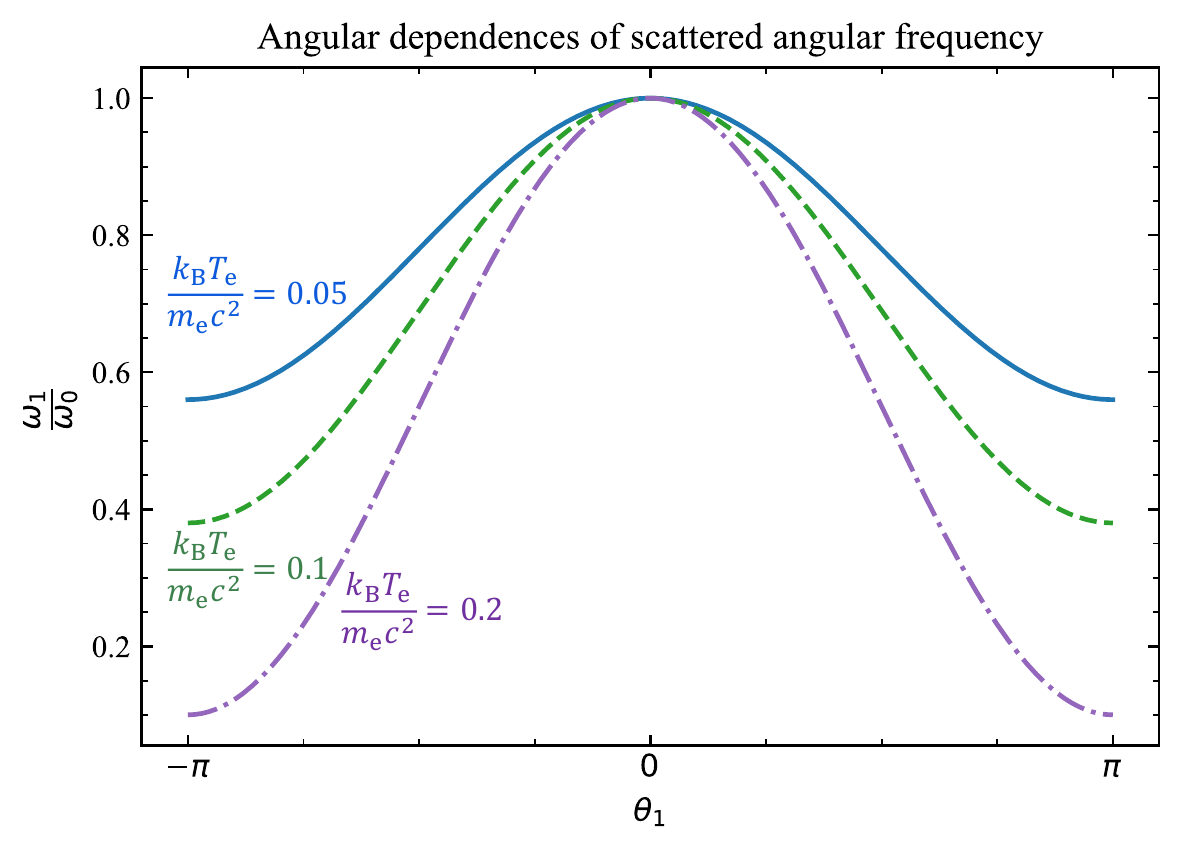}
\caption[Concept of Parametric Decay Instability]{\justifying The angular dependences of scattered angular frequency for a\proof{n} incident wave polarized perpendicular to the background magnetic field, $\bm{A}_{0\|} = \bm{0}$, when the condition \eqref{eq:Debye_length_is_very_large} is satisfied (Debye screening is ineffective). The blue solid line represents the case where $k_{\text{B}} T_{\text{e}} / (m_{\text{e}} c^{2})=0.05$, the green dashed line corresponds to $k_{\text{B}} T_{\text{e}} / (m_{\text{e}} c^{2})=0.1$, and the purple dash-dotted line illustrates the case where $k_{\text{B}} T_{\text{e}} / (m_{\text{e}} c^{2})=0.2$.}
\label{fig:induced_Compton_angular}
\end{figure}

For example, consider the case where $\theta_0=0$, corresponding to the induced Compton scattering of EM waves incident in the direction of the background magnetic field. The scattered wave’s angular frequency for different scattering angles $\theta_1$ is illustrated in Figure \ref{fig:induced_Compton_angular}. The incident wave transfers the maximum amount of energy to the plasma in the case of backward scattering ($\theta_1=\pm\pi$).

The maximum linear growth rate can be categorized based on the value of
\proof{\begin{equation}
\begin{aligned}
    \frac{\omega_{\mathrm{p}}^{2}}{k^{2} v_{\mathrm{th}}^{2}}(\nu = -1) &\simeq \frac{m_{\mathrm{e}} c^{2}}{8 k_{\mathrm{B}} T_{\mathrm{e}}}\left(\frac{\omega_{\mathrm{p}}}{\omega_0}\right)^2\left(1+\frac{\omega_{\text{p}}^2}{\omega_{\text{c}}^2}\right)^{-1} \\
    &=\frac{1}{8(2\pi)^2}\left(\frac{\lambda_0}{\lambda_{\text{De}}}\right)^2\left(1+\frac{\omega_{\text{p}}^2}{\omega_{\text{c}}^2}\right)^{-1},
\end{aligned}
\label{eq:Debye_length_ration}
\end{equation}}where $\lambda_0\equiv(2\pi c)/\omega_0$ is the wavelength of the incident wave, and we used equation \eqref{eq:approximate_wave_number_induced_Compton} for $k^2$. Depending on the quantity \eqref{eq:Debye_length_ration}, the maximum linear growth rate, \eqref{eq:growth_rate_induced_Compton_debye_sheelding} and \eqref{eq:maximum_growth_rate_magnetic2}, can be divided into the following cases,
\proof{\begin{equation}
\Gamma_{\mathrm{C}}^{\text{max}} = 
\begin{cases}
&\sqrt{\frac{\pi}{32 \mathrm{e}}} \left(\frac{\omega_{0}}{\omega_{\mathrm{c}}}\right)^{2}\left(1+\frac{\omega_{\text{p}}^2}{\omega_{\text{c}}^2}\right)^{-1} \frac{\omega_{\mathrm{p}}^{2} a_{\mathrm{e}}^2}{\omega_{0}} \frac{m_{\mathrm{e}} c^{2}}{k_{\mathrm{B}} T_{\mathrm{e}}},\\
& \quad\text{if}~\frac{8k_{\mathrm{B}} T_{\mathrm{e}}}{m_{\mathrm{e}} c^{2}}\left(\frac{\omega_0}{\omega_{\mathrm{p}}}\right)^2\left(1+\frac{\omega_{\text{p}}^2}{\omega_{\text{c}}^2}\right) \gg 1,\\
&\sqrt{\frac{32\mathrm{e}}{\pi}}\left(\frac{\omega_{0}}{\omega_{\mathrm{c}}}\right)^{2}\left(1+\frac{\omega_{\text{p}}^2}{\omega_{\text{c}}^2}\right)\left(\frac{\omega_{0}}{\omega_{\mathrm{p}}}\right)^{4} \frac{k_{\mathrm{B}} T_{\mathrm{e}}}{m_{\mathrm{e}} c^{2}} \frac{\left(a_{\mathrm{e}} \omega_{\mathrm{p}}\right)^{2}}{\omega_{0}},\\ 
& \quad\text{if}~\frac{8k_{\mathrm{B}} T_{\mathrm{e}}}{m_{\mathrm{e}} c^{2}}\left(\frac{\omega_0}{\omega_{\mathrm{p}}}\right)^2\left(1+\frac{\omega_{\text{p}}^2}{\omega_{\text{c}}^2}\right) \ll 1. 
\end{cases}
\label{eq:charged_mode_growth_rate_summary}
\end{equation}}Compared to the growth rate of induced Compton scattering in non magnetized $e^\pm$ plasma, the suppression effects can be expressed as follows:
\proof{\begin{equation}
\frac{\Gamma_{\mathrm{C}}^{\text{max}}}{\Gamma_{\mathrm{C}}^{\text{max}}(B_0=0)}= 
\begin{cases}
 &\left(\frac{\omega_{0}}{\omega_{\mathrm{c}}}\right)^{2}\left(1+\frac{\omega_{\text{p}}^2}{\omega_{\text{c}}^2}\right)^{-1},\\ &\quad\text{if}~\frac{8k_{\mathrm{B}} T_{\mathrm{e}}}{m_{\mathrm{e}} c^{2}}\left(\frac{\omega_0}{\omega_{\mathrm{p}}}\right)^2\left(1+\frac{\omega_{\text{p}}^2}{\omega_{\text{c}}^2}\right) \gg 1,\\
&\frac{\mathrm{e}}{2\pi}\left(\frac{\omega_{0}}{\omega_{\mathrm{c}}}\right)^{2}\left(1+\frac{\omega_{\text{p}}^2}{\omega_{\text{c}}^2}\right)\left(\frac{\omega_{0}}{\omega_{\mathrm{p}}}\right)^{4}\left(\frac{8k_{\mathrm{B}} T_{\mathrm{e}}}{m_{\mathrm{e}} c^{2}} \right)^2, \\
& \quad\text{if}~\frac{8k_{\mathrm{B}} T_{\mathrm{e}}}{m_{\mathrm{e}} c^{2}}\left(\frac{\omega_0}{\omega_{\mathrm{p}}}\right)^2\left(1+\frac{\omega_{\text{p}}^2}{\omega_{\text{c}}^2}\right) \ll 1.
\end{cases}
\end{equation}}When \proof{\(\frac{m_{\text{e}}c^2}{8k_{\text{B}}T_{e}}(\omega_{\text{p}}/\omega_0)^2(1 + \omega_{\mathrm{p}}^2 / \omega_{\mathrm{c}}^2)^{-1}=(\lambda_0/\lambda_{\mathrm{De}})^2(1 + \omega_{\mathrm{p}}^2 / \omega_{\mathrm{c}}^2)^{-1}/(32\pi^2)\)} is smaller than unity, the wavelength of the incident wave is sufficiently short compared to the Debye length, the growth rate of induced Compton scattering is suppressed only by the gyroradius effect \((\omega_0/\omega_{\text{c}})^2\). Conversely, when the wavelength ratio \proof{\((\lambda_0/\lambda_{\mathrm{De}})^2(1 + \omega_{\mathrm{p}}^2 / \omega_{\mathrm{c}}^2)^{-1}/(32\pi^2)\)} is significantly larger than unity, the growth rate of induced Compton scattering is not only suppressed by the gyroradius effect \((\omega_0/\omega_{\text{c}})^2\) but also further reduced by Debye screening \((\mathrm{e} / 2\pi) \left( \omega_{0} / \omega_{\mathrm{p}} \right)^{4} (\frac{8k_{\mathrm{B}} T_{\mathrm{e}}}{m_{\mathrm{e}} c^{2}})^{2}\), which hinders the charged modes. 
\proof{Additionally, the factor \((1 + \omega_{\mathrm{p}}^2 / \omega_{\mathrm{c}}^2)^{-1}\) is included, which arises due to the reduction of the phase velocity of the incident wave below $c$.\footnote{
\proof{In the regime of \(\omega_{0} \ll \omega_{\text{c}}\), the phase velocity of the growing transverse scattered wave is expressed as the Alfvén speed,
\begin{equation}
    v_{\text{A}}
    \equiv \frac{c}{\sqrt{1 + \frac{\omega_{\text{p}}^2}{\omega_{\text{c}}^2}}}.
    \nonumber
\end{equation}
Consequently, in equations \eqref{eq:growth_rate_induced_Compton_debye_sheelding}, \eqref{eq:maximum_growth_rate_magnetic2}, \eqref{eq:growth_rate_induced_Compton_subdominant}, and \eqref{eq:growth_rate_charged_induced_Comtpon}, the maximum growth rate can be rewritten by replacing the speed of light \(c\) with the phase velocity of the scattered wave.
}} This factor, referred to as the \textit{subluminal effect} in this study, is order unity in the magnetically dominated regime \((\omega_{\mathrm{c}} \gg \omega_{\mathrm{p}})\).}
See also Figure~\ref{fig:scattering_rate_summary}.

One of the application conditions for induced Compton scattering is that the energy given to the particles in the plasma by the ponderomotive potential $2n_{\text{e}0}\phi_{\pm}$ does not exceed the energy of the original 
EM field \rei{$\langle U_{\text{EM}}\rangle=2E_{0\|}^2$} \proof{$\langle U_{\text{EM}}\rangle=E_{0\perp}^2/(2\pi)$}. From equation \eqref{eq:pondero_drift}, which corresponds to the third term of equation \eqref{eq:ponderomotive_potential}, the condition for $\omega_{\text{c}}\gg\omega_0$ is given by
\begin{equation}
    \frac{1}{\sqrt{2}}\frac{\omega_{\text{p}}^2}{\omega_{\text{c}}}<\omega_0.
    \label{eq:application_condition_perp}
\end{equation}

If the density fluctuations propagate in a direction perpendicular to the \proof{background }magnetic field (for $k_{\|} = 0$), no Landau resonance occurs, except for cyclotron resonance. It is because the motion of charged particles perpendicular to the \proof{background }magnetic field is characterized by circular motion, which does not involve continual energy transfer from the EM wave to the particles (see section $55$ of the book \citep{1981phki.book.....L}). The detailed calculation is provided in Appendix \ref{Ap:wave_vector_perpendicular}.

\subsubsection{Neutral mode}
\begin{figure*}
\centering
\includegraphics[width=\textwidth]{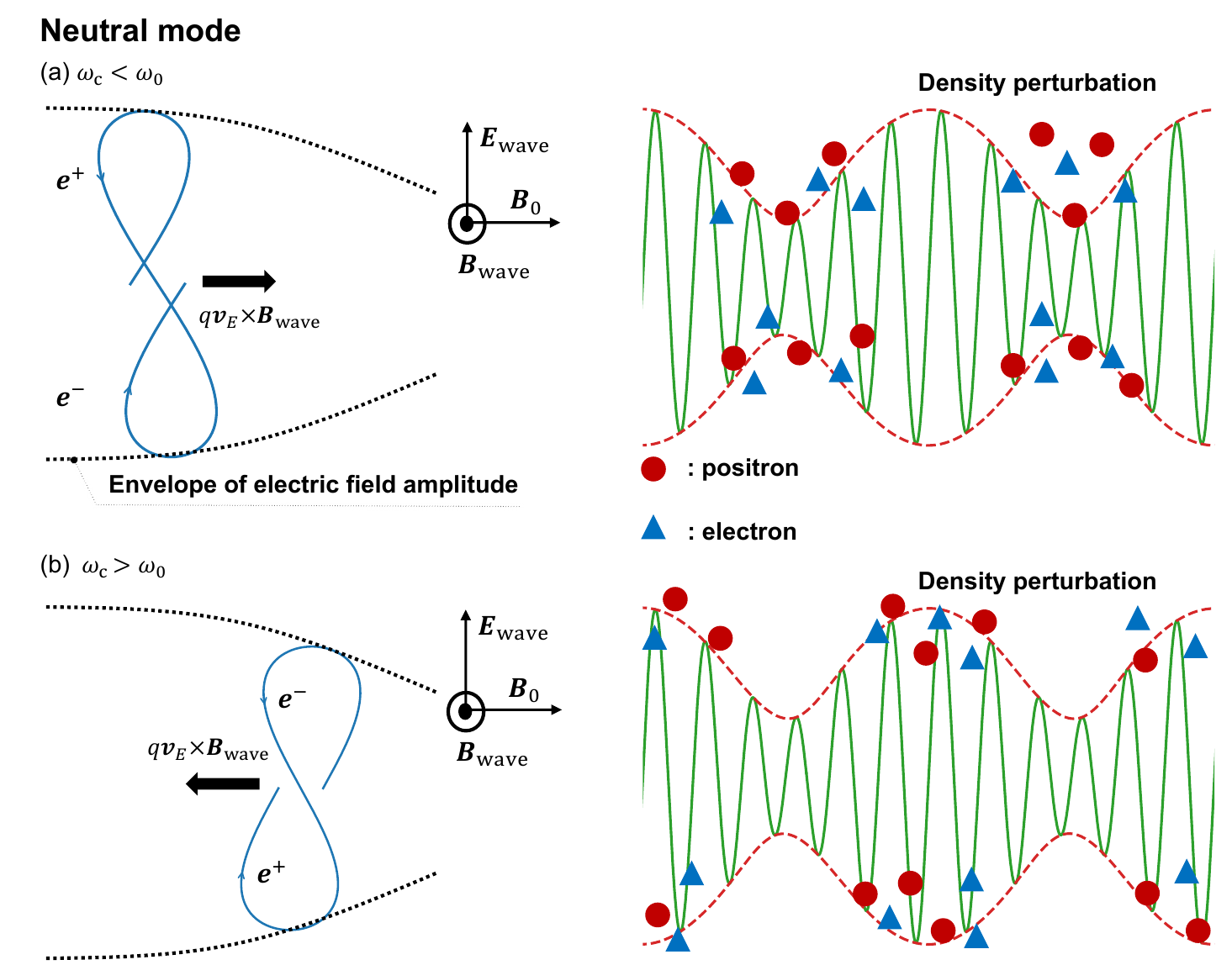}
\caption[Concept of Parametric Decay Instability]{\justifying The physical interpretation of ponderomotive force (left panel) and schematic picture of parametric instability (right panel) of the ``neutral mode" for an incident wave polarized perpendicular to the background magnetic field, $\bm{A}_{0\|}=\bm{0}$. In the left figure, the solid blue lines represent the trajectories of electrons and positrons perturbed by the beating EM wave. The black arrows depict the motion of the oscillation centers of these particle trajectories. Particles are perturbed in the direction of the electric field at the lowest order in \(v/c\), experiencing a force proportional to \(q\bm{E}_{\text{wave}} \omega_0^2/(\omega_0^2-\omega_{\text{c}}^2) \propto \bm{v}_{E}\) (\(\bm{v}_E\) corresponds to equation \eqref{eq:motion_magnetic_perp_no_drift}). Additionally,  at the next order in \(v/c\), the particles feel a Lorentz force from the oscillating magnetic field, given by  \((q/c)(\bm{v}_E \times \bm{B}_{\mathrm{wave}})\propto q^2(\bm{E}_{\text{wave}}\times\bm{B}_{\text{wave}})\omega_0^2/(\omega_0^2-\omega_{\text{c}}^2)\), which induces an oscillation at $2\omega_0$ and results in a figure-eight motion. (a) When \(\omega_{\text{c}} < \omega_0\), the factor \(\omega_0^2/(\omega_0^2-\omega_{\text{c}}^2)\) is positive, causing particles to experience a force toward smaller amplitudes, independent of their charge sign. This interpretation is similar to the case depicted in Figure \ref{fig:density_perturbation_nomg}. (b) When \(\omega_{\text{c}} > \omega_0\), the factor \(\omega_0^2/(\omega_0^2-\omega_{\text{c}}^2)\) becomes negative, reversing the sign of the Lorentz force compared to the \(\omega_{\text{c}} < \omega_0\) case, and causing particles to be pushed toward larger amplitudes. In the right panel: When \(\omega_{\text{c}} < \omega_0\), density fluctuations arise as particles, regardless of charge sign, accumulate in regions of smaller amplitude. Conversely, when \(\omega_{\text{c}} > \omega_0\), density fluctuations occur as particles accumulate in regions of larger amplitude, also independent of the charge sign.
EM field and the particle trajectories have a phase difference of \(\pi\), which \proof{is} derived from the equations of motion. \rei{(a) When \(\omega_{\text{c}} < \omega_0\), the orbit becomes smaller at larger amplitudes, and after one cycle, the oscillation centers shift in the direction of smaller amplitude. This behavior is similar to the case without a magnetic field. (b) When \(\omega_{\text{c}} > \omega_0\), the orbit becomes smaller at smaller amplitudes, and after one cycle, the oscillation centers shift in the direction of larger amplitude. This is due to the factor \(\omega_0^2/(\omega_0^2 - \omega_{\text{c}}^2)\), which causes the sign to reverse compared to the case without a magnetic field, resulting in both electrons and positrons advancing by a phase of \(\pi\).} 
}
\label{fig:neutral_mode}
\end{figure*}
We derive the  density fluctuations of the neutral mode. The ponderomotive force acting on the charged particles is independent of the charge sign, as represented by
\begin{equation}
\widetilde{\phi_{ \pm}}=-\frac{e^{2}}{m_{\mathrm{e}} c^{2}} \frac{\omega_{0}^{2}}{\omega_{\mathrm{c}}^{2}-\omega_{0}^{2}} \bm{A}_{1} \cdot \bm{A}_{0}^{*},
\label{eq:pondero_subdominant_magnetic}
\end{equation}
which comes from the second term of equation \eqref{eq:ponderomotive_potential}.
As shown Figure \ref{fig:nonlinear_current}, the nonlinear current \(\widetilde{\bm{j}_1}^{\text{nonlinear}}\) for the \rei{charged} \proof{neutral} mode\rei{, as given in equation \eqref{eq:definition_of_nonlinear_current_induced_Compton},} is obtained by combining the ponderomotive potential \eqref{eq:pondero_subdominant_magnetic} with the \rei{third} \proof{second} term of the particle velocity in equation \eqref{eq:general_velocity}. This velocity is expressed as
\begin{equation}
\bm{v}_{0\pm}^{(1)}=\mp \frac{e}{m_{\mathrm{e}} c} \frac{\omega_{0}^{2}}{\omega_{0}^{2}-\omega_{\mathrm{c}}^{2}} \bm{A}_0.
\label{eq:motion_magnetic_perp_no_drift}
\end{equation}
Using the dimensionless strength parameter from equation \eqref{eq:strength_parameter_no_magnetic}, the non-relativistic condition for the charged particle motion \eqref{eq:motion_magnetic_perp_no_drift} imposes the following constraint,
\begin{equation}
\left|\frac{\bm{v}_{0\pm}^{(1)}}{c}\right|=\frac{e\left|\bm{A}_0\right|}{m_{\text{e}} c^2}\left|\frac{\omega_{0}^{2}}{\omega_{0}^{2}-\omega_{\mathrm{c}}^{2}}\right| \ll 1 \Rightarrow a_{\text{e}}\left|\frac{\omega_{0}^{2}}{\omega_{0}^{2}-\omega_{\mathrm{c}}^{2}}\right|\ll 1.
\label{eq:strength_parameter_constraint_nonrela_neutral}
\end{equation}
In particular, in a strong \proof{background }magnetic field, the condition is given by 
\begin{equation}
    a_{\text{e}}\left(\frac{\omega_0}{\omega_{\mathrm{c}}}\right)^2 \ll 1.
\end{equation}
Substituting equation \eqref{eq:pondero_subdominant_magnetic} into equation \eqref{eq:density_fluctuation_pondero}, we obtain
\begin{equation}
\begin{aligned}
\widetilde{\delta n_{ \pm}}(\bm{k}, \omega) &= \frac{n_{\mathrm{e} 0} e^{2}}{m_{\mathrm{e}}^{2} c^{2}} \frac{\omega_{0}^{2}}{\omega_{\mathrm{c}}^{2}-\omega_{0}^{2}} \bm{A}_{1} \cdot \bm{A}_{0}^{*} \\
&\quad \times \sum_{\ell=-\infty}^{+\infty} \int \dd^{3} \bm{v} \frac{J_{\ell}^{2}\left(k_{\perp} r_{\mathrm{L}}\right)\bm{k} \cdot \frac{\partial f_{0\pm}}{\partial \bm{v}^{*}}}{\omega-k_{\|} v_{\|}-\ell \omega_{\mathrm{c}}}.
\end{aligned}
\label{eq:density_fluctuation_neutral}
\end{equation}
Unlike the charged mode, the density fluctuations in the neutral mode of electrons and positrons do not depend on the charge sign. Moreover, they do not depend on the longitudinal dielectric constant $\varepsilon_{\text{L}}$ and longitudinal electric susceptibility $H$. This is because, according to equation \eqref{eq:pondero_subdominant_magnetic}, \(\widetilde{\phi_{+}}\) and \(\widetilde{\phi_{-}}\) take the same value, causing the second term in equation \eqref{eq:density_fluctuation_pondero} to vanish. As illustrated in Figure \ref{fig:neutral_mode}, these ponderomotive forces do not lead to charge separation in the plasma.

Using equation \eqref{eq:density_fluctuation_neutral}, we derive the dispersion relation for the scattered wave. The $\omega_1$ component of the current induced by the beat between the incident and scattered waves is then given, using equation \eqref{eq:motion_magnetic_perp_no_drift}, by
\proof{\begin{equation}
\begin{aligned}
4\pi c\widetilde{\bm{j}}_1^{}(\bm{k}_1,\omega_1) &= 4\pi c\left\{e\left(n_{\mathrm{e0}}\bm{v}_{1+}^{(1)}+\widetilde{\delta n_{+}}\bm{v}_{0+}^{(1)}\right) \right.\\
&\quad \left. -e\left(n_{\mathrm{e0}}\bm{v}_{1-}^{(1)}+\widetilde{\delta n_{-}}\bm{v}_{0-}^{(1)}\right)\right\} \\
&= -\omega_{\mathrm{p}}^{2} \frac{\omega_{1}^{2}}{\omega_{1}^{2}-\omega_{\mathrm{c}}^{2}} \bm{A}_{1\perp} \\
&\quad - \frac{\omega_{\text{p}}^2}{2 n_{\mathrm{e0}}} \frac{\omega_{0}^{2}}{\omega_{0}^{2}-\omega_{\mathrm{c}}^{2}}\left(\widetilde{\delta n_{+}}+\widetilde{\delta n_{-}}\right) \bm{A}_0^{},
\end{aligned}
\label{eq:current1_neutral}
\end{equation}}\proof{where we first examine the case where \(\bm{A}_{1\|} = \bm{0}\). As will be shown later, if \(\bm{A}_{1\perp} = \bm{0}\), the scattered wave does not grow.} Substituting equation \eqref{eq:density_fluctuation_neutral} into equation \eqref{eq:current1_neutral} gives the source current. Then, substituting this current into the wave equation for the scattered wave \eqref{eq:wave_eq_for_scattered_wave_nonlinear}, the dispersion relation for the scattered wave is represented by
\proof{\begin{equation}
\begin{aligned}
c^{2}& k_{1}^{2}-\omega_{1}^{2}+\omega_{\mathrm{p}}^{2} \frac{\omega_{1}^{2}}{\omega_{1}^{2}-\omega_{\mathrm{c}}^{2}} =  \frac{1}{4} c^{2}\left(\omega_{\mathrm{p}} a_{\mathrm{e}} \mu\right)^{2} \\
&\times\left(\frac{\omega_{0}^{2}}{\omega_{0}^{2}-\omega_{\mathrm{c}}^{2}}\right)^{2} \sum_{\ell=-\infty}^{+\infty} \int \dd^{3} \bm{v} \frac{J_{\ell}^{2}\left(k_{\perp} r_{\mathrm{L}}\right)\bm{k} \cdot \frac{\partial f_{0}}{\partial \bm{v}^{*}}}{\omega-k_{\|} v_{\|}-\ell \omega_{\mathrm{c}}}.
\end{aligned}
\label{eq:dispersion_relation_pondero_subdominant}
\end{equation}}

The growth rate of the scattered wave can be derived from the dispersion relation \eqref{eq:dispersion_relation_pondero_subdominant}. The growth rate, \(\Gamma_{\mathrm{C}}\), reaches its maximum when the beating wave experiences Landau resonance with particles moving at velocities comparable to the phase velocity of the beat. As in previous discussions, we consider the weak coupling condition, \eqref{eq:condition_for_maximum_growth_rate_charged}. Under the approximation \(\omega_{\text{c}}\gg\omega_0\) and \(\omega_1\sim\omega_0\), by calculating the integral part of equation \eqref{eq:dispersion_relation_pondero_subdominant} according to the procedure in equation \eqref{eq:sekibun_no_keisan_induced_Compton}, the maximum growth rate is given by
\proof{\begin{equation}
\begin{aligned}
   \Gamma_{\mathrm{C}}^{\text{max}} \overset{\mu = \pm1}{=} \sqrt{\frac{\pi}{32 \mathrm{e}}} \frac{m_{\mathrm{e}} c^{2}}{k_{\mathrm{B}} T_{\mathrm{e}}} \frac{\left(\omega_{\mathrm{p}} a_{\mathrm{e}}\right)^{2}}{\omega_{0}}\left(\frac{\omega_{0}}{\omega_{\mathrm{c}}}\right)^{4}\left(1+\frac{\omega_{\text{p}}^2}{\omega_{\text{c}}^2}\right)^{-1}.
\end{aligned}
\label{eq:growth_rate_induced_Compton_subdominant}
\end{equation}}The most growing scattered wave has an electric field component that is parallel or anti-parallel to that of the incident wave,
\begin{equation}
\mu = \pm1.
\end{equation}
This result holds under the condition that the \proof{background }magnetic field is sufficiently strong, i.e., \(k_{\perp} v_{\mathrm{th}} / \omega_{\mathrm{c}} \ll k v_{\mathrm{th}} / |\omega| \sim 1\), considering only the \(\ell=0\) term in the infinite sum of the modified Bessel function. \rei{The angular frequency of the scattered wave is given by \eqref{eq:relation_omega1_omega0}.} \proof{The angular frequency of the scattered wave is given, from equation \eqref{eq:saidaiseichou_no_sannrannsyuuhasuu_Debye}, by
\begin{equation}
\begin{aligned}
\omega_1 &= \omega_0-\frac{1}{\sqrt{2}}k\cos\theta_{kB}v_{\text{th}}\\
&\sim \omega_0\left(1-\sqrt{ 2(1-\nu)\cos^2\theta_{kB}\frac{k_\text{B} T_{\text{e}}}{m_{\mathrm{e}} c^{2}}\left(1+\frac{\omega_{\text{p}}^2}{\omega_{\text{c}}^2}\right)}\right).    
\end{aligned}
\label{eq:omega1_omega0_relation_example}
\end{equation}
}The condition \(\omega_{1} \sim \omega_{0} \gg |\omega|\) imposes the \rei{temperature constraint \eqref{eq:temperature_constraint_induced_Compton}.}\proof{following constraint,
\begin{equation}
    \frac{k_{\mathrm{B}} T_{\mathrm{e}}}{m_{\mathrm{e}} c^2} \ll \frac{1}{2(1-\nu)\cos^2\theta_{kB}}\left(1+\frac{\omega_{\text{p}}^2}{\omega_{\text{c}}^2}\right)^{-1}.
\end{equation}}Additionally, from equation \eqref{eq:weak_coupling_condition}, we obtain the constraint by using equation\proof{s} \eqref{eq:growth_rate_induced_Compton_subdominant} and \rei{equation \eqref{eq:relation_omega1_and_omega_0}} \proof{\eqref{eq:omega1_omega0_relation_example}},
\proof{\begin{equation}
a_{\mathrm{e}}\frac{\omega_0}{\omega_{\text{c}}} \ll 4\left(\frac{2\text{e}}{\pi}\right)^{\frac{1}{4}}\left(\frac{k_{\mathrm{B}} T_{\mathrm{e}}}{m_{\mathrm{e}} c^2}\right)^{\frac{3}{4}} \frac{\omega_{\mathrm{c}}}{\omega_{\mathrm{p}}}\left(1+\frac{\omega_{\text{p}}^2}{\omega_{\text{c}}^2}\right)^{\frac{3}{4}}\left\{(1-\nu) \cos ^2 \theta_{kB}\right\}^{\frac{1}{4}}.
\label{eq:strength_parameter_limit_perp2-3}
\end{equation}}

\proof{When \(\bm{A}_{1\perp} = \bm{0}\), the dispersion relation for the scattered wave can be derived in the same manner as in the case of \(\bm{A}_{1\|} = \bm{0}\). In this case, \(\mu=0\) always holds, leading to the vanishing of the source term in the dispersion relation. Consequently, the scattered wave does not grow.}


Based on the discussion of charged mode and neutral mode, when \(\bm{A}_{0\|}=\bm{0}\), i.e., the electric field component of the incident EM wave is polarized perpendicular to the \proof{background }magnetic field, the maximum linear growth rate of induced Compton scattering can be concluded as follows:
\proof{\begin{equation}
\begin{aligned}
&\Gamma_{\mathrm{C}}^{\text{max}} = \sqrt{\frac{\pi}{32 \mathrm{e}}} 
\left(\frac{\omega_{0}}{\omega_{\mathrm{c}}}\right)^{2}
\frac{\omega_{\mathrm{p}}^{2} a_{\mathrm{e}}^2}{\omega_{0}} 
\frac{m_{\mathrm{e}} c^{2}}{k_{\mathrm{B}} T_{\mathrm{e}}}\left(1+\frac{\omega_{\text{p}}^2}{\omega_{\text{c}}^2}\right)^{{-1}} \\
&\times
\begin{cases}
1,\quad\text{if}~\frac{8k_{\mathrm{B}} T_{\mathrm{e}}}{m_{\mathrm{e}} c^{2}}
\left(\frac{\omega_0}{\omega_{\mathrm{p}}}\right)^2
\left(1+\frac{\omega_{\text{p}}^2}{\omega_{\text{c}}^2}\right) \gg 1,\\
\max\bigg\{
\left(\frac{\omega_{0}}{\omega_{\mathrm{c}}}\right)^{2}
,~\frac{\mathrm{e}}{2\pi}
\left(\frac{\omega_{0}}{\omega_{\mathrm{p}}}\right)^{4}
\left(\frac{8k_{\mathrm{B}} T_{\mathrm{e}}}{m_{\mathrm{e}} c^{2}} \right)^2
\left(1+\frac{\omega_{\text{p}}^2}{\omega_{\text{c}}^2}\right)^2
\bigg\}, \\
 \quad\text{if}~\frac{8k_{\mathrm{B}} T_{\mathrm{e}}}{m_{\mathrm{e}} c^{2}}
\left(\frac{\omega_0}{\omega_{\mathrm{p}}}\right)^2
\left(1+\frac{\omega_{\text{p}}^2}{\omega_{\text{c}}^2}\right) \ll 1.
\end{cases}
\end{aligned}
\label{eq:maximum_growth_rate_perp}
\end{equation}}Figure \ref{fig:scattering_rate_summary} summarises the linear growth rate of induced Compton scattering. 

\section{Discussion}
\label{sec:discussion_induced_Compton_magnetized}
\subsection{Scattering Rate for Incident Waves with a Broadband Spectrum}
In this section, we examine induced Compton scattering when the incident EM wave is not monochromatic but has a bandwidth $\Delta\omega$ and spectral shape $W(\omega)$. Observations indicate that some astrophysical emissions, including some FRBs, exhibit broad spectra with $\Delta\omega/\omega_0 \sim 1$ \citep{2021ApJ...923....1P}. Intuitively, when the incident radiative energy remains constant, increasing the bandwidth $\Delta\omega$ reduces the fraction of resonant energy transfer from the EM wave to the plasma. Consequently, the growth rate of induced Compton scattering decreases.

We briefly derive the linear growth rate of induced Compton scattering for a broadband incident EM wave. The method to derive the dispersion relation for the scattered wave is similar to that used in the monochromatic wave case. The vector potential is expressed as follows:
\begin{equation}
\begin{aligned}
\bm{A}(\bm{r}, t) &= \bm{A}_{0} \int_{-\infty}^{+\infty} \dd\omega_{0}^{\prime}~ W_0\left(\omega_{0}^{\prime}\right) 
\mathrm{e}^{i\left(\bm{k}_{0} \cdot \bm{r} - \omega_{0}^{\prime} t\right)} \\
&\quad + \bm{A}_{1} \int_{-\infty}^{+\infty} \dd\omega_{1}^{\prime}~ 
W_1\left(\omega_{1}^{\prime}\right) \mathrm{e}^{i\left(\bm{k}_{1} \cdot \bm{r} - \omega_{1}^{\prime} t\right)} + \text{c.c.}.
\end{aligned}
\end{equation}
Here, we assume that the spectral shape of the incident wave $W_0(\omega_0)$ and the scattered wave $W_1(\omega_1)$ are normalized as
\begin{equation}
\int_{-\infty}^{+\infty} W_i(x) ~\dd x = 1,\quad (i=0,1).
\end{equation}
The maximum linear growth rate of the scattered wave is obtained by averaging the growth rate at angular frequency $\omega_1^{\prime}$ weighted by the spectral shape of the scattered wave, $W_1(\omega_1)$:
\begin{equation}
\left\langle \Gamma_{\mathrm{C}, \max} \right\rangle \equiv \int_{-\infty}^{+\infty} W_1\left(\omega_{1}^{\prime}\right)2 \,\mathrm{Im}\, \omega_{1}^{\prime} 
~\dd\omega_{1}^{\prime}.
\end{equation}
To simplify the analysis, we consider the scattering of a flat spectrum with width $\Delta\omega$ in the following. The spectral shape $W(\omega)$ is given by
\begin{equation}
W_0(x)=W_1(x)=W(x) = \frac{1}{\Delta \omega},
\label{eq:flat_spectrum_induced}
\end{equation}
\begin{equation}
 \int_{\omega - \frac{1}{2} \Delta \omega}^{\omega + \frac{1}{2} \Delta \omega} \dd\omega~ W(\omega) = 1.
\end{equation}

First, we consider the case where the electric field component of the incident wave is polarized along the direction of the background magnetic field. The maximum linear growth rate of the ordinary mode is obtained by averaging the monochromatic result in equation \eqref{eq:maximum_growth_rate_no_magnetic_maedankai} with the weight \( W(\omega) \) of the incident and scattered waves. This is expressed as
\begin{equation}
\begin{aligned}
\left\langle \Gamma_{\mathrm{C}, \max}^{\|} \right\rangle &= -\frac{\sqrt{\pi}}{2} \left( \frac{c}{v_{\mathrm{th}}} \right)^{2}
\omega_{\mathrm{p}}^{2} a_{\mathrm{e}}^{2} \mu^{2} \int \dd\omega_{1}^{\prime} \frac{W\left(\omega_{1}^{\prime}\right)}{\omega_{1}^{\prime}} \\
&\times \int \dd\omega_{0}^{\prime}~ W\left(\omega_{0}^{\prime}\right) \frac{\omega^{\prime}}{k_{\parallel} v_{\mathrm{th}}}
\mathrm{e}^{-\left( \frac{\omega^{\prime}}{k_{\parallel} v_{\mathrm{th}}} \right)^{2}}.
\end{aligned}
\label{eq:broad_induced_Compton_parallel_maedankai}
\end{equation}
For an order estimation, the term \( 1/\omega_1^{\prime} \) does not change significantly over the integration interval from \( \omega_1 - (1/2)\Delta\omega \) to \( \omega_1 + (1/2)\Delta\omega \). Therefore, we can approximate \( 1/\omega_1^{\prime}\sim 1/\omega_1\sim 1/\omega_0\) and take it outside the integral. The growth rate then becomes
\begin{equation}
\begin{aligned}
\left\langle \Gamma_{\mathrm{C}, \max}^{\|} \right\rangle &\approx -\frac{\sqrt{\pi}}{2} \left( \frac{c}{v_{\mathrm{th}}} \right)^{2}
\frac{\omega_{\mathrm{p}}^{2} a_{\mathrm{e}}^{2} \mu^{2}}{\omega_0} \int \dd\omega_{1}^{\prime}~ W\left(\omega_{1}^{\prime}\right) \\
&\times \int \dd\omega_{0}^{\prime}~ W\left(\omega_{0}^{\prime}\right) \frac{\omega^{\prime}}{k_{\parallel} v_{\mathrm{th}}}
\mathrm{e}^{-\left( \frac{\omega^{\prime}}{k_{\parallel} v_{\mathrm{th}}} \right)^{2}}.
\end{aligned}
\label{eq:broad_induced_Compton_parallel_maedankai3}
\end{equation}
By substituting equation \eqref{eq:flat_spectrum_induced} into equation \eqref{eq:broad_induced_Compton_parallel_maedankai3}, we obtain the growth rate as
\begin{equation}
\left\langle \Gamma_{\mathrm{c}, \max}^{\|} \right\rangle = \frac{\sqrt{\pi}}{4} \left( \frac{c}{v_{\mathrm{th}}} \right)^{2}
\frac{\omega_{\mathrm{p}}^{2} a_{\mathrm{e}}^{2} \mu^{2}}{\omega_{0}} \left( \frac{k_{\|} v_{\mathrm{th}}}{\Delta \omega} \right)^{2} F(\alpha, \beta, \gamma).
\label{eq:broad_induced_Compton_parallel_maedankai2}
\end{equation}
Here, \( F(\alpha, \beta, \gamma) \) is a dimensionless integral defined by
\begin{equation}
F(\alpha, \beta, \gamma) \equiv \int_{\beta - \gamma}^{\beta + \gamma} \dd x \left\{ \mathrm{e}^{-(x - \alpha + \gamma)^{2}} - \mathrm{e}^{-(x - \alpha - \gamma)^{2}} \right\},
\end{equation}
with
\begin{equation}
\alpha \equiv \frac{\omega_{0}}{k_{\|} v_{\mathrm{th}}}, \quad
\beta \equiv \frac{\omega_{1}}{k_{\|} v_{\mathrm{th}}}, \quad
\gamma \equiv \frac{\Delta \omega}{2 k_{\|} v_{\mathrm{th}}}.
\end{equation}
Depending on the relative magnitude of the dimensionless bandwidth \( \gamma \) and the frequency difference \( |\alpha - \beta| \), the asymptotic values of \( F(\alpha, \beta, \gamma) \) are evaluated as
\begin{equation}
F(\alpha, \beta, \gamma) =
\begin{cases}
8 \mathrm{e}^{-(\alpha - \beta)^{2}} (\alpha - \beta) \gamma^{2}, & |\alpha - \beta| \gg \gamma \\
\sqrt{\pi}, & |\alpha - \beta| \ll \gamma
\end{cases}
.
\label{eq:muzigen_sekibun_induced_Compton}
\end{equation}
Substituting equation \eqref{eq:muzigen_sekibun_induced_Compton} into equation \eqref{eq:broad_induced_Compton_parallel_maedankai2} and using equations \eqref{eq:max_growth_dispersion} and \eqref{eq:relation_omega1_omega0}, the growth rate is evaluated as
\begin{equation}\proof{
\left\langle \Gamma_{\mathrm{c}, \max}^{\|} \right\rangle
= \left\{
\begin{array}{cc}
\sqrt{\frac{\pi}{32 \mathrm{e}}} \frac{\omega_{\mathrm{p}}^{2} a_{\mathrm{e}}^{2}}{\omega_{0}} \frac{m_{\mathrm{e}} c^{2}}{k_{\mathrm{B}} T_{\text{e}}}, &  \frac{\Delta \omega}{\omega_{0}} \ll \frac{ v_{\mathrm{th}}\sqrt{4(1-\nu)\cos^2\theta_{kB}}}{c} \\
\pi \frac{\omega_{\mathrm{p}}^{2} a_{\mathrm{e}}^{2}}{\omega_{0}} \left( \frac{\omega_{0}}{\Delta \omega} \right)^{2}, & \frac{\Delta \omega}{\omega_{0}} \gg \frac{v_{\mathrm{th}}\sqrt{4(1-\nu)\cos^2\theta_{kB}}}{c}
\end{array}
\right..}
\label{eq:growth_rate_parallel_induced_Comtpon}
\end{equation}

When the incident wave is broadband, that is, when \rei{$\Delta \omega/\omega_{0} \gg \sqrt{2}v_{\mathrm{th}}/c$} \proof{$\Delta \omega/\omega_{0} \gg 2\sqrt{2}v_{\mathrm{th}}/c$ for backward scattering}, the linear growth rate of the scattered wave exhibits a dependence of $\propto \Delta\omega^{-2}$. In particular, when $\omega_0 \sim \Delta\omega$, the growth rate agrees with \citet{2022ApJ...930..106G} and that evaluated using the quantum mechanical approach (see equation (65) of \citet{universe7030056}). This agreement is reasonable because the quantum mechanical approach naturally assumes that both the incident and scattered waves are broadband. On the other hand, in the narrowband case, i.e., when \rei{$\Delta \omega/\omega_{0} \ll \sqrt{2}v_{\mathrm{th}}/c$} \proof{$\Delta \omega/\omega_{0} \ll 2\sqrt{2}v_{\mathrm{th}}/c$ for backward scattering}, the growth rate is independent of the bandwidth and coincides with that of a monochromatic incident wave given by equation \eqref{eq:maximum_growth_rate_no_magnetic}.

Next, we calculate the linear growth rates of the charged and neutral modes respectively, i.e., when the incident wave is polarized perpendicular to the background magnetic field. \proof{We first discuss the maximum linear growth rate of the charged mode under two conditions: when Debye screening is effective \eqref{eq:Debye_screening_limit_induced_Compton} and when it is negligible \eqref{eq:Debye_length_is_very_large}. With Debye screening,} the maximum linear growth rate of the charged mode is obtained by multiplying equation \eqref{eq:charged_mode_totyuunosiki}\proof{, where the approximation $\omega_{\mathrm{c}}\gg\omega_0$ is applied,} for the monochromatic wave by the spectral shape \( W(\omega) \) and integrating over the angular frequency:
\proof{\begin{equation}
\begin{aligned}
\left\langle \Gamma_{\mathrm{c}, \mathrm{max}}^{\text{charged}} \right\rangle &= -\frac{\sqrt{\pi}}{2}
\left( \frac{c}{v_{\mathrm{th}}} \right)^{2} \left( \frac{\omega_{0}}{\omega_{\mathrm{c}}} \right)^{2}\left(1+\frac{\omega_{\text{p}}^2}{\omega_{\text{c}}^2}\right)^{-1}  \\
&\times \omega_{\mathrm{p}}^{2}
a_{\mathrm{e}}^{2}  \left( \frac{k v_{\mathrm{th}}}{\omega_{\mathrm{p}}} \right)^{4}\int \dd \omega_{1}^{\prime}~ \frac{W\left( \omega_{1}^{\prime} \right)}{\omega_{1}^{\prime}}\\
&\times \int \dd \omega_{0}^{\prime}~ W\left( \omega_{0}^{\prime} \right) \frac{\zeta \mathrm{e}^{\zeta^{2}}}{4 \left( 1 - 2 \zeta^{2} \right)^{2} \mathrm{e}^{2 \zeta^{2}} + 4 \zeta^{2} \pi}.
\end{aligned}
\end{equation}}Assuming that \( \zeta \) is small, we approximate the integrand as
\begin{equation}
\frac{\zeta \mathrm{e}^{\zeta^{2}}}{4 \left( 1 - 2 \zeta^{2} \right)^{2} \mathrm{e}^{2 \zeta^{2}} + 4 \zeta^{2} \pi} \approx \frac{1}{4} \zeta \mathrm{e}^{-\zeta^{2}}.
\label{eq:henna_kinzi_induced_Compton}
\end{equation}
Both sides of equation \eqref{eq:henna_kinzi_induced_Compton} exhibit similar exponential decay as \( \zeta \rightarrow \infty \), so the integral value does not change significantly. Therefore, an integral evaluation similar to that in equation \eqref{eq:broad_induced_Compton_parallel_maedankai} is possible. Using \rei{equations \eqref{eq:max_growth_dispersion}, \eqref{eq:relation_omega1_omega0}, and \eqref{eq:approximate_wave_number_induced_Compton}} \proof{equations \eqref{eq:approximate_wave_number_induced_Compton}, \eqref{eq:maximum_charged_Debye_induced_Compton}, and \eqref{eq:saidaiseichou_no_sannrannsyuuhasuu_Debye}}, we can evaluate the maximum linear growth rate of the charged mode as follows:
\proof{\begin{equation}
\begin{aligned}
\left\langle \Gamma_{\mathrm{c}, \max}^{\text{charged}} \right\rangle 
& \sim \frac{\pi}{4}  
\left( \frac{\omega_{0}}{\omega_{\mathrm{c}}} \right)^{2}
\left( 1 + \frac{\omega_{\text{p}}^2}{\omega_{\text{c}}^2} \right)^2 
\frac{\omega_{\mathrm{p}}^{2} a_{\mathrm{e}}^{2}}{\omega_{0}} \\
& \quad \times 
\left( \frac{8 k_{\mathrm{B}} T_{\mathrm{e}}}{m_{\mathrm{e}} c^{2}} \right)^{2} 
\left( \frac{\omega_{0}}{\omega_{\mathrm{p}}} \right)^{4}
\left( \frac{\omega_{0}}{\Delta \omega} \right)^{2}.
\end{aligned}
\label{eq:growth_rate_charged_induced_Comtpon}
\end{equation}
When Debye screening is negligible, the maximum linear growth rate of the charged mode is obtained by using equations \eqref{eq:cond_max} and \eqref{eq:charged_mode_growth_rate_without_debye_screening},
\begin{equation}
\left\langle \Gamma_{\mathrm{c}, \max}^{\text{charged}} \right\rangle\sim\pi \frac{\omega_{\mathrm{p}}^{2} a_{\mathrm{e}}^{2}}{\omega_{0}} 
\left( \frac{\omega_{0}}{\omega_{\mathrm{c}}} \right)^{2}
\left( \frac{\omega_{0}}{\Delta \omega} \right)^{2}.
\end{equation}
So, the maximum linear growth rate of the charged mode for a broadband incident wave is summarized as follows:
\begin{equation}
\begin{aligned}
\left\langle \Gamma_{\mathrm{c}, \max}^{\text{charged}} \right\rangle &\sim \pi \frac{\omega_{\mathrm{p}}^{2} a_{\mathrm{e}}^{2}}{\omega_{0}} \left( \frac{\omega_{0}}{\omega_{\mathrm{c}}} \right)^{2}
\left( \frac{\omega_{0}}{\Delta \omega} \right)^{2} \\
&\times \begin{cases}
\frac{1}{4}  
\left( 1 + \frac{\omega_{\text{p}}^2}{\omega_{\text{c}}^2} \right)^2 
\left( \frac{8 k_{\mathrm{B}} T_{\mathrm{e}}}{m_{\mathrm{e}} c^{2}} \right)^{2} 
\left( \frac{\omega_{0}}{\omega_{\mathrm{p}}} \right)^{4},\\
\quad\text{if}~\frac{8k_{\mathrm{B}} T_{\mathrm{e}}}{m_{\mathrm{e}} c^{2}}\left(\frac{\omega_0}{\omega_{\mathrm{p}}}\right)^2\left(1+\frac{\omega_{\text{p}}^2}{\omega_{\text{c}}^2}\right) \ll 1,\\
1,~\text{if}~\frac{8k_{\mathrm{B}} T_{\mathrm{e}}}{m_{\mathrm{e}} c^{2}}\left(\frac{\omega_0}{\omega_{\mathrm{p}}}\right)^2\left(1+\frac{\omega_{\text{p}}^2}{\omega_{\text{c}}^2}\right) \gg 1.
\end{cases}
\end{aligned}
\end{equation}}Similarly, the maximum linear growth rate of the neutral mode is obtained as
\begin{equation}
\left\langle \Gamma_{\mathrm{c}, \max}^{\text{neutral}} \right\rangle \sim \pi \frac{\omega_{\mathrm{p}}^{2} a_{\mathrm{e}}^{2}}{\omega_{0}} \left( \frac{\omega_{0}}{\omega_{\mathrm{c}}} \right)^{4}
\left( \frac{\omega_{0}}{\Delta \omega} \right)^{2}.
\label{eq:growth_rate_neutral_induced_Comtpon}
\end{equation}

\subsection{Application of Scattering Theory to FRB Propagation During Magnetar Bursts}
We consider induced Compton scattering during the propagation of FRBs in a magnetar magnetosphere. Previously, this was discussed based on theories without \proof{background }magnetic fields. Our study reveals the significant suppression of the linear growth rate due to both the gyroradius and Debye screening effects. Particularly, during X-ray bursts from magnetars, high-density fireball plasma is formed \citep{1995MNRAS.275..255T}, through which FRBs may propagate \citep{2020ApJ...904L..15I,2023MNRAS.519.4094W}. Therefore, the suppression effect due to Debye screening could be crucial.

To estimate the linear growth rate of induced Compton scattering for an FRB in a typical magnetar magnetosphere region (\(r \sim 10^8~\mathrm{cm}\)), we begin by evaluating the plasma frequency, cyclotron frequency, and the strength parameter of the FRB. Assuming a dipole magnetic field for the magnetar, the \proof{background }magnetic field strength is given by
\begin{equation}
B(r) \sim B_{\mathrm{p}} \frac{R^{3}}{r^{3}} \sim 10^{8} \, \mathrm{G}~ \, r_{8}^{-3} \, R_6^{3} \, B_{\mathrm{p,14}},
\end{equation}
where \(R\sim10^6~\mathrm{cm}\) is the radius of the magnetar. The plasma density in the magnetar magnetosphere can be characterized by the Goldreich-Julian density \citep{1969ApJ...157..869G},
\begin{equation}
n_{\mathrm{GJ}} \sim \frac{B(r)}{c e P} \sim 6.9 \times 10^{6} \, \mathrm{cm}^{-3} \, P_{\mathrm{sec}}^{-1} \, r_{8}^{-3} \, R_6^{3} \, B_{\mathrm{p,14}},
\end{equation}
where \(P\) is the spin period of the magnetar. The plasma frequency for $e^\pm$ plasma is expressed as follows:
\begin{equation}
\begin{aligned}
\omega_{\mathrm{p}} &= \sqrt{\frac{8 \pi e^{2}\mathcal{M} n_{\mathrm{GJ}}}{m_{\mathrm{e}}}} \\
&\sim 2.1 \times 10^{11} \, \mathrm{Hz}~ \mathcal{M}_{6}^{\frac{1}{2}} \, P_{\mathrm{sec}}^{-\frac{1}{2}} \, r_{8}^{-\frac{3}{2}} \, R_6^{\frac{3}{2}} \, B_{\mathrm{p,14}}^{\frac{1}{2}},
\end{aligned}
\label{eq:omegap_FRB_application_induced_Compton}
\end{equation}
where \(\mathcal{M}\) denotes the multiplicity. The plasma density can be significantly enhanced by the $e^\pm$ pair production from high energy photons \citep{1983ApJ...273..761D}. For instance, in a fireball associated with an X-ray burst, \(\mathcal{M}\) can
increase up to \(\sim 10^{7}\)
\citep{1995MNRAS.275..255T,2020ApJ...904L..15I} (see also \citep{2000ApJ...543..340T,2002ApJ...574..332T}). The electron cyclotron frequency is represented by
\begin{equation}
\omega_{\mathrm{c}} \sim 1.8 \times 10^{15} \, \mathrm{Hz}~ \, r_{8}^{-3} \, R_6^{3} \, B_{\mathrm{p,14}}.
\label{eq:omegac_FRB_application_induced_Compton}
\end{equation}
\proof{According to equations \eqref{eq:omegap_FRB_application_induced_Compton} and \eqref{eq:omegac_FRB_application_induced_Compton}, the ratio of plasma frequency to cyclotron frequency is given by
\begin{equation}
    \frac{\omega_{\mathrm{p}}}{\omega_{\mathrm{c}}} \sim 1.2 \times 10^{-4} \frac{\mathcal{M}_6^{\frac{1}{2}} r_8^{\frac{3}{2}}}{P_{\mathrm{sec}}^{\frac{1}{2}} R_6^{\frac{3}{2}} B_{\mathrm{p,14}}^{\frac{1}{2}}}.
\end{equation}
Thus, in applying induced Compton scattering to FRBs, the subluminal effect \((1 + \omega_{\mathrm{p}}^2 / \omega_{\mathrm{c}}^2) \sim 1\) can be neglected.}
The dimensionless strength parameter \eqref{eq:strength_parameter_no_magnetic} of the FRB, given the Poynting luminosity \(L \sim 2c r^{2} E^{2}\), can be evaluated as follows:
\begin{equation}
\begin{aligned}
a_{\mathrm{e}} &= \frac{2e E}{m_{\mathrm{e}} c \omega_{0}} \sim \frac{e}{m_{\mathrm{e}} c \omega_{0}} \left(\frac{2L}{c r^{2}}\right)^{\frac{1}{2}} \\
&\sim 2.3 \times 10^{3} \, \nu_9^{-1} \, L_{38}^{\frac{1}{2}} \, r_{8}^{-1}.
\end{aligned}
\label{eq:FRB_strength_no_magnetic}
\end{equation}
Here, we use the isotropic luminosity of Galactic FRB 20200428, and $\omega=2\pi\nu$. It should be noted that the isotropic luminosity of a typical cosmological FRB is \(\sim10^{42}\) erg/s, which is several orders of magnitude higher than that of Galactic FRB 20200428 (see the review paper \citet{2023RvMP...95c5005Z} for details). 

We compare the growth rate of the induced Compton scattering derived in this study with the duration of FRBs. Observations indicate that FRB 20200428 have durations of approximately $0.5~\text{ms}$ \citep{2020Natur.587...54C}, and typical FRBs also have millisecond durations. Therefore, we assume 
\begin{equation}
    \Delta t^{-1} = 10^{3}~\text{s}^{-1}.
\end{equation}
The inverse of the dynamical time at $r \sim 10^8~\mathrm{cm}$ is given by
\begin{equation}
t_{\mathrm{dyn}}^{-1} = \frac{c}{r} \sim 3 \times 10^{2} \, \mathrm{s}^{-1}\, r_{8}^{-1}.
\end{equation}
Since $t_{\mathrm{dyn}}^{-1} < \Delta t^{-1}$, the characteristic timescale of the FRB in a magnetar magnetosphere is determined by its duration rather than the dynamical time. Observations also indicate that the bandwidth of FRB 20200428 ranges from at least $400~\mathrm{MHz}$ to $1468~\mathrm{MHz}$ \citep{2020Natur.587...54C,2020Natur.587...59B,2020ATel13687....1Z,2020ApJ...898L..29M,2020ATel13686....1T,2020ATel13688....1R}. Assuming $\nu \sim \Delta\nu = 10^9~\mathrm{Hz}$, we evaluate the growth time scale of induced Compton scattering using the broadband equations \eqref{eq:growth_rate_parallel_induced_Comtpon}, \eqref{eq:growth_rate_charged_induced_Comtpon} and \eqref{eq:growth_rate_neutral_induced_Comtpon}.


\proof{When the electric field of FRBs is aligned with the background magnetic field, the angular frequency must not exceed the plasma frequency, as described by equation \eqref{eq:cutoff_frequency_induced_Compton}. Therefore, FRB waves usually cannot propagate. In the following, we consider the case where the plasma density is extremely low ($\mathcal{M}\sim1$) enough for propagation.}
For the electric field component of FRBs along the background magnetic field, it is shown by equation \eqref{eq:FRB_strength_no_magnetic} that the condition for the particles to remain non-relativistic, described by equation \eqref{eq:strength_parameter_constraint_nonrela_para}, is not satisfied. Consequently, the linear growth rate for the ordinary mode derived in this study, given by equation \eqref{eq:growth_rate_parallel_induced_Comtpon}, cannot be applied. 
The growth rate of induced Compton scattering for strong incident waves ($a_{\mathrm{e}}>1$) without a background magnetic field has been derived \proof{(in the oscillation-center frame of particles)} \citep{2019MNRAS.490.1474L,universe7030056}. Since particles can move freely along the direction of the background magnetic field, it \rei{is} \proof{may be} valid to apply the growth rate derived for the case without a magnetic field\footnote{\proof{In a strong incident wave, the $\bm{E}_{\text{wave}} \times \bm{B}_{\text{wave}}$ drift of particles in the $e^\pm$ plasma becomes significant along the propagation direction of the wave \citep{1975ctf..book.....L,1970PhRvD...1.2738S,1971ApJ...165..523G,2003PhyU...46..645B}. However, when the background magnetic field is sufficiently stronger than the magnetic component of the wave, i.e., $|\bm{B}_{\text{wave}}| \ll |\bm{B}_0|$, the drift motion is suppressed, and the particles oscillate while remaining frozen to the background magnetic field. Consequently, the growth rate derived in the nonmagnetized plasma oscillation-center frame is expected to be applicable to a perpendicularly propagating O-mode incident wave in the plasma rest frame.
}}. The growth rate is described by equation (66) of \citet{universe7030056} and can be estimated as follows:
\begin{equation}
\begin{aligned}
\left( t_{\mathrm{c}, \|}^{\text{broad}} \right)^{-1} &\sim  \frac{\omega_{\mathrm{p}}^{2} }{\omega_{0}a_{\mathrm{e}}}  \\
&= 3.1 \times 10^{3} \mathrm{~s}^{-1} \frac{\mathcal{M}_{0} R_{6}^{3} B_{\mathrm{p}, 14} }{P_{\mathrm{sec}} r_{8}^{2} \nu_{9}L_{38}^{\frac{1}{2}}} \sim \Delta t^{-1}
\end{aligned}
\end{equation}
\rei{This result indicates that the electric field component along the background magnetic field should be significantly attenuated by induced Compton scattering for out parameters.}
\proof{Only in the case of low plasma density can the electric field component parallel to the background magnetic field propagate without being affected by induced Compton scattering.}

For the FRB polarized perpendicular to the \proof{background }magnetic field, the applicability condition for the growth rate of induced Compton scattering, given by equation \eqref{eq:strength_parameter_constraint_nonrela_perp}, is
\begin{equation}
a_{\mathrm{e}} \frac{\omega_{0}}{\omega_{\mathrm{c}}} \sim 8.0 \times 10^{-3} \, L_{38}^{1/2} \, r_{8}^{2} \, B_{\mathrm{p},14}^{-1}R_{\mathrm{NS,6}}^{-3} \ll 1,
\label{eq:nonrelativistic_condition_for_X-mode_FRB}
\end{equation}
which is well satisfied. 
Instead, the distance from the center of the magnetar corresponding to this condition is expressed as follows:
\begin{equation}
r<1.1\times10^9~\mathrm{cm}~L_{38}^{-\frac{1}{2}}B_{\mathrm{p,14}}R_{\mathrm{NS,6}}^{\frac{3}{2}}.
\label{eq:relativistic_condition_for_induced_scattering_X-mode}
\end{equation}
The boundary of this region is comparable to the radius of the magnetar's light cylinder,
\begin{equation}
R_{\text{L}}=\frac{cP}{2\pi}\sim5\times10^9~\text{cm}~P_{\text{sec}}.
\end{equation}

The condition for Debye screening to be effective in induced Compton scattering is \( 8k_{\text{B}}T_{\text{e}}/(m_{\text{e}}c^2)(\omega_0/\omega_{\text{p}})^2 < 1 \). Converting this into a condition on the multiplicity, we obtain
\begin{equation}
\mathcal{M} > 1.1 \times 10^{3} \frac{r_{8}^{3} \nu_{9}^{2} P_{\mathrm{sec}} T_{\mathrm{80keV}}}{B_{\mathrm{p},14} R_{\mathrm{NS},6}^{3}}.
\end{equation}
This condition can be achieved through $e^{\pm}$ pair production by high energy photons. Instead, the condition can be expressed as a constraint on the distance from the center of the magnetar, given by the following equation,
\begin{equation}
r<9.1\times10^8~\mathrm{cm}~\frac{\mathcal{M}_6^{\frac{1}{3}}B_{\mathrm{p,14}}^{\frac{1}{3}}R_{\mathrm{NS,6}}}{\nu_9^{\frac{2}{3}}P_{\mathrm{sec}}^{\frac{1}{3}}T_{\mathrm{80keV}}^{\frac{1}{3}}}.
\end{equation}
This result suggests that Debye screening can operate effectively against induced Compton scattering over a wide range of the magnetar magnetosphere.
The growth rates of the charged and neutral modes are evaluated as follows\proof{, assuming that the subluminal effect can be neglected}:
\begin{equation}
\proof{\begin{aligned}
&\left( t_{\text{charged}}^{\text{broad}} \right)^{-1} = \frac{\pi}{4}  \left( \frac{\omega_{0}}{\omega_{\mathrm{c}}} \right)^{2}
\frac{\omega_{\mathrm{p}}^{2} a_{\mathrm{e}}^{2}}{\omega_{0}} \left( \frac{8k_{\mathrm{B}} T_{\mathrm{e}}}{m_{\mathrm{e}} c^{2}} \right)^{2}
\left( \frac{\omega_{0}}{\omega_{\mathrm{p}}} \right)^{4} \left( \frac{\omega_{0}}{\Delta \omega} \right)^{2} \\
&= 4.6 \times 10^{2} \, \mathrm{s}^{-1} \frac{P_{\mathrm{sec}} r_{8}^{7} L_{38}T_{\mathrm{80keV}}^2}{\mathcal{M}_{6} \nu_{9}^{2} R_{6}^{9} B_{\mathrm{p},14}^{3}}
\left( \frac{ \Delta \nu/\nu_0}{1} \right)^{-2} \sim \Delta t^{-1},
\end{aligned}}
\label{eq:charged_Compton_FRB}
\end{equation}
\begin{equation}
\begin{aligned}
&\left( t_{\text{neutral}}^{\text{broad}} \right)^{-1} = \pi \frac{\omega_{\mathrm{p}}^{2} a_{\mathrm{e}}^{2}}{\omega_{0}}
\left( \frac{\omega_{0}}{\omega_{\mathrm{c}}} \right)^{4} \left( \frac{\omega_{0}}{\Delta \omega} \right)^{2} \\
&= 1.8 \times 10^{-2} \, \mathrm{s}^{-1} \frac{\mathcal{M}_{6} L_{38} \nu_{9}^{2} r_{8}^{7}}{P_{\mathrm{sec}} R_{6}^{9} B_{\mathrm{p},14}^{3}}
\left( \frac{ \Delta \nu/\nu_0}{1} \right)^{-2} \ll \Delta t^{-1}.
\end{aligned}
\label{eq:neutral_Compton_FRB}
\end{equation}
\proof{FRBs can propagate without being significantly affected by induced Compton scattering at least in the region of \( r = 10^8 \) cm.}
Notably, the growth rate of the charged mode \rei{increases} \proof{decreases} with higher plasma density. This is because the suppression effect of Debye screening on induced Compton scattering becomes more significant as plasma density increases. This dependence is opposite to that of the neutral mode and the ordinary mode. 
In the previous studies, the linear growth rate for induced Compton scattering in strongly magnetized $e^\pm$ plasma has taken only the gyroradius effect \((\omega_0/\omega_{\text{c}})^2\) into account \citep{1976MNRAS.174...59B,2020MNRAS.494.1217K,2024PhRvD.109d3048N} (see Figure \ref{fig:scattering_rate_summary}). In this case, the linear growth rate is expressed as
\begin{equation}
\begin{aligned}
    t_{\text{C},\text{wrong}}^{-1} &\sim \frac{\left(\omega_{\mathrm{p}} a_{\mathrm{e}}\right)^{2}}{\omega_{0}}\left(\frac{\omega_{0}}{\omega_{\mathrm{c}}}\right)^{2} \\
    &\sim 4.5 \times 10^{8} \, \mathrm{s}^{-1}~\frac{r_{8} \, L_{38} \, \mathcal{M}_{6} \, R_{\mathrm{NS,6}}^{3}}{B_{\mathrm{p,14}} \, \nu_9 \, P_{\mathrm{sec}} \,} \gg \Delta t^{-1}
\end{aligned}
\end{equation}
which results in a large overestimation.

In addition to induced Compton scattering, it is necessary to consider other interactions between EM waves and plasma and to compare how these processes compete. For instance, there is an instability that
fast magnetosonic waves (or X-mode waves) decay into Alfvén waves 
in force-free magnetohydrodynamics (MHD) \citep{1998PhRvD..57.3219T,2019ApJ...881...13L,2019MNRAS.483.1731L,2023ApJ...957..102G}. 
Their study shows that in neutron star magnetospheres, this decay occurs efficiently when both the cyclotron frequency and plasma frequency exceed the angular frequency of the incident wave. However, determining how this instability competes with other induced scattering processes requires more detailed analysis.

In this study, we restricted our discussion to regions in the magnetosphere where the FRB amplitude satisfies the condition $a_{\mathrm{e}} \omega_{0}/\omega_{\mathrm{c}}<1$. However, in regions where the condition $a_{\mathrm{e}} \omega_{0}/\omega_{\mathrm{c}}>1$ holds, various nonlinear interactions become significant. These regions are strongly influenced by the interaction of the wave with plasma, potentially leading to wave attenuation or the suppression of wave propagation.

In a low density, where the test-particle approximation holds, electrons and positrons may be accelerated to relativistic speeds by the incident wave. 
\proof{
The test-particle approximation condition is expressed by equation \eqref{eq:Debye_length_is_very_large}, and in the case of maximum growth, it requires equation \eqref{eq:Debye_length_ration}. 
}This process could trigger avalanche pair production and cause significant wave attenuation \citep{2021ApJ...922L...7B,2022PhRvL.128y5003B}. However, this argument relies on the cross section for test-particle scattering and may change substantially when plasma effects are included. Additionally, when the plasma moves in bulk at relativistic speeds, the wave vector of the incident wave aligns with the background magnetic field in the plasma rest frame. This alignment reduces the scattering cross section per particle, which can suppress wave attenuation \citep{2022MNRAS.515.2020Q}.

In contrast, in high-density plasma where the test-particle approximation breaks down, collective behaviors dominate.
\proof{
This condition is expressed by equation \eqref{eq:Debye_screening_limit_induced_Compton}, and in the case of maximum growth, it requires equation \eqref{eq:Debye_length_ration}. 
}One such mechanism is a nonlinear shock by the incident wave, which could lead to significant energy dissipation of the wave \citep{2022arXiv221013506C,2023ApJ...959...34B,2024ApJ...975..223B,2024arXiv240715076V}.
Still the propagation of superluminal EM waves is not completely understood in the regime $\omega_{\mathrm{c}}/a_{\mathrm{e}}<\omega_{0}<a_{\mathrm{e}}\omega_{\mathrm{c}}$
\citep{2024A&A...690A.332S}. 
\rei{Note also that, when GHz FRBs propagate in association with large EM pulses, the FRB amplitude remains within the condition $a_{\mathrm{e}} \omega_{0}/\omega_{\mathrm{c}}<1$, making the discussion of our study applicable \citep{2020ApJ...897....1L}.}
\proof{We showed in equations \eqref{eq:charged_Compton_FRB} and \eqref{eq:neutral_Compton_FRB} that FRB waves can propagate through the magnetosphere in the linear regime, but at the same time, our results suggest that the waves may be scattered as they approach the nonlinear regime. In reality, they may exit the magnetosphere together with low-frequency magnetic pulses in the kHz-MHz range, which could keep their amplitude within the linear regime \citep{2020ApJ...897....1L}. However, it is necessary to consider the dynamics of how these magnetic pulses escape from the magnetosphere, which we leave for future studies.}

If the plasma is moving relativistically in bulk, it is necessary to reestimate the above order evaluations. The evaluations must be reassessed in the comoving frame of the plasma, with each physical quantity Lorentz transformed to the rest frame. This will be addressed in future work.

\section{Summary}
\begin{figure*}
\centering
\includegraphics[width=\textwidth]{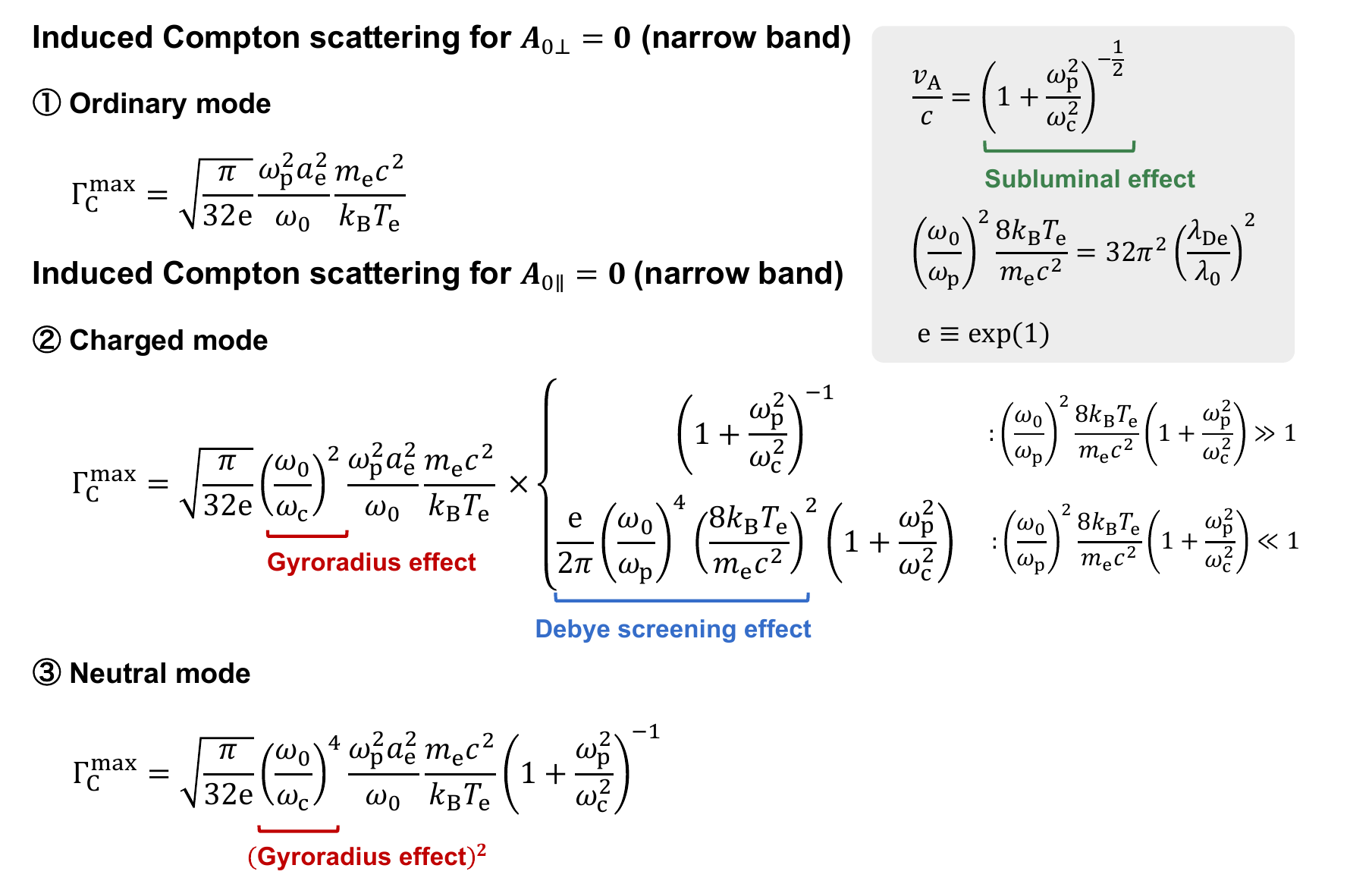}
\caption[Concept of Parametric Decay Instability]{\justifying Summary of the growth rate for induced Compton scattering. Our study has identified induced Compton scattering corresponding to three types of unstable modes: ordinary mode, charged mode, and neutral mode. When the incident wave is polarized along the \proof{background }magnetic field ($\bm{A}_{\perp} = \bm{0}$), it is scattered according to the growth rate of the ordinary mode given by equation \eqref{eq:maximum_growth_rate_no_magnetic}. In contrast, when the incident wave is polarized perpendicular to the \proof{background }magnetic field ($\bm{A}_{\parallel} = \bm{0}$), it is scattered according to the growth rates of the charged mode and neutral mode, as described by equations \eqref{eq:charged_mode_growth_rate_summary} and \eqref{eq:growth_rate_induced_Compton_subdominant}, respectively. In particular, the growth rate \rei{is determined by the comparison between} \proof{depends on the Debye factor \(\frac{8k_{\text{B}}T_{\text{e}}}{m_{\text{e}}c^2}\left(\omega_0/\omega_{\mathrm{p}}\right)^2(1 + \omega_{\mathrm{p}}^2 / \omega_{\mathrm{c}}^2)\)} from equations \eqref{eq:Debye_screening_limit_induced_Compton} and \eqref{eq:Debye_length_is_very_large}. When \proof{\(\frac{8k_{\text{B}}T_{\text{e}}}{m_{\text{e}}c^2}\left(\omega_0/\omega_{\mathrm{p}}\right)^2(1 + \omega_{\mathrm{p}}^2 / \omega_{\mathrm{c}}^2) \gg 1\)}, the growth rate is determined by the charged mode and suppressed by the gyroradius effect \((\omega_0/\omega_{\text{c}})^2\). On the other hand, when \proof{\(\frac{8k_{\text{B}}T_{\text{e}}}{m_{\text{e}}c^2}\left(\omega_0/\omega_{\mathrm{p}}\right)^2(1 + \omega_{\mathrm{p}}^2 / \omega_{\mathrm{c}}^2) \ll 1\)}, i.e., when the plasma density is sufficiently high or the plasma temperature is sufficiently low, the growth rate of the charged mode is further suppressed by Debye screening in addition to the gyroradius effect \((\omega_0/\omega_{\text{c}})^2\), by a factor of \((\mathrm{e} / 2\pi) \left( \omega_{0} / \omega_{\mathrm{p}} \right)^{4} \left(8 k_{\mathrm{B}} T_{\mathrm{e}} / m_{\mathrm{e}} c^{2} \right)^{2}\). Consequently, the overall \rei{reaction} \proof{growth} rate is determined by the larger rate of the charged mode with Debye screening or neutral mode.

}
\label{fig:scattering_rate_summary}
\end{figure*}
In this study, we analytically formulated the linear growth rate for induced Compton scattering
in magnetized $e^\pm$ plasma,
as summarized in Figure \ref{fig:scattering_rate_summary}, for the first time. As a result, we demonstrated that the gyroradius effect and the Debye screening effect can significantly impact the linear growth rate of induced Compton scattering in magnetized $e^\pm$ plasma. \proof{Additionally, the subluminal effect is negligible in magnetically dominated environments (e.g., within magnetar magnetospheres).} We also derived the density fluctuations in magnetized $e^\pm$ plasma caused by the ponderomotive potential created by the beat of incident and scattered waves
in equation (\ref{eq:density_fluctuation_pondero}).

In addition to the ordinary mode for O-mode waves, two types of density fluctuation modes determine the linear growth rate of induced Compton scattering, i.e, the charged mode and the neutral mode, for EM waves polarized perpendicular to the background magnetic field. In particular, in a strong \proof{background }magnetic field, the charged mode is the first-order density fluctuation in \(\omega_0/\omega_{\text{c}}\), as represented by equation \eqref{eq:density_fluctuation_perp}. This mode induces charge separation in $e^\pm$ plasma. On the other hand, the neutral mode is the second-order density fluctuation in \(\omega_0/\omega_{\text{c}}\), as described by equation \eqref{eq:density_fluctuation_neutral}. This mode does not induce charge separation in $e^\pm$ plasma.

The overall linear growth rate 
for X-mode waves
is determined by whether the charged mode or the neutral mode dominates, depending on the value \proof{\(\frac{8k_{\text{B}}T_{\text{e}}}{m_{\text{e}}c^2} \left( \omega_0 / \omega_{\mathrm{p}} \right)^2(1 + \omega_{\mathrm{p}}^2 / \omega_{\mathrm{c}}^2)\)}. As shown in equation \eqref{eq:Debye_length_ration}, this quantity is the square of the ratio of the Debye length \(\lambda_{\text{De}}\) to the wavelength of the incident wave \(\lambda_0\). When \proof{\(\frac{8k_{\text{B}}T_{\text{e}}}{m_{\text{e}}c^2} \left( \omega_0 / \omega_{\mathrm{p}} \right)^2(1 + \omega_{\mathrm{p}}^2 / \omega_{\mathrm{c}}^2) \gg 1\)}, meaning that the wavelength of the incident wave is shorter than the Debye length, the charged mode is not subject to Debye screening. In this case, as shown by equation \eqref{eq:maximum_growth_rate_magnetic2}, the linear growth rate of induced Compton scattering is determined by the charged mode, which dominates at the order of \(\omega_0/\omega_{\text{c}}\). On the other hand, when \proof{\(\frac{8k_{\text{B}}T_{\text{e}}}{m_{\text{e}}c^2} \left( \omega_0 / \omega_{\mathrm{p}} \right)^2(1 + \omega_{\mathrm{p}}^2 / \omega_{\mathrm{c}}^2) \ll 1\)}, meaning that the wavelength of the incident wave is much longer than the Debye length, the charged mode is subject to Debye screening. In this case, as shown by equation \eqref{eq:maximum_growth_rate_perp}, the linear growth rate of induced Compton scattering is determined by the mode with the higher growth rate, either the charged mode with Debye screening or the neutral mode (the next order of \(\omega_0/\omega_{\text{c}}\)). The Debye screening effect and the charged and neutral modes have been newly identified in this study.
It is also the first time the gyroradius effect has been correctly derived from the kinetic formulation.

We applied our findings to the Galactic FRB 20200428.
Firstly, since the ordinary mode grows quickly, O-mode waves are scattered and cannot propagate \proof{unless the plasma density is low}. This suggests that FRBs are primarily waves polarized perpendicular to the \proof{background }magnetic field.
Secondly, it has been found that, among the modes of induced Compton scattering, the charged mode can become dominant compared to the neutral mode. The linear growth rate of the charged mode can be comparable to the inverse of the FRB duration time, i.e.,
FRBs could escape from the magnetosphere.
The linear growth rates of induced Compton scattering for both the charged and neutral modes heavily depend on the distance from the magnetar's center and the $e^\pm$ plasma density. Therefore, in future works, we need to accurately constrain the emission regions of FRBs by considering the parameter dependencies. 

Our identification of induced Compton scattering associated with three unstable modes—the ordinary mode, the charged mode, and the neutral mode— 
has significant implications for
the polarization states of FRBs. 
Let us consider the polarization of EM waves emitted from a strongly magnetized $e^\pm$ plasma (magnetosphere or outflow),
assuming that the initial waves are unpolarized, i.e., a mixture of X-mode and O-mode.
When induced scattering occurs, the ordinary mode becomes effective first, scattering the electric field component parallel to the \proof{background }magnetic field.
This results in the wave propagating as an X-mode wave, and consequently, the FRB 
can approach
100\% linear polarization. 
Next,
if the charged and neutral modes are also effective, the scattered waves tend to have polarization perpendicular to the background magnetic field. However, since these modes also produce scattered waves with other polarizations, the FRB's degree of linear polarization decreases to less than 100\%. Therefore, if the last scattering surfaces of the charged and neutral modes are located much deeper inside the magnetosphere than that of the ordinary mode, the FRB 
can achieve
100\% linear polarization. If these surfaces are in regions similar to the ordinary mode, the degree of linear polarization 
becomes less than 100\%. This discussion can be consistent with the fact that many observed FRBs exhibit nearly 100\% linear polarization \citep{2018ApJ...863....2G,2018Natur.553..182M,2019MNRAS.488..868O}.

Other parametric instabilities, such as induced Brillouin scattering and induced Raman scattering can be derived considering the magnetized $e^\pm$ plasma effects. The treatment up to the equation for the density fluctuation caused by the ponderomotive potential \eqref{eq:density_fluctuation_pondero} remains the same as this study. Specifically, in $e^\pm$ plasma, induced Brillouin scattering is closely related to induced Compton scattering among parametric instabilities. If both the Stokes and anti-Stokes waves are considered in equation \eqref{eq:energy_momentum_conservation_Compton}, the linear growth rates of parametric modulation instabilities, such as the filamentation instability, can also be derived within the same framework.

A detailed comparison between Particle-in-Cell (PIC) simulations and our theoretical predictions is needed, and we are currently pursuing this \footnote{This research was initiated in response to an unknown suppression effects observed in our group’s numerical simulations of induced Compton scattering. Our formulation has identified that these suppressions are due to the Debye screening and the neutral mode.}.  In plasma, various parametric instabilities, such as induced Brillioin scattering, occur concurrently. Therefore, it is necessary to verify whether each analytic linear growth rate is available using numerical simulation. Moreover, discussing the saturation of growth in the nonlinear regime is also necessary. When the energy of the incident and scattered waves and the plasma becomes comparable, simulations are essential to understand how the system evolves.

The findings of this study are expected to have significant implications for both astrophysics and experimental plasma physics in the future. Several mechanisms have been proposed for the FRBs generations, including not only magnetars but also interacting neutron star binaries
\citep{2020ApJ...893L..26I,2020ApJ...893L..39L,2021ApJ...920...54W}, binary neutron star mergers \citep{2013PASJ...65L..12T}, stellar-mass black holes \citep{2017MNRAS.471L..92K}, 
and supermassive black holes \citep{2016PhRvD..93b3001R}. However, the exact origin of extragalactic FRBs remains unclear. By estimating the regions where FRBs are attenuated due to parametric instability, our findings can potentially constrain their generation sites and progenitors.
They also provide new theoretical insights into 
free electron laser with a guiding magnetic field \citep{1979PhFl...22.1089K,1980PhFl...23.2376F},
experiments on
$e^\pm$ plasma \citep{2024NatCo..15.5029A}, and the experimental verification of plasma phenomena in magnetic fields \citep{2022PhRvE.105b5203Y,2022PhRvE.106b5205M}.

\begin{acknowledgments}
We are deeply grateful to the anonymous referees for their constructive feedback. 
We want to express our gratitude to Yasuhiro Nariyuki for invaluable discussions on our study including the parametric instability of broadband incident waves. We are grateful to Keiichi Maeda, Wataru Ishizaki and Takahiro Tanaka for his invaluable feedback on our discussions. RN is supported by JST SPRING, Grant Number JPMJSP2110. KI is supported by MEXT/JSPS KAKENHI Grant
No.23H01172, 23H05430, 23H04900, 22H00130, 20H01901, 20H01904. KI, SK, and MI is supported by MEXT/JSPS KAKENHI Grant No.20H00158.
\end{acknowledgments}
\appendix
\section{Detailed Derivation of Density Fluctuations Induced by the Ponderomotive Potential in a Magnetic Field}
\label{Ap:pondero_fluctuation}
To derive the density fluctuations induced by the ponderomotive potential, we start with the perturbed Vlasov equation \eqref{eq:Vlasov_pondero},
\begin{equation}
\begin{aligned}
    &\frac{\partial \delta f_{ \pm}}{\partial t}+\bm{v} \cdot \frac{\partial \delta f_{ \pm}}{\partial \bm{r}}+\left(-\frac{1}{m_{e}} \bm{\nabla} \phi_{ \pm} \pm \frac{e}{m_{e}} \bm{E}\right) \cdot \frac{\partial f_{0\pm}}{\partial \bm{v}}\\
    &\pm \frac{e}{m_{\text{e}} c}\left(\bm{v} \times \bm{B}_{0}\right) \cdot \frac{\partial \delta f_{ \pm}}{\partial \bm{v}}=0.
    \label{eq:Vlasov_pondero_appendix}
\end{aligned}
\end{equation}
We express the velocity derivative term of equation \eqref{eq:Vlasov_pondero_appendix} using cylindrical coordinates for the velocity coordinates, given by
\begin{equation}
\bm{v} \equiv\left(v_{\|}, v_{\perp} \cos \varphi, v_{\perp} \sin \varphi\right),
\end{equation}
as follows:
\begin{equation}
\left(\bm{v} \times \bm{B}_{0}\right) \cdot \frac{\partial \delta f_{ \pm}}{\partial \bm{v}}=-B_{0} \frac{\partial \delta f_{ \pm}}{\partial \varphi}.
\end{equation}
Thus, the fluctuation of the distribution function must satisfy the following equation,
\begin{equation}
\begin{aligned}
\frac{\partial \delta f_{ \pm}}{\partial t}
& + \bm{v} \cdot \frac{\partial \delta f_{ \pm}}{\partial \bm{r}}
- \frac{1}{m_{\mathrm{e}}}\left(\bm{\nabla} \phi_{ \pm}\right) \cdot \frac{\partial f_{0 \pm}}{\partial \bm{v}} \\
& + \omega_{\mathrm{c} \mp} \frac{\partial \delta f_{ \pm}}{\partial \varphi} 
\pm \frac{e}{m_{\mathrm{e}}} \bm{E} \cdot \frac{\partial f_{0 \pm}}{\partial \bm{v}} = 0.
\end{aligned}
\label{eq:pondero_Vlasov_henkan}
\end{equation}

Next, we perform a Fourier transform on the distribution function fluctuation \(\delta f_{\pm}\), the ponderomotive potential \(\phi\), and the fluctuating electric field \(\bm{E}\) as equation \eqref{eq:Fourier_transformation_definition2}.
By Fourier transforming equation \eqref{eq:pondero_Vlasov_henkan}, we get
\begin{equation}
\begin{aligned}
-\mathrm{i}(\omega - \bm{k} \cdot \bm{v}) \widetilde{\delta f_{\pm}} 
& + \left(-\mathrm{i} \frac{\bm{k}}{m_{\mathrm{e}}} \widetilde{\phi_{\pm}} 
\pm \frac{e}{m_{\mathrm{e}}} \widetilde{\bm{E}}\right) \cdot \frac{\partial f_{0 \pm}}{\partial \bm{v}} \\
& + \omega_{\mathrm{c} \mp} \frac{\partial \widetilde{\delta f_{\pm}}}{\partial \varphi} = 0.
\end{aligned}
\label{eq:pondero_Vlasov_Fourier}
\end{equation}
Equation \eqref{eq:pondero_Vlasov_Fourier} is an inhomogeneous first-order differential equation with respect to \(\varphi\), which can be solved using the method of variation of constants as
\begin{equation}
\begin{aligned}
\widetilde{\delta f_{\pm}} = &\frac{1}{\omega_{\mathrm{c} \pm}} \int \mathrm{d} \varphi' \exp \left[-\mathrm{i} \left( \frac{\omega - k_{\|} v_{\|}}{\omega_{\mathrm{c}\mp}} \varphi' - k_{\perp} r_{\mathrm{L}\mp} \sin \varphi' \right) \right] \\
& \times \left(-\mathrm{i} \frac{\bm{k}}{m_{\mathrm{e}}} \widetilde{\phi_{\pm}} \pm \frac{e}{m_{\mathrm{e}}} \widetilde{\bm{E}}\right) \cdot \frac{\partial f_{0 \pm}}{\partial \bm{v}} \\
& \times \exp \left[\mathrm{i} \left( \frac{\omega - k_{\|} v_{\|}}{\omega_{\mathrm{c} \mp}} \varphi - k_{\perp} r_{\mathrm{L}\mp} \sin \varphi \right) \right].
\end{aligned}
\end{equation}
Using the Bessel function identity,
\begin{equation}
\text{e}^{ \pm \mathrm{i} z \sin \varphi} = \sum_{\ell=-\infty}^{+\infty} J_{\ell}( \pm z) ~\text{e}^{\mathrm{i} \ell \varphi},
\end{equation}
we can write
\begin{equation}
\begin{aligned}
\widetilde{\delta f_{\pm}} &= \frac{1}{\omega_{\mathrm{c} \pm}} 
\sum_{\ell=-\infty}^{+\infty} \int \mathrm{d} \varphi' \exp \left( -\mathrm{i} \frac{\omega - k_{\|} v_{\|} + \ell \omega_{\mathrm{c}\mp}}{\omega_{\mathrm{c}\mp}} \varphi' \right) \\
&\quad \times J_{\ell}\left(k_{\perp} r_{\mathrm{L}\pm}\right) 
\left(-\mathrm{i} \frac{\bm{k}}{m_{\mathrm{e}}} \widetilde{\phi_{\pm}} \pm \frac{e}{m_{\mathrm{e}}} \widetilde{\bm{E}}\right) 
\cdot \frac{\partial f_{0 \pm}}{\partial \bm{v}} \\
&\quad \times \exp \left[\mathrm{i} \left( \frac{\omega - k_{\|} v_{\|}}{\omega_{\mathrm{c} \mp}} \varphi 
- k_{\perp} r_{\mathrm{L}\mp} \sin \varphi \right) \right].
\end{aligned}
\label{eq:distribution_fluctuation_teisuuhennka}
\end{equation}
Assuming that the electric field fluctuation, induced by the beating wave, only has a longitudinal component, we have
\begin{equation}
\widetilde{\bm{E}} = -\frac{4 \pi \mathrm{i}}{k^{2}} \bm{k} \widetilde{\rho},
\end{equation}
where $\widetilde{\rho}$ is defined as the Fourier coefficient of charge density.
Considering the integral part of equation \eqref{eq:distribution_fluctuation_teisuuhennka}, we decompose the velocity derivative into components parallel and perpendicular to the \proof{background }magnetic field,
\begin{equation}
\begin{aligned}
&\sum_{\ell=-\infty}^{+\infty}\int \mathrm{d} \varphi' \exp \left( -\mathrm{i} \frac{\omega - k_{\|} v_{\|} + \ell \omega_{\mathrm{c}\mp}}{\omega_{\mathrm{c}\mp}} \varphi' \right)  J_{\ell}\left(k_{\perp} r_{\mathrm{L}\pm}\right) \\
&\quad \times (-\mathrm{i}) \left( \frac{\widetilde{\phi_{\pm}}}{m_{\mathrm{e}}} \pm \frac{4 \pi e}{m_{\mathrm{e}} k^{2}} \widetilde{\rho} \right) \bm{k} \cdot \frac{\partial f_{0 \pm}}{\partial \bm{v}} \\
&= \sum_{\ell=-\infty}^{+\infty} \frac{\omega_{\mathrm{c}\mp} J_{\ell}\left(k_{\perp} r_{\mathrm{L}\pm}\right)}{\omega - k_{\|} v_{\|} + \ell \omega_{\mathrm{c}\mp}} \exp \left( -\mathrm{i} \frac{\omega - k_{\|} v_{\|} + \ell \omega_{\mathrm{c}\mp}}{\omega_{\mathrm{c}\mp}} \varphi \right) \\
&\quad \times \left( \frac{\widetilde{\phi_{\pm}}}{m_{\mathrm{e}}} \pm \frac{4 \pi e}{m_{\mathrm{e}} k^{2}} \widetilde{\rho} \right) k_{\|} \frac{\partial f_{0 \pm}}{\partial v_{\|}} \\
&- \sum_{\ell=-\infty}^{+\infty} \int \mathrm{d} \varphi' \exp \left( -\mathrm{i} \frac{\omega - k_{\|} v_{\|} + \ell \omega_{\mathrm{c}\mp}}{\omega_{\mathrm{c}\mp}} \varphi' \right) \\
&\times \frac{J_{\ell+1}\left(k_{\perp} r_{\mathrm{L}\pm}\right) + J_{\ell-1}\left(k_{\perp} r_{\mathrm{L}\pm}\right)}{2} \mathrm{i} \left( \frac{\widetilde{\phi_{\pm}}}{m_{\mathrm{e}}} \pm \frac{4 \pi e}{m_{\mathrm{e}} k^{2}} \widetilde{\rho} \right) k_{\perp} \frac{\partial f_{0 \pm}}{\partial v_{\perp}},
\end{aligned}
\label{eq:Bessel_function_no_yabai_henkei}
\end{equation}
where we used the following relation,
\begin{equation}
\bm{k} \cdot \frac{\partial f_{0\pm}}{\partial \bm{v}} = k_{\|} \frac{\partial f_{0\pm}}{\partial v_{\|}} + k_{\perp} \cos \varphi^{\prime} \frac{\partial f_{0\pm}}{\partial v_{\perp}},
\end{equation}
along with the identity $\cos\varphi^\prime = \left( \text{e}^{\text{i}\varphi^{\prime}} + \text{e}^{-\text{i}\varphi^\prime} \right)/2$.
Using the Bessel function identity,
\begin{equation}
J_{\ell+1}(z) + J_{\ell-1}(z) = \frac{2 \ell}{z} J_{\ell}(z),
\end{equation}
and the following differential operator definition,
\begin{equation}
k_{\|} \frac{\partial f_{0 \pm}}{\partial v_{\|}} + \frac{\ell}{r_{\mathrm{L}\pm}} \frac{\partial f_{0 \pm}}{\partial v_{\perp}} \equiv \bm{k} \cdot \frac{\partial f_{0 \pm}}{\partial \bm{v}^{\ast}},
\end{equation}
the fluctuation of the distribution function can be calculated as
\begin{equation}
\begin{aligned}
\widetilde{\delta f_{\pm}} = &-\sum_{\ell=-\infty}^{+\infty} \sum_{m=-\infty}^{+\infty} 
\frac{J_{\ell}\left(k_{\perp} r_{\mathrm{L}\pm}\right) J_{m}\left(k_{\perp} r_{\mathrm{L}\pm}\right)}
{\omega - k_{\|} v_{\|} + \ell \omega_{\mathrm{c}\mp}} \\
& \times \text{e}^{-\mathrm{i}(\ell-m) \varphi} \left( \frac{\widetilde{\phi_{\pm}}}{m_{\mathrm{e}}} 
\pm \frac{4 \pi e}{m_{\mathrm{e}} k^{2}} \widetilde{\rho} \right) 
\bm{k} \cdot \frac{\partial f_{0 \pm}}{\partial \bm{v}^{\ast}}.
\end{aligned}
\end{equation}
The density fluctuation can then be obtained by integrating $\widetilde{\delta f_{\pm}}$ over velocity space,
\begin{equation}
\begin{aligned}
\widetilde{\delta n_{\pm}} &= n_{\mathrm{e}0} \int \widetilde{\delta f_{\pm}} \, \mathrm{d}^{3} \bm{v} \\
&= -n_{\mathrm{e}0} \left( \frac{\widetilde{\phi_{\pm}}}{m_{\mathrm{e}}} 
\pm \frac{4 \pi e}{m_{\mathrm{e}} k^{2}} \widetilde{\rho} \right) \\
&\quad \times \sum_{\ell,m} \int \mathrm{d}^{3} \bm{v} 
\frac{J_{\ell}\left(k_{\perp} r_{\mathrm{L}\pm}\right) J_{m}\left(k_{\perp} r_{\mathrm{L}\pm}\right)}
{\omega - k_{\|} v_{\|} + \ell \omega_{\mathrm{c}\mp}} \\
&\quad \times \text{e}^{-\mathrm{i}(\ell-m) \varphi} 
\bm{k} \cdot \frac{\partial f_{0 \pm}}{\partial \bm{v}^{\ast}},
\end{aligned}
\end{equation}
where each variable in the velocity integral is integrated over the following ranges,
\begin{equation}
\int \mathrm{d}^{3} \bm{v} = \int_{0}^{\infty} v_{\perp} \mathrm{d} v_{\perp} \int_{-\infty}^{+\infty} \mathrm{d} v_{\|} \int_{0}^{2 \pi} \mathrm{d} \varphi.
\end{equation}
For the integral over the angle variable, we use the following relation,
\begin{equation}
\int_{0}^{2\pi} \text{e}^{-\mathrm{i}(\ell-m) \varphi} \mathrm{d}\varphi = \begin{cases}
2\pi & \text{if } \quad \ell=m, \\
0 & \text{if } \quad \ell \neq m.
\end{cases}
\end{equation}
Thus, we can write
\begin{equation}
\begin{aligned}
\widetilde{\delta n_{\pm}} = &-n_{\mathrm{e}0} \left( \frac{\widetilde{\phi_{\pm}}}{m_{\mathrm{e}}} 
\pm \frac{4 \pi e}{m_{\mathrm{e}} k^{2}} \widetilde{\rho} \right) \\
&\times \sum_{\ell} \int \mathrm{d}^{3} \bm{v} 
\frac{J_{\ell}^2\left(k_{\perp} r_{\mathrm{L}\pm}\right)}
{\omega - k_{\|} v_{\|} + \ell \omega_{\mathrm{c}\mp}} \,
\bm{k} \cdot \frac{\partial f_{0 \pm}}{\partial \bm{v}^{\ast}}.
\end{aligned}
\end{equation}
Defining the longitudinal electric susceptibility as equation \eqref{eq:longitudinal_electric_susceptibility_induced_Compton},
the density fluctuation can be expressed as
\begin{equation}
\widetilde{\delta n_{\pm}} = -\frac{n_{\mathrm{e}0}}{m_{\mathrm{e}}} \widetilde{\phi_{\pm}} \sum_{\ell} \int \mathrm{d}^{3} \bm{v} \frac{J_{\ell}^2\left(k_{\perp} r_{\mathrm{L}\pm}\right)}{\omega - k_{\|} v_{\|} + \ell \omega_{\mathrm{c}\mp}} \bm{k} \cdot \frac{\partial f_{0 \pm}}{\partial \bm{v}^{\ast}} \mp \frac{H_{\pm}}{e} \widetilde{\rho}.
\label{eq:density_fluctuation_pondero_totyuu}
\end{equation}

The charge density can be expressed as a combination of density fluctuations, allowing for the self-consistent determination of the density fluctuation,
\begin{equation}
\widetilde{\rho} = e \widetilde{\delta n_{+}} - e \widetilde{\delta n_{-}}.
\end{equation}
Using the longitudinal dielectric constant \eqref{eq:longitudinal_dielectric_constant_induced_Compton}, the charge density can be expressed as
\begin{equation}
\begin{aligned}
\widetilde{\rho} = &-e \frac{n_{\mathrm{e}0}}{m_{\mathrm{e}} \varepsilon_{\mathrm{L}}} \left\{ \widetilde{\phi}_{+} \sum_{\ell} \int \mathrm{d}^{3} \bm{v} \frac{J_{\ell}^{2}\left(k_{\perp} r_{\mathrm{L}+}\right)}{\omega - k_{\|} v_{\|} + \ell \omega_{\mathrm{c}-}} \bm{k} \cdot \frac{\partial f_{0+}}{\partial \bm{v}^{*}} \right. \\
& \left. - \widetilde{\phi_{-}} \sum_{\ell} \int \mathrm{d}^{3} \bm{v} \frac{J_{\ell}^{2}\left(k_{\perp} r_{\mathrm{L}-}\right)}{\omega - k_{\|} v_{\|} + \ell \omega_{\mathrm{c}+}} \bm{k} \cdot \frac{\partial f_{0-}}{\partial \bm{v}^{*}} \right\}.
\end{aligned}
\label{eq:charge_density_pondero}
\end{equation}
By substituting equation \eqref{eq:charge_density_pondero} into equation \eqref{eq:density_fluctuation_pondero_totyuu}, the density fluctuation \eqref{eq:density_fluctuation_pondero} can be derived as follows:
\begin{equation}
\begin{aligned}
\widetilde{\delta n_{\pm}} &= -\frac{n_{\mathrm{e}0}}{m_{\mathrm{e}}} \left\{ \widetilde{\phi_{\pm}} \sum_{\ell} \int \mathrm{d}^{3} \bm{v} \frac{J_{\ell}^{2}\left(k_{\perp} r_{\mathrm{L}\pm}\right)}{\omega - k_{\|} v_{\|} + \ell \omega_{\mathrm{c}\mp}} \bm{k} \cdot \frac{\partial f_{0\pm}}{\partial \bm{v}^{*}} \right\} \\
&\quad \pm \frac{n_{\mathrm{e}0} H_{\pm}}{m_{\mathrm{e}} \varepsilon_{\mathrm{L}}} \left\{ \widetilde{\phi_{+}} \sum_{\ell} \int \mathrm{d}^{3} \bm{v} \frac{J_{\ell}^{2}\left(k_{\perp} r_{\mathrm{L}+}\right)}{\omega - k_{\|} v_{\|} + \ell \omega_{\mathrm{c}-}} \bm{k} \cdot \frac{\partial f_{0+}}{\partial \bm{v}^{*}} \right. \\
&\quad \left. - \widetilde{\phi_{-}} \sum_{\ell} \int \mathrm{d}^{3} \bm{v} \frac{J_{\ell}^{2}\left(k_{\perp} r_{\mathrm{L}-}\right)}{\omega - k_{\|} v_{\|} + \ell \omega_{\mathrm{c}+}} \bm{k} \cdot \frac{\partial f_{0-}}{\partial \bm{v}^{*}} \right\}.
\end{aligned}
\label{eq:density_fluctuation_pondero_Appendix}
\end{equation}

\section{Scattering with density fluctuations having wave vector perpendicular to the background magnetic field}
\label{Ap:wave_vector_perpendicular}
We derive induced Compton scattering when the density fluctuations have a wave vector component perpendicular to the background magnetic field.
We can use the asymptotic expansion of the plasma dispersion function for large arguments \citep{1961pdf..book.....F},
\begin{equation}
Z(\xi)=\mathrm{i} \sqrt{\pi} \mathrm{e}^{-\xi^{2}}\sigma-\frac{1}{\xi}\left(1+\frac{1}{2 \xi^{2}}+\frac{3}{4 \xi^{4}}+\cdots\right)\quad \text{for}~\xi\gg 1,
\end{equation}
where $\sigma=0,1,2$ for $\text{Im}\,\xi>0,=0,<0$, respectively.
Expanding up to the first order in $k_{\|} v_{\text{th}} / |\omega|$, we get
\begin{equation}
Z\left(\frac{\omega+\mathrm{i}\epsilon}{k_{\|} v_{\text{th}}}\right) \simeq  - \frac{k_{\|} v_{\text{th}}}{\omega}.
\label{eq:plasma_dispersion_function_dekai_tenkai}
\end{equation}
We can calculate the longitudinal electric susceptibility \eqref{eq:longitudinal_electric_susceptibility} and the integral in the dispersion relation \eqref{eq:dispersion_relation_pondero_drift} for the scattered wave by using equation \eqref{eq:plasma_dispersion_function_dekai_tenkai}. For sufficiently \rei{large} \proof{strong background} magnetic fields where $k_{\perp} v_{\mathrm{th}} / \omega_{\mathrm{c}} \ll 1$, considering only the $\ell = 0$ term in the modified Bessel function's infinite sum, we get
\begin{equation}
\begin{aligned}
H(\bm{k}, \omega) &\simeq \frac{\omega_{\text{p}}^2}{k^2v_{\text{th}}^2}\left[1+\frac{\omega}{k_{\|} v_{\text{th}}}
\left\{1-\frac{1}{2}\left(\frac{k_{\perp}v_{\text{th}}}{\omega_{\text{c}}}\right)^2\right\} \right. \\
&\quad \left. \times Z\left(\frac{\omega}{k_{\|} v_{\text{th}}}\right) \right] \\
&\overset{k_{\|}\rightarrow 0}{\longrightarrow} \frac{1}{2} \frac{\omega_{\mathrm{p}}^{2}}{\omega_{\mathrm{c}}^{2}},    
\end{aligned}
\label{eq:longitudinal_electric_susceptibility_perp}
\end{equation}
\begin{equation}
\begin{aligned}
   \sum_{\ell=-\infty}^{+\infty} \int \dd^{3} \bm{v} \frac{J_{\ell}^{2}\left(k_{\perp} r_{\mathrm{L}}\right)\bm{k} \cdot \frac{\partial f_{0}}{\partial \bm{v}^{*}}}{\omega-k_{\|} v_{\|}-\ell \omega_{\mathrm{c}}}&=\frac{m_{\text{e}}k^2}{4\pi e^2n_{\text{e}0}}H\\
   &\overset{k_{\|}\rightarrow 0}{\longrightarrow}\frac{k_{\perp}^{2}}{\omega_{\mathrm{c}}^{2}}.
\end{aligned}
\label{eq:sekibun_perp}
\end{equation}
Therefore, the dispersion relation for the scattered wave \eqref{eq:dispersion_relation_pondero_drift} becomes
\begin{equation}
c^{2} k_{1}^{2} - \omega_{1}^{2}+\omega_{\mathrm{p}}^{2} \frac{\omega_{1}^{2}}{\omega_{1}^{2}-\omega_{\mathrm{c}}^{2}} \simeq \frac{c^{2}}{4} \left(\frac{\omega_{0}}{\omega_{\mathrm{c}}}\right)^{2} \left(\bar{a}_{\mathrm{e}} \omega_{\mathrm{p}}\right)^{2} \frac{\omega_{\mathrm{c}}^{2}}{\omega_{\mathrm{c}}^{2} + \omega_{\mathrm{p}}^{2}} \frac{k_{\perp}^{2}}{\omega_{\mathrm{c}}^{2}}.
\label{eq:dispersion_relation_stable}
\end{equation}
According to equation \eqref{eq:dispersion_relation_stable}, the imaginary part of \(\omega_1\) is zero, \(\text{Im}~\omega_1 = 0\), indicating that this is a stable mode. Thus, density fluctuations propagating perpendicular to the \proof{background }magnetic field do not produce induced Compton scattering. Note that if the \proof{background }magnetic field is not large enough, considering finite $\ell$ terms in the Bessel function's infinite sum could allow the incident wave to resonate with Bernstein waves, but this case is not considered in this paper.
\nocite{*}
\newpage
\bibliographystyle{apsrev4-2}
\bibliography{apssamp}

\end{document}